\documentclass [12pt,eqsecnum,amsfonts,aps]{revtex4}

\input epsf
\usepackage{graphicx}

\newcommand{\beq}{\begin{equation}}
\newcommand{\eeq}{\end{equation}}
\newcommand{\beqs}{\begin{eqnarray}}
\newcommand{\eeqs}{\end{eqnarray}}

\begin{document}

\baselineskip 6.0mm

\title{The Structure of Chromatic Polynomials of 
Planar Triangulation Graphs and Implications for Chromatic Zeros and 
Asymptotic Limiting Quantities} 

\bigskip

\author{Robert Shrock}
%\email{robert.shrock@stonybrook.edu}

\author{Yan Xu}
%\email{yan.xu@stonybrook.edu}

\affiliation{ C. N. Yang Institute for Theoretical Physics \\
Stony Brook University \\ 
Stony Brook, N. Y. 11794}

\begin{abstract}

We present an analysis of the structure and properties of chromatic polynomials
$P(G_{pt,\vec m},q)$ of one-parameter and multi-parameter families of planar
triangulation graphs $G_{pt,\vec m}$, where ${\vec m} = (m_1,...,m_p)$ is a
vector of integer parameters.  We use these to study the ratio of
$|P(G_{pt,\vec m},\tau+1)|$ to the Tutte upper bound $(\tau-1)^{n-5}$, where
$\tau=(1+\sqrt{5} \ )/2$ and $n$ is the number of vertices in $G_{pt,\vec
m}$. In particular, we calculate limiting values of this ratio as $n \to
\infty$ for various families of planar triangulations. We also use our
calculations to study zeros of these chromatic polynomials.  We study a large
class of families $G_{pt,\vec m}$ with $p=1$ and $p=2$ and show that these have
a structure of the form $P(G_{pt,m},q) = c_{_{G_{pt}},1}\lambda_1^m +
c_{_{G_{pt}},2}\lambda_2^m + c_{_{G_{pt}},3}\lambda_3^m$ for $p=1$, where
$\lambda_1=q-2$, $\lambda_2=q-3$, and $\lambda_3=-1$, and $P(G_{pt,\vec m},q) =
\sum_{i_1=1}^3 \sum_{i_2=1}^3 c_{_{G_{pt}},i_1 i_2}
\lambda_{i_1}^{m_1}\lambda_{i_2}^{m_2}$ for $p=2$. We derive properties of the
coefficients $c_{_{G_{pt}},\vec i}$ and show that $P(G_{pt,\vec m},q)$ has a
real chromatic zero that approaches $(1/2)(3+\sqrt{5} \ )$ as one or more of
the $m_i \to \infty$.  The generalization to $p \ge 3$ is given. Further, we
present a one-parameter family of planar triangulations with real zeros that
approach 3 from below as $m \to \infty$.  Implications for the ground-state
entropy of the Potts antiferromagnet are discussed.

\end{abstract}


\maketitle

\newpage

\pagestyle{plain}
\pagenumbering{arabic}

\section{Introduction}
\label{intro}

In this paper we present exact results on the structure and properties of
chromatic polynomials $P(G_{pt,\vec m},q)$ of one-parameter and multi-parameter
families of planar triangulation ($pt$) graphs $G_{pt,\vec m}$, where ${\vec m}
= (m_1,...,m_p)$ is a vector of integer parameters.  Our results substantially
generalize our previous study in \cite{tub}. In standard notation we let
$G=(V,E)$ be a graph with vertex and edge sets $V$ and $E$, and denote the
number of vertices and edges as $n=n(G)=|V|$ and $e(G)=|E|$, respectively. The
resultant set of faces is denoted $F(G)$, with cardinality $f(G)=|F(G)|$.
We recall the definitions of a planar graph $G$ as one that can be drawn in a
plane without any crossing edges, and a triangulation as a graph all of whose
faces are triangles.  For an arbitrary graph $G$, the chromatic polynomial
$P(G,q)$ enumerates the number of ways of associating $q$ colors with the
vertices of $G$, subject to the constraint that adjacent vertices have
different colors (called a proper $q$-coloring of $G$) \cite{dong}.  The
minimum number of colors for a proper $q$-coloring of a graph $G$ is the
chromatic number of $G$, denoted $\chi(G)$. Without loss of generality, we
restrict here to graphs $G$ that are connected, have no multiple edges, and
have no loops (where a loop is defined as an edge that connects a vertex back
to itself).  Multiple edges are also excluded by our restriction here to
triangulation graphs, since they produce faces that are not triangles.  An
important identity is $P(G,q)=Z_{PAF}(G,q,0)$, where $Z_{PAF}(G,q,T)$ denotes
the partition function of the Potts antiferromagnet (PAF) on the graph $G$ at
temperature $T$. Because of this identity, properties of chromatic polynomials
are of interest both for mathematical graph theory and for statistical physics.
A particularly significant feature of the Potts antiferromagnet is the fact
that it generically exhibits nonzero ground-state (i.e., zero-temperature)
entropy per vertex for sufficiently large $q$ on a given graph. 

The chromatic polynomial of a graph $G$ may be computed via the
deletion-contraction relation $P(G,q)=P(G-e,q)-P(G/e,q)$, where $G-e$ denotes
$G$ with an edge $e$ deleted and $G/e$ denotes the graph obtained by deleting
$e$ and identifying the vertices that it connected.  From this follows the
cluster formula $P(G,q) = \sum_{G' \subseteq G} (-1)^{e(G')} \ q^{k(G')}$,
where $G'=(V,E')$ with $E' \subseteq E$ and $k(G')$ denotes the number of
connected components of $G'$.  Via this relation, one can generalize $q$ from
positive integers to real and complex numbers, as is necessary when analyzing
zeros of $P(G,q)$, called chromatic zeros.  For a general graph $G$ or family
of graphs $G_{\vec m}$, the calculation of the chromatic polynomial takes an
exponentially long time.  Planar triangulations comprise a class that is
particularly amenable to analysis, as we shall show. Since a triangulation
graph $G_t$ (whether planar or not) contains at least one triangle, $P(G_t,q)$
always contains the factor $P(K_3,q) = q(q-1)(q-2)$. (Here, $K_n$ is the
complete graph with $n$ vertices, defined as the graph such that each vertex is
adjacent to every other vertex by an edge.)

An interesting upper bound on an evaluation of a chromatic polynomial of a
planar triangulation was derived by Tutte \cite{tutte70}
(see also \cite{tutte69},\cite{tutte84}), namely
\beq
0 < |P(G_{pt}, \tau+1)| \le U(n(G_{pt})) \ ,
\label{tub}
\eeq
where $\tau=(1+\sqrt{5} \, )/2$ is the golden ratio and
\beq
U(n) = \tau^{5-n} = (\tau-1)^{n-5} \ .
\label{u}
\eeq
As in \cite{tub}, it is natural to define the ratio
\beq
r(G_{pt}) \equiv \frac{|P(G_{pt},\tau+1)|}{U(n(G_{pt}))} \ ,
\label{rpg}
\eeq
which is bounded above by 1.  In this paper we shall present calculations of
this ratio, and its limit as $n \to \infty$, for a number of different families
of planar triangulations. In \cite{tub} we showed that if $p=1$, then, with
$\vec m \equiv m$, if $P(G_{pt,m},q)$ involves only a single power of a
polynomial, $(\lambda_{G_{pt}})^m$ as in (\ref{pgform1}) below, it follows that
$r(G_{pt,m})$ approaches zero exponentially rapidly as $m \to \infty$. In
\cite{tub} we also constructed one-parameter families $G_{pt,m}$ for which
$P(G_{pt,m},q)$ is a sum of powers of certain terms $\lambda_i$ with $i=1,2,3$,
given below in (\ref{lamform}), then $r(G_{pt,m})$ may approach a finite
nonzero constant as $m \to \infty$.  We generalize these results here to the
families $G_{pt,\vec m}$ with $p \ge 2$.  We also exhibit a $p=1$ family,
denoted $F_m$, for which $P(F_m,q)$ is a sum of nonpolynomial $\lambda_{F,i}$s,
$i=1,2,3$, and show that for this family, $r(F_m)$ decreases to zero
(exponentially rapidly) as $m \to \infty$.  It should be noted that the Tutte
upper bound is satisfied as an equality by the triangle, $K_3$, but for planar
triangulations with higher $n$, the upper bound is realized as a strict
inequality.

Part of our work concerns zeros of chromatic polynomials $P(G_{pt,\vec m},q)$
for families of planar triangulations $G_{pt,\vec m}$, i.e., chromatic zeros of
these graphs. An interesting aspect of this study relates to an empirical
observation made by Tutte in connection with his upper bound, namely that for a
planar triangulation $G_{pt}$, $P(G_{pt},q)$ typically has a real zero close to
$q=\tau+1 = 2.6180339887...$.  This observation has been somewhat mysterious
over the years, for at least two reasons.  First, Tutte's upper bound
(\ref{tub}) does not imply that a $P(G_{pt},q)$ need have a zero near to
$\tau+1$. Second, it is known that for an arbitrary (loopless) graph $G$,
$P(G,q)$ cannot vanish exactly at $q=\tau+1=(3+\sqrt{5} \ )/2$. We recall the
proof. If $P(G,(3+\sqrt{5} \ )/2)$ were zero, then, since $P(G,q)$ is a
polynomial in $q$, it would have the factor $[q-(3+\sqrt{5} \ )/2]$. But since
$P(G,q)$ has rational coefficients (actually, integer coefficients, but all we
use here is the property that they are rational), it would therefore also have
to contain a factor involving the algebraic conjugate root, namely
$[q-(3-\sqrt{5} \ )/2]$.  However, this would imply that $P(G,q)$ would also
vanish at $q=(3-\sqrt{5} \ )/2 = 0.381966...$, but this is impossible, since
this point lies in an interval (0,1) where $P(G,q)$ cannot vanish
\cite{woodall77,rtrev}.  As part of our analysis here, we shed some light on
this mystery by constructing families of planar triangulation graphs
$G_{pt,\vec m}$ with $p=1$ and $p=2$ for which the chromatic polynomials have
the respective forms $P(G_{pt,m},q) = c_{_{G_{pt}},1}\lambda_1^m +
c_{_{G_{pt}},2}\lambda_2^m + c_{_{G_{pt}},3}\lambda_3^m$ for $p=1$ (with $m_1
\equiv m$), where $\lambda_1=q-2$, $\lambda_2=q-3$, and $\lambda_3=-1$, and
$P(G_{pt,\vec m},q) = \sum_{i_1=1}^3 \sum_{i_2=1}^3 c_{_{G_{pt}},i_1 i_2}
\lambda_{i_1}^{m_1}\lambda_{i_2}^{m_2}$ for $p=2$. We derive properties of the
coefficients $c_{_{G_{pt}},\vec i}$ and show that $P(G_{pt,\vec m},q)$ has a
real chromatic zero that approaches $(1/2)(3+\sqrt{5} \ )$ as one or more of
the $m_i \to \infty$.  We give the generalization of this result to $p \ge
3$. Further, we construct a one-parameter family of planar triangulations of
this type with real zeros that approach 3 from below as $m \to \infty$.

\section{General Properties of One-Parameter Families of Planar 
Triangulations}
\label{recursive}

\subsection{General} 

We have constructed and studied various one-parameter families of planar
triangulations $G_{pt,m}$ that can be built up in an interative (recursive)
manner. (Here and below, for families where ${\vec m}$ is one-dimensional, we
set $m_1 \equiv m$ to simplify the notation.)  In this section we derive
general properties of the chromatic polynomials of these families of planar
triangulations. For our families, the number of vertices is linearly related to
$m$,
\beq
n(G_{pt,m}) = \alpha \, m + \beta \ , 
\label{nmrel}
\eeq
where $\alpha$ and $\beta$ are constants that depend on the type of family. We
recall the Euler relation $|V(G)|-|E(G)|+|F(G)|=\chi_E=2$ for a graph $G$
embedded on a surface of genus 0, such as the plane, where $\chi_E$ is the
Euler characteristic.  In general, for a planar graph each of whose faces has
$p$ sides, $n(G)$, $e(G)$, and $f(G)$ satisfy the relations
$e(G)=p(n(G)-2)/(p-2)$ and $f(G)=2(n(G)-2)/(p-2)$.  For the case of interest
here, namely planar triangulation graphs, where each face is a triangle, it
follows that
\beq
e(G_{pt}) = 3(n(G_{pt})-2)
\label{entri}
\eeq
and
\beq
f(G_{pt}) = 2(n(G_{pt})-2) \ ,
\label{fntri}
\eeq
so that $e(G_{pt})=(3/2)f(G_{pt})$.

In our present study we will make use of several results that we derived in
\cite{tub}.  First, since $U(n) \to 0$ as $n \to \infty$ and since $m$ is
proportional to $n$, it follows that for these families of planar
triangulations,
\beq
\lim_{m \to \infty} P(G_{pt,m},\tau+1) = 0 \ . 
\label{ptp1limit}
\eeq
Second, given the upper bound (\ref{tub}) and the fact that $U(n)$ approaches
zero exponentially fast as $n \to \infty$, it follows that 
\beq
P(G_{pt,m},\tau+1) \quad {\rm approaches \ \ zero \ \ exponentially \ \ fast
  \ \ as} \ \ m \to \infty . 
\label{ptp1approach}
\eeq

We recall two definitions that apply to any graph: (i) the degree
$d(v_i)$ of a vertex $v_i \in V$ is the number of edges that connect to it,
and (ii) a $k$-regular graph is a graph for which all vertices have degree $k$.
Since a triangulation graph $G_t$ is not, in general, $k$-regular, it is useful
to define an effective vertex degree in the limit $|V| \to \infty$.  For this
purpose, we introduce, as in our earlier work, the notation $\{G\}$ for the
formal limit $n \to \infty$ of a family of graphs $G$. We define
\beqs
d_{eff} (\{ G \}) & = & \lim_{|V| \to \infty} \frac{2|E|}{|V|} \cr\cr
                       & = & \frac{\sum_i n_i d_i}{|V|} \ ,
\label{deltaeff}
\eeqs
where for a given $G$, $n_i$ denotes the number of vertices with degree
$d_i$ and $n(G) \equiv |V|$.  Substituting (\ref{entri}) in
(\ref{deltaeff}), we obtain 
\beq
d_{eff} (\{ G_{pt} \}) = 6 \ . 
\label{deltaefft}
\eeq
We will use this below, in Sect. \ref{gsentropy}.

\subsection{Families with $P(G_{pt,m},q)$ Consisting of a Power of a Single
  Polynomial} 

There are several ways of constructing one-parameter families of planar
triangulations.  One method that we have used is the following, which produces
families for which the chromatic polynomial involves a single power of a
polynomial in $q$. Start with a basic graph $G_{pt,1}$, drawn in the usual
explicitly planar manner.  The outer edges of this graph clearly form a
triangle, $K_3$.  Next pick an interior triangle in $G_{pt,1}$ and place a copy
of $G_{pt,1}$ in this triangle so that the intersection of the resultant graph
with the original $G_{pt,1}$ is the triangle chosen.  Denote this as
$G_{pt,2}$. Continuing in this manner, one constructs $G_{pt,m}$ with $m \ge
3$.  The chromatic polynomial $P(G_{pt,2},q)$ is calculated from
$P(G_{pt,1},q)$ by using the $s=3$ special case of the complete-graph 
intersection theorem.  This theorem states that if for two
graphs $G$ and $H$ (which are not necessarily planar or triangulations), the
intersection $G \cap H = K_s$ for some $s$, then
$P(G \cup H,q) = P(G,q)P(H,q)/P(K_s,q)$. (Note that $P(K_s,q) = 
\prod_{j=0}^{s-1}(q-j)$.) 

It follows that for planar triangulations formed in this interative manner,
the chromatic polynomial has the form
\beq
P(G_{pt,m},q) = c_{G_{pt}} \, (\lambda_{G_{pt}})^m \ .
\label{pgform1}
\eeq
where the coefficient $c_{G_{pt}}$ and the term $\lambda_{G_{pt}}$ are
polynomials in $q$ that do not depend on $m$.  Here and below, it is implicitly
understood that $m \ge m_{min}$, where $m_{min}$ is the minimal value of $m$
for which the family $G_{pt,m}$ is well defined.  Since $G_{pt}$ contains at
least one triangle, $K_3$, the coefficient $c_{G_{pt}}$ contains (and may be
equal to) $P(K_3,q)=q(q-1)(q-2)$.  The chromatic number of $G_{pt}$ may be 3 or
4. In the case of a planar triangulation which is a strip of the triangular
lattice of length $m$ vertices with cylindrical boundary conditions, to be
discussed below, an alternate and equivalent way to construct the $(m+1)$'th
member of the family is simply to add a layer of vertices to the strip at one
end.

\subsection{Families with $P(G_{pt,m},q)$ Consisting of Powers of Several 
Functions} 

We have also devised methods to obtain families of planar triangulations
$G_{pt,\vec m}$ with the property that the chromatic polynomial is a sum of
more than one power of a function of $q$.  We begin with the simplest case,
$p=1$, i.e., one-parameter families and then discuss families with $p \ge
2$. For one-parameter families, we find the general structure 
\beq 
P(G_{pt,m},q) = \sum_{j=1}^{j_{max}} c_{_{G_{pt}},j} \, 
(\lambda_{G_{pt},j})^m \ ,
\label{pgform}
\eeq
where $m \ge m_{min}$ and the $c_{_{G_{pt}},j}$ and $\lambda_{G_{pt},j}$ are
certain coefficients and functions depending on $q$ but not on $m$. Here we use
the label $G_{pt}$ to refer to the general family of planar triangulations
$G_{pt,m}$.  We will describe these methods below.  Parenthetically, we recall
that the form (\ref{pgform}) is a general one for one-parameter recursive
families of graphs, whether or not they are planar triangulations \cite{bds},
\cite{w}.  For (\ref{pgform}) evaluated at a given value $q=q_0$, as $m \to
\infty$, and hence $n \to \infty$, the behavior of $P(G_{pt,m},q)$ is
controlled by which $\lambda_{G_{pt},j}$ is dominant at $q=q_0$, i.e., which of
these has the largest magnitude $|\lambda_{G_{pt},j}(q_0)|$.  For our purposes,
a $q_0$ of major interest is $\tau+1$, since the Tutte upper bound (\ref{tub})
applies for this value. We denote the $\lambda_{G_{pt},j}$ that is dominant at
$q=\tau+1$ as $\lambda_{G_{pt},dom}$. Clearly, if $P(G_{pt,m},q)$ involves only
a single power, as in (\ref{pgform1}), then
$\lambda_{G_{pt},dom}=\lambda_{G_{pt}}$.

As in earlier works \cite{w,hs,a}, it can be convenient to obtain the chromatic
polynomials $P(G_{pt,m},q)$ via a Taylor series expansion, in an auxiliary
variable $x$, of a generating function $\Gamma(G_{pt},q,x)$.  Below, we will
have occasion to use this method for the family $F_m$ (see (\ref{gamtaylor})).
Both the form (\ref{pgform}) and the expression via a generating function are
equivalent to the property that $P(G_{pt,m},q)$ satisfies a recursion relation,
for $m \ge j_{max}+m_{min}$:
\beq
P(G_{pt,m},q) + \sum_{j=1}^{j_{max}} b_{G_{pt},j} \, P(G_{pt,m-j},q) = 0 \ , 
\label{recursionrel}
\eeq
where the $b_{G_{pt},j}$, $j=1,...j_{max}$, are given by 
\beq
1 + \sum_{j=1}^{j_{max}} b_j x^j = \prod_{j=1}^{j_{max}}
(1-\lambda_{G_{pt},j} \, x) \ .
\label{blam}
\eeq
Thus, 
\beq
b_{G_{pt},1} = -\sum_{j=1}^{j_{max}} \lambda_{G_{pt},j} \ , 
\label{b1}
\eeq
\beq
b_{G_{pt},2} = \sum_{j=1, \ k=1, \ j \ne k}^{j_{max}} \lambda_{G_{pt},j}
 \lambda_{G_{pt},k} \ , 
\label{b2}
\eeq
and so forth, up to 
\beq
b_{G_{pt},j_{max}} = (-1)^{j_{max}} \, 
\prod_{j=1}^{j_{max}} \lambda_{G_{pt},j} \ . 
\label{bjmax}
\eeq

\subsection{Asymptotic Behavior as $m \to \infty$} 

In \cite{tub} we discussed the asymptotic behavior of the chromatic polynomials
as $m \to \infty$. In both the cases of Eq. (\ref{pgform1}) and
(\ref{pgform}), a single power $[\lambda_{G_{pt},j}]^m$ dominates the sum as
$m \to \infty$.  For a member of a one-parameter family of planar
triangulations, $G_{pt,m}$, we use the notation $r(G_{pt,m})$ for the ratio 
(\ref{rpg}), and we define 
\beq 
r(G_{pt,\infty}) \equiv \lim_{m \to \infty} r(G_{pt,m})  \ . 
\label{rinf}
\eeq
We define the (real, non-negative) constant $a_{G_{pt}}$ as \cite{tub} 
\beq
a_{G_{pt}} = \lim_{n \to \infty} [ r(G_{pt,m}) ]^{1/n} = 
\frac{|\lambda_{G_{pt},dom}(\tau+1)|^{1/\alpha}}{\tau-1} \ . 
\label{a}
\eeq

We showed in \cite{tub} that if $j_{max}=1$, then $a_{G_{pt}} < 1$ and
hence for the classes of $G_{pt,m}$ under consideration, (i)
$r(G_{pt,\infty})=0$ and (ii) $r(G_{pt,m})$ decreases toward zero exponentially
rapidly as a function of $m$ and $n$ as $m \to \infty$. Note that this does not
imply that $P(G_{pt,m},q)$ has a zero that approaches $q=\tau+1$ as $m, \ n \to
\infty$.

For one-parameter families of planar triangulation graphs $G_{pt,m}$ where
$P(G_{pt,m},q)$ has the form (\ref{pgform}) with $j_{max} \ge 2$, 
a consequence of the Tutte upper bound (\ref{tub})
is that as $m \to \infty$, any contribution
$c_{_{G_{pt}},j}(\lambda_{G_{pt,j}})^m$ in (\ref{pgform}), when evaluated
at $q=\tau+1$, must be less than or equal in magnitude to $(\tau-1)^{n-5}$. 
Therefore, for a given $j$ in this case, either the
coefficient $c_{_{G_{pt}},j}$ vanishes for $q=\tau+1$ or, if this
coefficient does not vanish at $q=\tau+1$, then, taking into account the
relation (\ref{nmrel}), it follows that 
\beq
\frac{|\lambda_{G_{pt},j}|^{1/\alpha}}{\tau-1} \le 1 \quad
\quad {\rm at} \ \ q = \tau + 1 \quad \forall \ j \ . 
\label{lamupper}
\eeq
If this inequality is realized as an equality, then $r(G_{pt,\infty})$ is a
nonzero constant, which necessarily lies in the interval $(0,1)$, so that
$a_{G_{pt}}=1$.  For the families $G_{pt,m}$ for which $m$ and $n$ are linearly
related, as specified in (\ref{nmrel}), this type of behavior occurs if and
only if, when $P(G_{pt,m},q)$ is evaluated at $q=\tau+1$, the (necessarily)
dominant $\lambda_{G_{pt},j}$ (with nonvanishing coefficient $c_{G_{pt},j}$),
is equal to $\tau-1$ in magnitude, i.e., $|\lambda_{G_{pt},dom}|= \tau-1$ at
$q=\tau+1$. This is true, in particular, if this $\lambda_{G_{pt},j} = q-2$.
In the structural form (\ref{pgfacform}) below, we shall label this as the
$j=1$ term.

It is a general property that if a graph $G$ contains a complete graph $K_p$ as
a subgraph, then $P(G,q)$ contains the factor $P(K_p,q)$.  In particular, a
triangulation graph, whether planar or not, has the factor $P(K_3,q)$ and a
planar triangulation may also contain a $K_4$.  (However, by Kuratowski's
Theorem, a planar graph may not contain a $K_p$ with $p \ge 5$.)
Thus, for a planar triangulation $G_{pt}$, $P(G_{pt},q)=0$ for $q=0, \ 1, \ 2$.
If $G_{pt} \supseteq K_4$, then $P(G_{pt},q)$ also vanishes at $q=3$.  If
$P(G_{pt},q)$ has the form of a single power, given as Eq. (\ref{pgform1}),
then these factors are explicit.  If, however, $P(G_{pt},q)$ has the form of a
sum of $j_{max} \ge 2$ powers $[\lambda_{_{G_{pt}},j}]^m$, then the conditions
that $P(G_{pt},q)$ vanish at $q=0, \ 1, \ 2$ imply relations between the
various terms.  Moreover, the condition that $P(G_{pt,m},\tau+1)$ obeys the
Tutte upper bound (\ref{tub}) also implies conditions on the structure of this
chromatic polynomial.  We derive these next.

\section{Properties of a Class of $G_{pt,m}$ with $P(G_{pt,m},q)$ Having 
$j_{max}=3$ and Certain $\lambda_{G_{pt},j}$}

\subsection{Structure of Coefficients $c_{_{G_{pt}},j}$ in $P(G_{pt,m},q)$} 

For a large class of one-parameter families of planar triangulations that we
have constructed and studied, for which $P(G_{pt,m},q)$ has the form
(\ref{pgform}), we find that (i) $j_{max}=3$ and (ii) the
$\lambda_{_{G_{pt}},j} \equiv \lambda_j$ with $j=1, \ 2, \ 3$ have the form
\beq
\lambda_1=q-2, \quad \lambda_2=q-3, \quad \lambda_3=-1 \ . 
\label{lamform}
\eeq
For this class of planar triangulations, we can derive some general results
concerning the functional form of the coefficients $c_{_{G_{pt}},j}$ (where we
will often suppress the subscript $pt$ on $G_{pt}$ where the meaning is
obvious).  Using the general form (\ref{pgform}) with $j_{max}=3$ and these
$\lambda_j$'s, we can derive the following identities.  The fact that
$P(G_{pt},0)=0$ implies that
\beq
c_{G,1} \, (-2)^m + c_{G,2} \, (-3)^m + c_{G,3} \, (-1)^m = 0 \ . 
\label{pgq0}
\eeq
where for ease of notation we suppress the subscript $pt$ on $G_{pt}$ here and
in related equations.  Since this equation must hold for arbitrary $m$
(understood implicitly to be an integer in the range $m \ge m_{min}$, where
$m_{min}$ is the minimal value for which the family $G_{pt,m}$ is well
defined), it implies that $c_{G,j} = 0$ for all $j$.  Hence, for these
families,
\beq 
c_{G,j} \quad {\rm contains \ the \ factor} \ q \ {\rm for} \ 
j=1, \ 2, \ 3 \ . 
\label{cjqfactor}
\eeq

The evaluation $P(G_{pt},1)=0$ reads 
\beq
c_{G,1} \, (-1)^m + c_{G,2} \, (-2)^m + c_{G,3} \, (-1)^m = 0 
\quad {\rm at} \ \ q=1 \ . 
\label{pgq1}
\eeq
Since this equation must hold for arbitrary $m \ge m_{min}$, it implies two
conditions on the evaluation of the coefficients at $q=1$, namely
\beq
c_{G,2} = 0 \quad {\rm and} \quad 
c_{G,1} + c_{G,3} = 0 \quad {\rm at} \ \ q=1 \ . 
\label{cgj13q1}
\eeq
In particular, (\ref{cgj13q1}) implies that
\beq
c_{G,2} \quad {\rm contains \ the \ factor} \ q-1 \ . 
\label{cgj2qm1factor}
\eeq
The evaluation $P(G_{pt},2)=0$ reads 
\beq
c_{G,1} \, 0^m + [c_{G,2} + c_{G,3}](-1)^m = 0 \quad {\rm at} \ \ q=2 \ . 
\label{cgj23q2}
\eeq
Since this equation holds for arbitrary $m \ge m_{min}$, it implies that
\beq
c_{G,2} + c_{G,3} = 0 \quad {\rm at} \ \ q = 2 \ . 
\label{cgj23q2cond}
\eeq
If a family $G_{pt,m}$ which has $m_{min}=0$, (\ref{cgj23q2}) and 
(\ref{cgj23q2cond}) together would also imply that $c_{G,1}=0$ at
$q=2$. 

Continuing with $P(G_{pt,m},q)$ of the form (\ref{pgform}) with
(\ref{lamform}), we next analyze the evaluation of $P(G_{pt,m},q)$ at
$q=\tau+1$, viz., $P(G_{pt,m},\tau+1)$. Since $\tau-1 = 0.61803...$ and $\tau-2
= -0.381966$ are smaller than unity in magnitude, the first two terms in
$P(G_{pt,m},\tau+1)$ vanish exponentially rapidly as $m$ increases.  As regards
the ratio $r(G_{pt,m})$, as $m$ increases, the contribution of the first term
to this upper bound approaches a constant, while the contribution of the second
term vanishes exponentially rapidly. Given the relation (\ref{nmrel}), the
Tutte upper bound also vanishes exponentially rapidly as a function of $m$.
Therefore, in order for $P(G_{pt,m},\tau+1)$ to satisfy the Tutte upper bound
(\ref{tub}), it is necessary and sufficient that
\beq
c_{G,3} = 0 \quad {\rm at} \ \ q=\tau+1 \ . 
\label{c3tp1zero}
\eeq
This means that 
\beq
c_{G,3} \quad {\rm contains \ the \ factor} \quad 
q-\bigg (\frac{3+\sqrt{5}}{2} \ \bigg ) \ . 
\label{c3factor1}
\eeq
Given that a chromatic polynomial has rational (actually integer) coefficients
as a polynomial in $q$, this means that $c_{G,3}$ must also contain a factor
involving the algebraically conjugate root, i.e.,
\beq
c_{G,3} \quad {\rm contains \ the \ factor} \quad q -
 \bigg ( \frac{3-\sqrt{5}}{2} \ \bigg ) \ . 
\label{c3factor2}
\eeq
Combining these, we derive the result that 
\beq
c_{G,3} \quad {\rm contains \ the \ factor} \quad q^2-3q+1 \ . 
\label{c3factor}
\eeq

Having proved these results, it is thus convenient to extract the factors
explicitly and define 
\beq
\kappa_{G,1} \equiv \frac{c_{G,1}}{q} \ , 
\label{kappa1}
\eeq
\beq
\kappa_{G,2} \equiv \frac{c_{G,2}}{q(q-1)} \ , 
\label{kappa2}
\eeq
and
\beq
\kappa_{G,3} \equiv \frac{c_{G,3}}{q(q^2-3q+1)} \ . 
\label{kappa3}
\eeq

For this class of planar triangulation graphs $G_{pt,m}$, we thus have the
general structural formula
\beqs
P(G_{pt,m},q) & = & q \bigg [ \kappa_{G,1}(q-2)^m + 
\kappa_{G,2}(q-1)(q-3)^m \cr\cr
 & + & \kappa_{G,3}(q^2-3q+1)(-1)^m \bigg ] \ , 
\label{pgfacform}
\eeqs
where $m$ and $n$ are related by (\ref{nmrel}). We observe that
the form (\ref{pgfacform}) satisfies the general results that we derived above
for the evaluation at $q=\tau+1$. Thus, if 
$P(G_{pt,m},q)$ has this form (\ref{pgfacform}) with (\ref{nmrel}) and
$\alpha=1$, then
\beq
r(G_{pt,\infty}) = [q \, \kappa_{G,1}] \bigg |_{q=\tau+1} 
\label{rinfpgfacform}
\eeq
and hence
\beq
a({G_{pt}}) = 1 \ . 
\label{a1}
\eeq

The conditions on the coefficients $c_{G,j}$'s evaluated at $q=1$ and
$q=2$ that we have derived, (\ref{cgj13q1}), together
with the definitions (\ref{kappa1})-(\ref{kappa3}), are equivalent to the
following relations:
\beq
\kappa_{G,1} = \kappa_{G,3} \quad {\rm at} \ \ q=1
\label{kgj13q1}
\eeq
and
\beq
\kappa_{G,2} = \kappa_{G,3} \quad {\rm at} \ \ q=2 \ . 
\label{kgj23q2}
\eeq

For certain families of planar triangulations $G_{pt,m} \equiv G_m$, the
chromatic number $\chi(G_m)$ is 3 for even $m$ and 4 for odd $m$ or vice
versa.  In these cases, we can also derive another relation between the
coefficients.  Thus, if $\chi(G_m) = 3$ for even $m$ and $\chi(G_m) =
4$ for odd $m$, then $\kappa_{G,1} = \kappa_{G,3}$ at $q=3$. On the other hand,
if $\chi(G_m) = 3$ for odd $m$ and $\chi(G_m) = 4$ for even $m$, then
$\kappa_{G,1} = -\kappa_{G,3}$ at $q=3$.  In the case of families $G_m$
for which $\chi(G_m)=4$ for all $m$, we have
\beq
\kappa_{G,1} + \kappa_{G,3} \, (-1)^m = 0 \quad {\rm at} \ \ 
q=3 \quad {\rm if} \ \ \chi(G_m)=4 \ , 
\label{kchi4}
\eeq
which implies 
\beq
\kappa_{G,1} = \kappa_{G,3} = 0  \quad {\rm at} \ \ q = 3 \ \ 
{\rm if} \ \ \chi(G_m) = 4 \ . 
\label{chi4result}
\eeq

\subsection{Properties of Real Chromatic Zeros} 

Here we derive some properties of chromatic zeros of planar triangulation
graphs $G_{pt,m}$ for which the chromatic polynomial has the form
(\ref{pgform}).  It is appropriate first to review some relevant properties of
chromatic zeros of general graphs and planar triangulation graphs.  For a
general graph $G$, it is elementary that there are no negative chromatic zeros
and that there are no chromatic zeros in the intervals (0,1)
\cite{woodall77,rtrev}.  The property that (0,1) is a zero-free interval for
the chromatic polynomial implies that $q=\tau+1$ cannot be a chromatic zero for
any graph $G$, as noted above (independent of whether it is a planar
triangulation or not).  Another interval that has been proved to be free of
chromatic zeros is $(1,32/27]$ \cite{jackson93,thomassen97}.  For an arbitrary
graph $G$, let us denote the total number of subgraphs $H \subseteq G$ that
are triangles as $N_t$, and let $n=n(G)$ and $e = e(G)$. It has been proved
\cite{jasonbrown98} that for an arbitrary graph $G$ with $n \ge 4$ vertices, if
$N_t < u(G)$, where
\beq
u(G) = \frac{e(e-n)+n-1}{2(n-2)} \ , 
\label{ntu}
\eeq
then $P(G,q)$ has complex zeros. 

Specializing now to planar triangulation graphs, it has been proved that
$G_{pt}$ has no chromatic zeros in the interval $(2,q_w)$
\cite{woodall92,woodall04}, where $q_w$ is the unique real zero of
\beq
\lambda_{TC} = q^3-9q^2+29q-32 \ ,
\label{lamtc}
\eeq
i.e.,
\beq
q_w = 3-\frac{[12(9+\sqrt{177} \ )]^{1/3}}{6} + 4[12(9+\sqrt{177} \ )]^{-1/3}
= 2.546602..
\label{qw}
\eeq
We remark that $q_w$ occurs as a chromatic zero of some planar triangulation
graphs, in particular, the family comprised of cylindrical sections of the
triangulation lattice with $L_y=3$, or equivalently iterated octahedra. In 1992
Woodall conjectured that a planar triangulation has no chromatic zeros in the
interval $(q_m,3)$, where $q_m = 2.6778146..$ is the unique real zero of
$q^3-9q^2+30q-35$, \cite{woodall92}, but later he gave counterexamples to his
conjecture involving one-parameter families of planar triangulations each of
which has a real zero that approaches 3 from below as this parameter goes to
infinity \cite{woodall97}.  We also note that for a planar triangulation,
substituting (\ref{entri}) into (\ref{ntu}), 
\beq
u(G) = \frac{(3n-7)(2n-5)}{2(n-2)} \quad {\rm for} \ \ G = G_{pt} \ . 
\label{ugpt}
\eeq

Here we present some further results on chromatic zeros of planar
triangulations. First, if $P(G_{pt,m},q)$ has the form (\ref{pgform1})
involving only a single power of a $\lambda_{G_{pt}}$, then its zeros are
fixed, independent of $m$, and hence although it typically has a zero close to
$\tau+1$, this zero does not move as a function of $m$. However, if
$P(G_{pt,m},q)$ has the multi-term form (\ref{pgfacform}) with $j_{max}=3$ and
the $\lambda_j$'s in (\ref{lamform}), then it necessarily has a zero in the
interval $[q_w,3)$ that approaches $\tau+1$ as $m \to \infty$.  The proof of
this is as follows.  Let us assume that $q$ is a real number in this interval
$[q_w,3)$.  In the limit as $m \to \infty$, the first two terms, which are
proportional to $(q-2)^m$ and $(q-3)^m$, respectively, vanish (exponentially
fast), so that
\beq
P(G_{pt,m},q) \sim \epsilon_m + q(q^2-3q+1) \kappa_{G_{pt},3}(-1)^m \ , 
\label{pgmeps}
\eeq
where $\epsilon_m$ denotes the contribution of these first two terms. 
If $\kappa_{G_{pt},3}$ happens to vanish at 
$q=\tau+1$, then the result follows, since $\epsilon_m \to 0$ as $m \to
\infty$.  If $\kappa_{G_{pt},3} \ne 0$ at $q=\tau+1$, then consider the
limit as $q \to \tau+1$, where we can write 
\beqs 
\frac{(-1)^mP(G_{pt,m},q)}{(\tau+1) \kappa_{G_{pt},3}|_{q=\tau+1}} & = & 
\frac{(-1)^m\epsilon_m}{(\tau+1) \kappa_{G_{pt},3}|_{q=\tau+1}} 
+ q^2-3q+1 \cr\cr & \equiv & \delta_m + q^2-3q+1 \ . 
\label{auxeq1}
\eeqs
To show that $P(G_{pt},q)$ has a zero that approaches $q=\tau+1$ as $m \to
\infty$, we use the fact that $P(G_{pt,m},q)$ is a continuous function of $q$
and solve (\ref{auxeq1}) for $q$, subject to the condition that 
$q \in [q_w,3)$, obtaining a consistent result with 
\beq
q = \frac{1}{2} \bigg [ 3 + \sqrt{5-4\delta_m} \ \bigg ] \ , 
\label{qzero}
\eeq
which approaches $q=\tau+1$ as $m \to \infty$.  Note that the other zero at
$q=(1/2)(3 - \sqrt{5-4\delta_m} \ )$, is irrelevant because we assumed at the
outset that $q$ is in the interval $[q_w,3)$ and this other zero is outside
this interval; in the vicinity of this other zero, the analysis does not apply
because the terms proportional to $(q-2)^m$ and $(q-3)^m$ do not vanish as $m
\to \infty$.

In \ref{tub} we have exhibited two one-parameter families of planar
triangulation graphs with this property, namely $B_m$ and $H_m$.  We construct
and analyze several more families of this type here.  Thus, for these families,
we have provided an understanding of why $P(G_{pt,m},q)$ has a chromatic zero
near to $\tau+1$ and, furthermore, have proved that this zero approaches
$\tau+1$ as $m \to \infty$.  From the derivation above, it is evident that our
result requires, for a given family $G_{pt,m}$, that $m$ be sufficiently large.
As is illustrated from numerical results presented below, for specific families
of planar triangulations that we have studied, $P(G_{pt,m},q)$ has a real zero
reasonably close to $\tau+1$ even for moderate values of $m$.

Our second result follows immediately from this analysis. With the same
assumptions, we have observed that in the limit $m \to \infty$, $P(G_{pt,m},q)$
has a real zero in the interval $q \in [q_w,3)$ if and only if
$c_{_{G_{pt}},3}$ has a real zero in this interval, $q \in [q_w,3)$. We know
that there is at least one such zero, namely the one arising from the factor
$q^2-3q+1$ in $c_{_{G_{pt}},3}$.  Therefore, in the limit $m \to \infty$,
$P(G_{pt,m},q)$ has another real zero in the interval $q \in [q_w,3)$ in
addition to the one approaching $\tau+1$ if and only if $\kappa_{G,3}$ has a
real zero in this interval $q \in [q_w,3)$.  We will present several
applications of these results below. 

We remark that, with the same assumptions as above, 
\beq
\lim_{m \to \infty} |P(G_{pt,m},q)| = c_{G_{pt},3} \ . 
\label{lim1}
\eeq
Note that the limit $\lim_{m \to \infty} P(G_{pt,m},q)$ itself does not exist,
because the term $\lambda_3^m = (-1)^m$ factor has no limit as $m \to \infty$.

In \cite{tub}, we investigated the question of whether for a planar
triangulation graph $G_{pt}$ it is true that the chromatic zero of $G_{pt}$
nearest to $\tau+1$ is always real. We exhibited an example, with a graph we
denoted $G_{CM,1}$, which is, to our knowledge, the first case for which the
zero closest to $\tau+1$ is not real but instead the zeros closest to $\tau+1$
form a complex-conjugate pair.  Our result is in agreement with a previous
observation by Woodall that this graph has no real zero near to $\tau+1$
\cite{woodallpriv}.  Since $P(G_{CM,m},q)$ has the form of (\ref{pgform1})
with a single $\lambda$, as $m$ increases, its zeros are fixed and just
increase in multiplicity, in contrast to the motion of chromatic zeros for
families $G_{pt,m}$ whose chromatic polynomials are of the form (\ref{pgform})
with $j_{max} > 1$.

The value of $q$ where the Tutte upper bound applies, namely $q=\tau+1=(3 +
\sqrt{5} \ )/2$ is also a member of a sequence of numbers related to roots of
unity, namely the Tutte-Beraha numbers, $q_r$.  Thus, for a root of unity of
the form $z_r=e^{\pi i/r}$, one defines $q_r = (z_r+z_r^*)^2 =
4\cos^2(\pi/r)$. One has $\tau+1 = q_5$.  Parenthetically, we note that
chromatic zeros have been studied for sections of triangular lattices with
various boundary conditions that are not planar triangulations, either because
they have at least one face that is not a triangle or because they are not
planar (e.g., have toroidal or Klein-bottle boundary conditions).  We refer the
reader to \cite{tub} for references to some of these papers; here, in view of
our focus on the Tutte upper bound (\ref{tub}) we restrict to planar
triangulations.

\subsection{Properties of Complex Chromatic Zeros}

As before, we consider a one-parameter of planar triangulation graphs
$G_{pt,m}$ such that $P(G_{pt,m},q)$ has the form (\ref{pgfacform}).  Here we
give a general determination of the continuous accumulation set ${\cal B}$ of
chromatic zeros of $P(G_{pt,m},q)$ in the complex $q$ plane in the limit $m \to
\infty$ (and, hence, owing to (\ref{nmrel}), $n \to \infty$).  If the zeros
form a discrete set (some with multiplicities that go to infinity as $n \to
\infty$), then this locus is null.  As in earlier work, we denote the formal
limit of the family $G_m$ as $m \to \infty$ as $\{ G \}$. In general, the locus
${\cal B}$ may or may not intersect the real $q$ axis.  If it does, the maximal
point where it intersects the real axis is denoted $q_c(\{ G \})$.  For a
$P(G,q)$ of the form (\ref{pgform}), the curves comprising the locus ${\cal B}$
are determined as the solutions of the equality in magnitude of the dominant
$\lambda_{G_{pt},j}$, in accordance with general results for recursive
functions \cite{bkw}. These curves extend infinitely far from the origin if and
only if such an equality can be satisfied as $|q| \to \infty$, as was discussed
in \cite{w},\cite{wa}-\cite{wa2}. 

We now consider families of planar triangulations $G_{pt,m}$ whose chromatic
polynomials have the form (\ref{pgfacform}).  For these families, first,
because $\lambda_1=q-2$ is dominant for large $q$ and is equal in magnitude to
$\lambda_3=-1$ at $q=3$, it follows that $q_c(\{G_{pt} \})=3$. Second, as a
consequence of the fact that the equality $|q-2|=|q-3|$ holds for the infinite
line ${\rm Re}(q)=5/2$, part of the boundary ${\cal B}$ extends infinitely far
away from the origin in the $q$ plane, i.e., passes through the origin of the
$1/q$ plane.  As is evident for specific families $G_{pt,m}$, as $m$ and hence
$n$ go to infinity, the degree of one or more vertices also goes to infinity,
so the fact that the magnitudes of zeros diverge is in accord with the upper
bound $|q| < b \Delta(G)$ obtained in \cite{sokalbound} (with $b \simeq 7.96$)
and strengthened slightly in \cite{fp} (with $b \simeq 6.91$), where
$\Delta(G)$ is the maximal degree of any vertex in $G$. Note, however, that the
property that the degree of a vertex diverges as $n \to \infty$ does not, by
itself, imply that ${\cal B}$ passes through the origin of the $1/q$ plane.
This is clear from the $n \to \infty$ limit of wheel graphs, for which the
central vertex $v_{cent.}$ has degree $d(v_{cent.}) \to \infty$, but the locus
${\cal B}$ has bounded support in the $q$ plane \cite{w,dg}.  Continuing with
our analysis of families of planar triangulations $G_{pt,m}$ whose chromatic
polynomials have the form (\ref{pgfacform}), the ${\cal B}$ separates the $q$
plane into three regions, which we denote as $R_j$, $j=1, \ 2, \ 3$.  These are
defined as follows:
\beq
R_1: Re(q) > \frac{5}{2} \quad {\rm and} \quad |q-2| > 1 \ , 
\label{br1}
\eeq
\beq
R_2: Re(q) < \frac{5}{2} \quad {\rm and} \quad |q-3| > 1 \ , 
\label{br2}
\eeq
and
\beq
R_3: |q-2| < 1 \quad {\rm and} \quad |q-3| < 1 \ . 
\label{br3}
\eeq
The boundaries between these regions are thus the two circular arcs
\beq
{\cal B}(R_1,R_3):
q = 2+e^{i\theta} \ , -\frac{\pi}{3} < \theta < \frac{\pi}{3}
\label{leftarc}
\eeq
and
\beq {\cal B}(R_2,R_3): q=3+e^{i\phi} \ , \frac{2\pi}{3} < \phi < 
\frac{4\pi}{3} \ ,
\label{rightarc}
\eeq
together with the semi-infinite vertical line segments
\beq
{\cal B}(R_1,R_2) = \{q\}: \quad
{\rm Re}(q)=\frac{5}{2} \quad {\rm and} \quad 
|{\rm Im}(q)| > \frac{\sqrt{3}}{2} \ . 
\label{lines}
\eeq
These meet at the triple points
\beq
q_t, \ q_t^* = \frac{5 \pm i\sqrt{3}}{2} \ . 
\label{qt}
\eeq
A specific example of a locus ${\cal B}$ that extends infinitely far from the
origin in the $q$ plane was studied in \cite{rr,w} for the family $B_m$.  Here
we have generalized the result to all families of planar triangulations whose
chromatic polynomials have the form (\ref{pgfacform}).

We have constructed and studied a two-parameter family of planar triangulations
$D_{m_1,m_2}$.  By keeping one of the two indices $m_1$ or $m_2$ fixed, we have
obtained a number of one-parameter families and have analyzed the chromatic
polynomials for these. We have also considered the special cases where one
allows both $m_1$ and $m_2$ to vary, such that one is a linear function of the
other.  We have, in particular, analyzed the diagonal case where $m_1=m_2$, 
$D_{m,m}$, to be discussed below.  For this family the chromatic polynomial
$P(D_{m,m},q)$ has the form (\ref{pgform}) with $j_{max}=6$ and a set of
$\lambda$'s that are squares and cross products of those in (\ref{pgfacform}).
Since the equations defining the equality of magnitude of dominant $\lambda$'s
can again be satisfied as $1/q \to 0$, it again follows, by the criteria of
\cite{w,wa,w3} that the locus ${\cal B}$ extends infinitely far from the origin
of the complex $q$ plane.

In a different direction, we have also studied a family $F_m$ (see below) for
which the chromatic polynomial $P(F_m,q)$ is of the form (\ref{pgfacform}) with
$j_{max}=3$, but with terms $\lambda_{F,j}$, $j=1,2,3$, that are not simple
polynomials, but instead are roots of a cubic equation, (\ref{eqf}).  In the $m
\to \infty$ limit, the continuous accumulation set ${\cal B}$ for this family
has bounded magnitude (does not extend infinitly far away from the origin in
the $q$ plane), as can be seen because the condition that defines ${\cal B}$,
namely the equality in magnitude of two dominant $\lambda_{F,j}$'s, cannot be
satisfied for arbitrarily great $|q|$.

\section{The Family $B_m$}

In Ref. \cite{tub}, we constructed and studied several one-parameter families
of planar triangulations with chromatic polynomials of the multi-term form
(\ref{pgfacform}).  (Some graphs in these families are listed in 
\cite{readwilson}.) In view of the general factorizations for the coefficients
$c_{_{G_{pt,m}},j}$ that we have proved here, it is useful to express the
results in terms of the reduced coefficients $\kappa_{G,j}$.  For the
family of bipyramid graphs $B_n$, Eqs. (6.1)-(6.4) of Ref. \cite{tub} are
equivalently expressed as Eq. (\ref{pgfacform}), with $n(B_m)=m+2$, which is
well-defined for $m \ge m_{min}=3$, with the reduced coefficients
\beq
\kappa_{B,1} = \kappa_{B,2} = \kappa_{B,3} = 1 \ . 
\label{kappabj123}
\eeq
For this family, $r(B_\infty)=(-1+\sqrt{5} \ )/2 = 0.6180..$ and, as a special
case of (\ref{a1}), $a_B=1$.  Here and below we will index the members of the
various families of planar triangulations as $G_{pt,m}$, equivalent to the
notation $G_{pt,n}$ that we used for some families in \cite{tub} via the
relation (\ref{nmrel}).

\section{The Family $H_m$}

\begin{figure}
  \begin{center}
    \includegraphics[height=6cm]{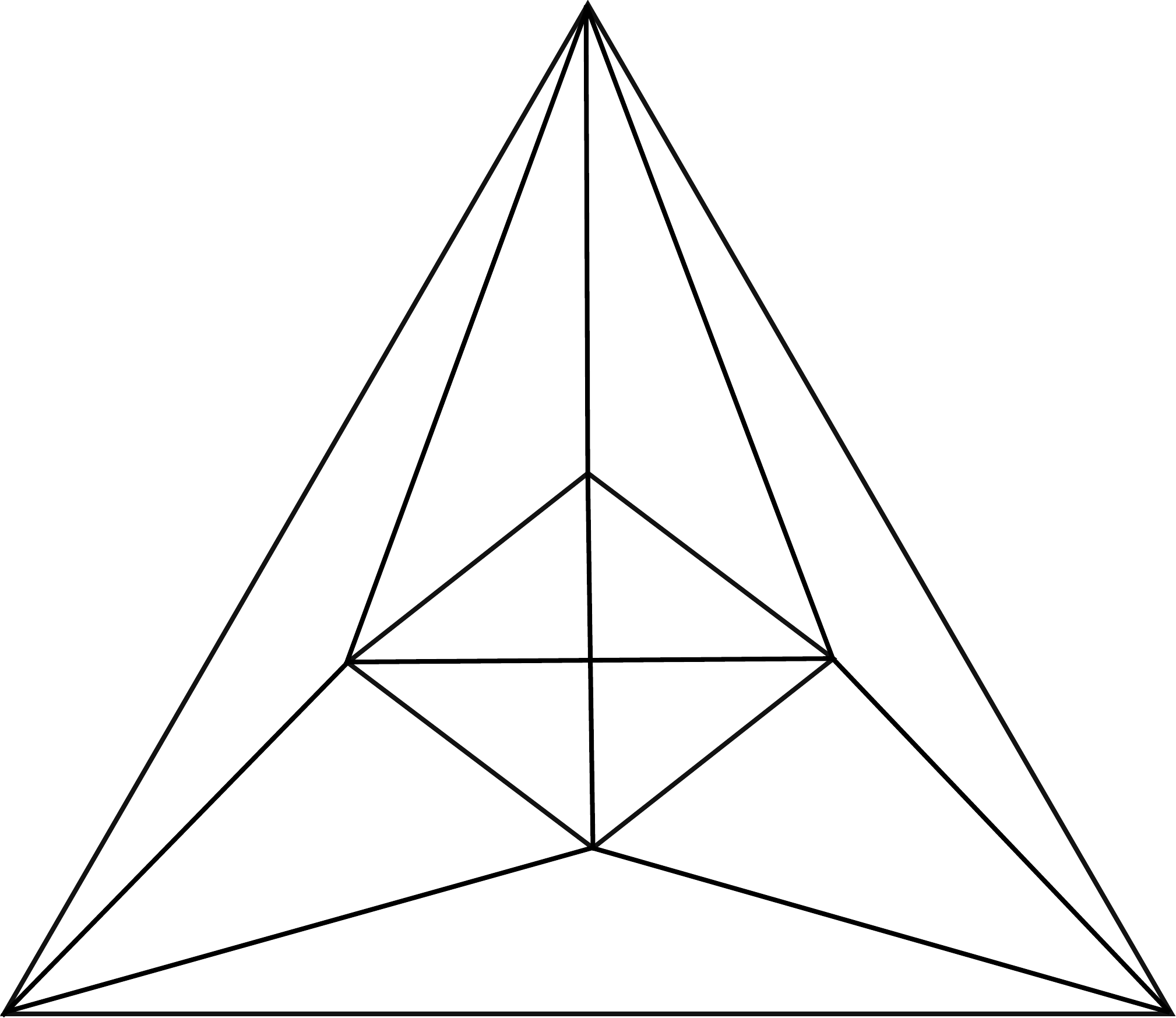}
  \end{center}
\caption{Graph $H_3$.}
\label{gvepm3n8}
\end{figure}

\begin{figure}
  \begin{center}
    \includegraphics[height=6cm]{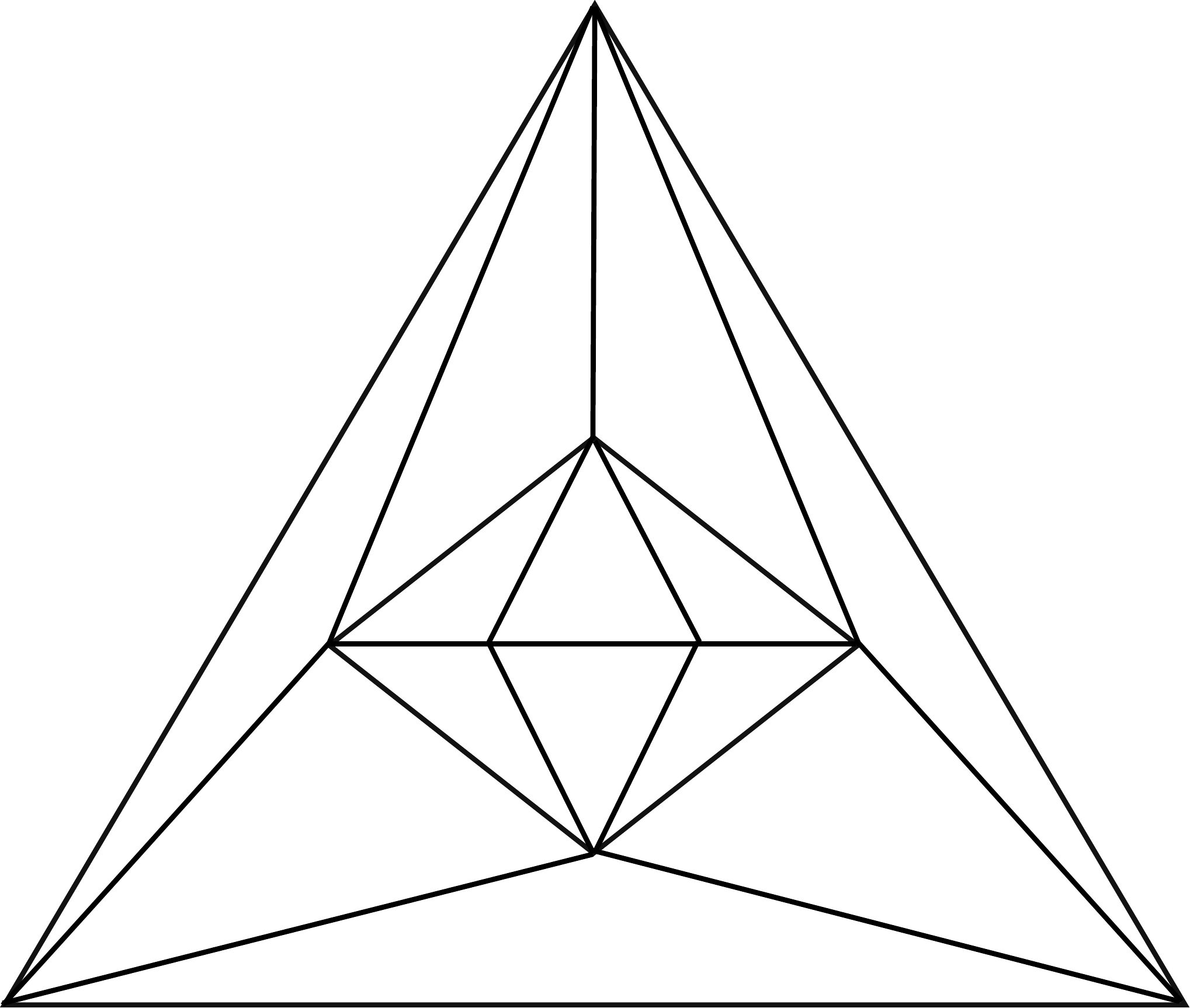}
  \end{center}
\caption{Graph $H_4$.}
\label{gvepm4n9}
\end{figure}

Here we remark on another family of planar triangulations $H_m$ \cite{tub},
which is well-defined for $m \ge m_{min}=3$ and has $n(H_m)=m+5$.  In
Fig. \ref{gvepm3n8} we show the lowest member of the family, $H_3$. The next
higher member, $H_4$, is constructed by adding a vertex and associated edges in
the central diamond-like subgraph, as shown in Fig. \ref{gvepm4n9}, and so
forth for higher members.
We have given the chromatic polynomial for this family in Ref. \cite{tub} and
have analyzed its properties there.  In our present notation, 
$P(H_m,q)$ has the form (\ref{pgfacform}) with 
\beq
\kappa_{H,1} = (q-3)^3 \ , 
\label{kgvepj1} 
\eeq
\beq
\kappa_{H,2} = q^3-9q^2+30q-35 \ , 
\label{kgvepj2}
\eeq
and
\beq
\kappa_{H,3} = -(q-3)(q-5) \ . 
\label{kgvepj3}
\eeq
For this family, $r(H_\infty)=(7-3\sqrt{5} \ )/2 = 0.145898..$ \cite{tub}.

\section{The Family $L_m$}

\begin{figure}
  \begin{center}
    \includegraphics[height=6cm]{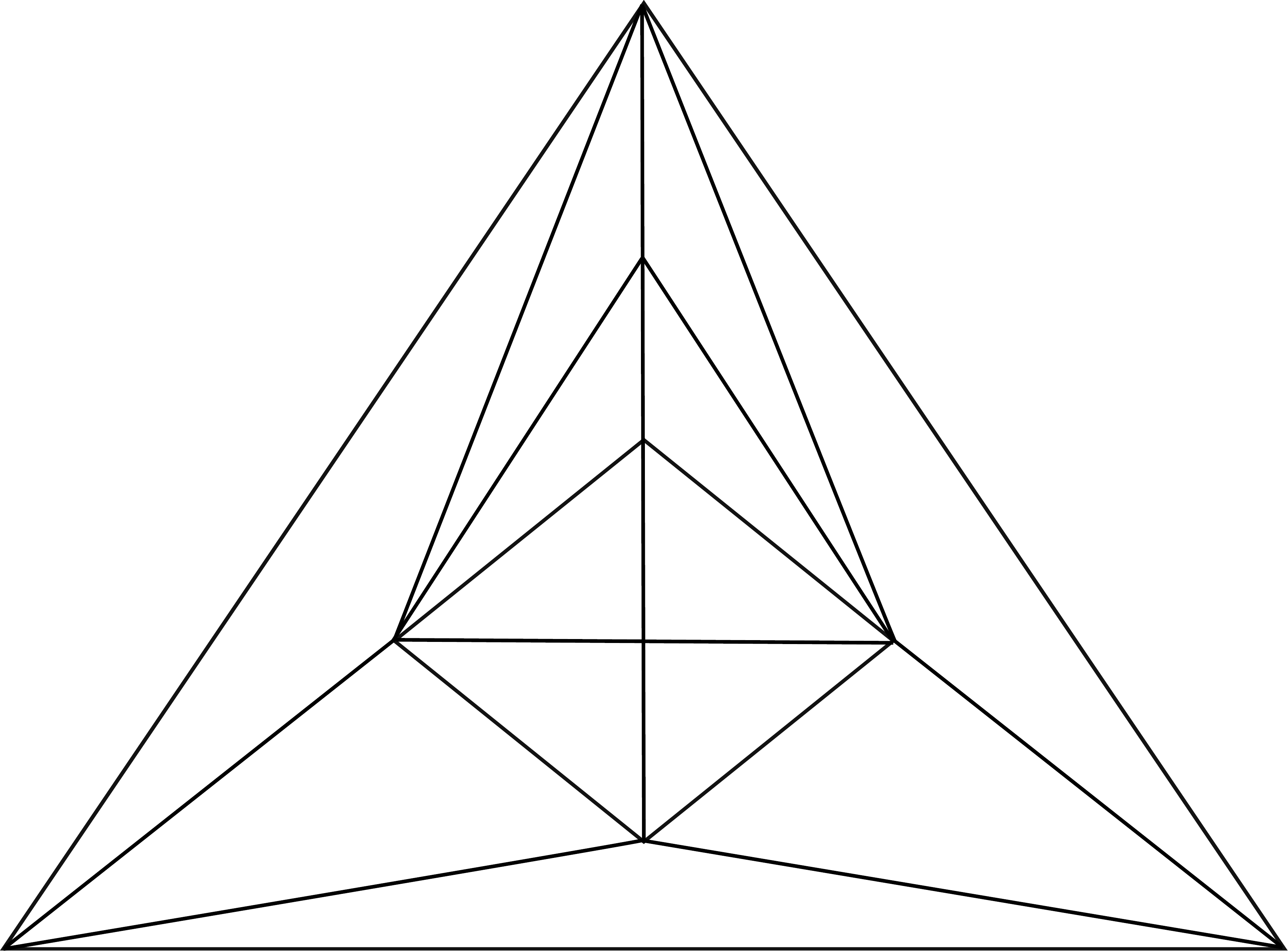}
  \end{center}
\caption{Graph $L_4$.}
\label{ghepm4n9}
\end{figure}

For comparative purposes, it is useful to study another family of planar
triangulations with chromatic polynomials of the multi-term form
(\ref{pgform}).  Here we denote this family as $L_m$.  It is well-defined
for $m \ge m_{min}=3$ and has $n(L_m)=m+5$. The lowest member
of this family, $L_3$, is the same as $H_3$, shown in
Fig. \ref{gvepm3n8}. The next higher member, $L_4$, is constructed by
adding a vertex and associated edges in the central ``diamond'', as shown in
Fig. \ref{ghepm4n9}, and so forth for higher members.  

For this family of planar triangulations we calculate the chromatic polynomial
$P(L_m,q)$ to be of the form (\ref{pgfacform}) with 
\beq
\kappa_{L,1} = (q-2)(q-3)^2 \ , 
\label{khepj1}
\eeq
\beq
\kappa_{L,2} = q^3-9q^2+29q-32 \ , 
\label{khepj2}
\eeq
(equal to $\lambda_{TC}$) and
\beq
\kappa_{L,3} = 2(q-3) \ . 
\label{khepj3}
\eeq
$P(L_m,q)$ contains the factor $P(K_4,q)$ and has $\chi(L_m)=4$. 

Evaluating $P(L_m,q)$ at $q=\tau+1$, we find 
\beq
P(L_m,\tau+1) = (-2+\sqrt{5} \ ) \bigg [ (\tau-1)^m + 
2(\tau-2)^m \bigg ] \ . 
\label{pheptp1}
\eeq
Consequently,
\beq
r(L_m) = ( -2+\sqrt{5} \ ) 
\bigg [ 1 + 2 \bigg ( \frac{1-\sqrt{5}}{2} \ \bigg )^m \bigg ] \ . 
\label{rhepm}
\eeq
As before, since $|(1-\sqrt{5} \, )/2| < 1$, the second term in
Eq. (\ref{rhepm}) vanishes (exponentially fast) as $m \to \infty$, so 
\beq
r(L_\infty) = -2+\sqrt{5} = 0.236068
\label{rghepninf}
\eeq
(to the indicated accuracy) and, as a special case of Eq. (\ref{a1}), 
$a_L = 1$. 

We proved in general above that for any family of planar triangulations
$G_{pt,m}$ with chromatic polynomials $P(G_{pt,m},q)$ of the form
(\ref{pgfacform}), $P(G_{pt,m},q)$ has a zero that approaches $\tau+1$ as $m
\to \infty$.  The families $B_m$, $H_m$, and $L_m$ (as well as others to be
discussed below) illustrate this general result.  In the present case, for odd
$m$ and hence even $n=m+5$, this zero of $P(L_m,q)$ is slightly less than
$q=\tau+1$, while for even $m$ and hence odd $n$, the nearby zero is slightly
greater than $\tau+1$. We list these zeros in Table \ref{ghepnzerotable} for
$m$ from 4 to 15 (i.e., $n$ from 9 to 20).

\begin{table}
\caption{\footnotesize{Location of zero $q_z$ of $P(L_m,q)$ closest to
$\tau+1$, as a function of the number of vertices, $n=m+5$. Notation
$a$e-$\nu$ means $a \times 10^{-\nu}$ here and in tables below.}}
\begin{center}
\begin{tabular}{|c|c|c|}
\hline\hline
 $n$ &  $q_z$    & $q_z-(\tau+1)$  \\
\hline\hline
  9  &  2.630048  &   0.01201     \\
 10  &  2.614750  & $-0.003284$   \\
 11  &  2.621594  &   0.003560    \\
 12  &  2.616447  & $-0.0015865$  \\
 13  &  2.619250  &   0.001216    \\
 14  &  2.617375  & $-0.6588$e-3  \\
 15  &  2.618477  &   0.4432e-3   \\
 16  &  2.617774  & $-2.598$e-4   \\
 17  &  2.618200  &   1.661e-4    \\
 18  &  2.6179335 & $-1.005$e-4   \\
 19  &  2.618097  &   0.6294e-4   \\
 20  &  2.617995  & $-3.8574$e-5  \\
\hline\hline
\end{tabular}
\end{center}
\label{ghepnzerotable}
\end{table}
If and only if $m$ is odd, i.e., $n$ is even, $P(L_m,q)$ has another real
zero somewhat larger than 3, which decreases monotonically toward 3 from above
as $m \to \infty$. As examples, for $n=8$, 12, 16, 20, and 24, this zero occurs
at approximately $q=3.61$, 3.37, 3.25, 3.19, and 3.16, respectively.

\section{Two-Parameter Families of Planar Triangulations, $G_{pt,m_1,m_2}$}

In this section we introduce a substantial generalization to a two-parameter
family $G_{pt,m_1,m_2}$ of planar triangulations involving the three
$\lambda_j$'s in (\ref{lamform}), with a chromatic polynomial of the form 
\beq
P(G_{pt,m_1,m_2},q) = \sum_{i_1=1}^3 \, \sum_{i_2=1}^3 \, 
c_{G,i_1 i_2} \, \lambda_{i_1}^{m_1} \, \lambda_{i_2}^{m_2}  \ . 
\label{pgm1m2}
\eeq
Explicitly, 
\beqs
& & P(G_{pt,m_1,m_2},q) = 
  c_{G,11}(q-2)^{m_1+m_2}
+ c_{G,22}(q-3)^{m_1+m_2}
+ c_{G,33}(-1)^{m_1+m_2} \cr\cr
& + & c_{G,12}(q-2)^{m_1}(q-3)^{m_2} 
+ c_{G,21}(q-3)^{m_1}(q-2)^{m_2} \cr\cr
& + & c_{G,13}(q-2)^{m_1}(-1)^{m_2} 
+ c_{G,31}(-1)^{m_1}(q-2)^{m_2} \cr\cr
& + & c_{G,23}(q-3)^{m_1}(-1)^{m_2} 
+ c_{G,32}(-1)^{m_1}(q-3)^{m_2} \ . 
\label{pdm1m2lam}
\eeqs
(Below we shall often take $i_1 = i$, $i_2 = j$ to simplify the notation.) 
Clearly, if one keeps one of the indices $m_1$ or $m_2$ fixed and varies the
other, this defines an infinite set of one-parameter families of planar
triangulations.  For specific families $G_{pt,m_1,m_2}$ we will show how the
general structure (\ref{pgm1m2}) reduces, in such cases, to the form
(\ref{pgfacform}) considered above for a class of one-parameter planar
triangulations, with $m$ being equal to the variable index, up to an
appropriate integer shift.

As before for the one-parameter families, an equivalent way to obtain the
$P(G_{pt,m_1,m_2},q)$ is via a Taylor series expansion, in the auxiliary
variables $x_1$ and $x_2$, of a generating function $\Gamma(G_{pt},q,x_1,x_2)$.
Equivalent to both of these is the property that
$P(G_{pt,m_1,m_2},q)$ satisfies a two-dimensional recursion relation, for
$m_1 \ge (m_1)_{min}+3$ and $m_2 \ge (m_2)_{min}+3$,
\beq
P(G_{pt,m_1,m_2},q) +
\sum_{i_1=1}^3 \sum_{i_2=1}^3 b_{G_{pt},i_1 i_2} \, P(G_{pt,m_1-i_1,m_2-i_2},q)
 = 0 \ .
\label{2drecursionrel}
\eeq
The coefficients in this recursion relation are given by
\beq
 1 + \sum_{i_1=1}^3 \sum_{i_2=1}^3 b_{G_{pt},i_1 i_2} \, x_1^{i_1} x_2^{i_2} =
\bigg [ \prod_{i_1=1}^3 (1-\lambda_{i_1} \, x_1) \bigg ]
\bigg [ \prod_{i_2=1}^3 (1-\lambda_{i_2} \, x_2) \bigg ] \ .
\label{blam2d}
\eeq
Note that they satisfy the symmetry property
\beq
b_{G_{pt},i_1 i_2} = b_{G_{pt},i_2 i_1} \ .
\label{bsym}
\eeq

We first derive a number of restrictions on the coefficients 
$c_{G,i_1 i_2}$.  As is
true of any triangulation, $P(G_{pt,m_1,m_2},q)=0$ for $q=0$, $q=1$, and
$q=2$. The evaluation $P(G_{pt,m_1,m_2},0)=0$ reads
\beqs
& & c_{G,11}(-2)^{m_1+m_2}+c_{G,22}(-3)^{m_1+m_2}+c_{G,33}(-1)^{m_1+m_2} \cr\cr
& + & c_{G,12}(-2)^{m_1}(-3)^{m_2} + c_{G,21}(-3)^{m_1}(-2)^{m_2} \cr\cr
& + & c_{G,13}(-2)^{m_1}(-1)^{m_2} + c_{G,31}(-1)^{m_1}(-2)^{m_2} \cr\cr
& + & c_{G,23}(-3)^{m_1}(-1)^{m_2} + c_{G,32}(-1)^{m_1}(-3)^{m_2} = 0 \quad 
{\rm at} \ \ q=0 \ . 
\label{pgm1m2q0}
\eeqs
Since this equation applies for arbitrary $m_1$ and $m_2$ in their respective
ranges, it implies that $c_{G,i_1 i_2}=0$ for all $i_1, \ i_2$ at $q=0$ and
hence that
\beq
c_{G,i_1 i_2} \quad {\rm contains \ the \ factor} \ \ q \quad \forall \ i_1, \
i_2 \ .
\label{cgijqfactor}
\eeq
It will often be convenient to extract this common factor, via the definition
\beq
\bar c_{G,i_1 i_2} = \frac{c_{G,i_1 i_2}}{q} \ .
\label{cijbar}
\eeq

The evaluation $P(G_{pt,m_1,m_2},1)=0$ reads
\beqs
& & [c_{G,11}+c_{G,33}+c_{G,13}+c_{G,31}](-1)^{m_1+m_2} 
 + c_{G,22}(-2)^{m_1+m_2} \cr\cr
&+& [c_{G,12}+c_{G,32}](-1)^{m_1}(-2)^{m_2} \cr\cr
&+& [c_{G,21}+c_{G,23}](-2)^{m_1}(-1)^{m_2} = 0 \quad {\rm at} \ \ q=1 \ . 
\label{pgm1m2q1}
\eeqs
Since this equation applies for arbitrary $m_1$ and $m_2$, it implies the
conditions 
\beqs
& & c_{G,11}+c_{G,33}+c_{G,13} + c_{G,31} = 0, \quad c_{G,22}=0, \cr\cr
& & c_{G,12}+c_{G,32} = 0, \quad 
    c_{G,21}+c_{G,23} = 0  \quad {\rm at} \ \ q=1 \ . 
\label{pgm1m2q1conditions}
\eeqs
In particular, this implies that 
\beq
c_{G,22} \quad {\rm contains \ the \ factor} \ \ q-1 \ . 
\label{cg22qm1factor}
\eeq

The evaluation $P(G_{pt,m_1,m_2},2)=0$ reads
\beqs
& & c_{G,11}0^{m_1+m_2} + 
[c_{G,22}+c_{G,33}+c_{G,23}+c_{G,32}](-1)^{m_1+m_2} \cr\cr
& + & [c_{G,12}+c_{G,13}]0^{m_1}(-1)^{m_2} + 
      [c_{G,21}+c_{G,31}](-1)^{m_1}0^{m_2} = 0 \quad {\rm at} \ \ q=2 \ . 
\label{pgm1m2q2}
\eeqs
Since this equation applies for arbitrary $m_1$ and $m_2$, including
$m_1=m_2=0$, it implies the conditions 
\beqs
& & c_{G,11}=0, \quad c_{G,22}+c_{G,33}+c_{G,23}+c_{G,32}=0, \cr\cr
& & c_{G,12}+c_{G,13}=0, \quad c_{G,21}+c_{G,31}=0 \quad {\rm at} \ \ q=2 \ . 
\label{pgm1m2q2conditions}
\eeqs
In particular, this implies that
\beq
c_{G,11} \quad {\rm contains \ the \ factor} \ \ q-2 \ .
\label{cg11qm2factor}
\eeq
For families $G_{pt,m_1,m_2}$ with $\chi(G_{pt,m_1,m_2})=4$ for certain 
values of $m_1$ and $m_2$, further conditions hold, as we shall discuss
below. 

We next derive some further restrictions on the coefficients $c_{G,i_1 i_2}$
from the condition that $P(G_{pt,m_1,m_2},q)$ must obey the Tutte upper bound
when evaluated at $q=\tau+1$.  We consider families such that
\beq
n(G_{pt,m_1,m_2}) = m_1+m_2 + \beta
\label{nmmrel}
\eeq
Carrying out this evaluation and calculating the
ratio $r(G_{pt,m_1,m_2})$, we have
\beqs
r(G_{pt,m_1,m_2}) & = & (\tau-1)^{5-\beta} \bigg | c_{G,11} + 
c_{G,22} \bigg ( \frac{\tau-2}{\tau-1} \bigg )^{m_1+m_2} + 
c_{G,33} \bigg ( \frac{-1}{\tau-1} \bigg )^{m_1+m_2} \cr\cr
& + & c_{G,12}\bigg ( \frac{\tau-2}{\tau-1} \bigg )^{m_2} + 
      c_{G,21}\bigg ( \frac{\tau-2}{\tau-1} \bigg )^{m_1} \cr\cr
& + & c_{G,13}\bigg ( \frac{-1}{\tau-1} \bigg )^{m_2} + 
      c_{G,31}\bigg ( \frac{-1}{\tau-1} \bigg )^{m_1} \cr\cr
& + & c_{G,23}\frac{(\tau-2)^{m_1}(-1)^{m_2}}{(\tau-1)^{m_1+m_2}}
     +c_{G,32}\frac{(-1)^{m_1}(\tau-2)^{m_2}}{(\tau-1)^{m_1+m_2}} \bigg | \ . 
\label{rgm1m2tp1}
\eeqs
The condition that $r(G_{pt,m_1,m_2}) \le 1$ for arbitrary $m_1$ and $m_2$
implies that
\beq
c_{G,33} = c_{G,13} = c_{G,31} = c_{G,23} = c_{G,32} = 0 \quad {\rm at} \ \ 
q=\tau+1 \ . 
\label{cijqtp1}
\eeq
By the same argument that we used above for the analysis of the coefficients of
chromatic polynomials of one-parameter planar triangulation graphs,
(\ref{cijqtp1}) implies that
\beq
c_{G,i_1 i_2} \quad 
{\rm contains \ the \ factor} \ \ q^2-3q+1 \quad {\rm if} \ \
i_1=3 \ {\rm or} \ i_2=3 \ . 
\label{ci3orj3factor}
\eeq
By taking either $m_1 \to \infty$ or $m_2 \to \infty$, and requiring that the
resultant ratio $r(G_{pt,\infty,m_2})$ or $r(G_{pt,m_1,\infty})$ must
obey the Tutte upper bound, one deduces the inequality
\beq
(\tau-1)^{5-\beta}|c_{G,11}| < 1 \quad {\rm at} \ \ q=\tau+1 \ . 
\label{cg11tpq}
\eeq

We next generalize our result on a real chromatic zero that approaches
$q=\tau+1$ for one-parameter planar triangulations to these two-parameter
planar triangulations with $P(G_{pt,m_1,m_2},q)$ of the form (\ref{pgm1m2}).
As in the $p=1$ case, we assume that $q$ is a real number in the interval
$[q_w,3)$.  We will actually obtain two results, corresponding to 
$m_1 \to \infty$ for fixed $m_2$ and $m_2 \to \infty$ for fixed $m_1$.  For the
first of these limits, the six terms proportional to $c_{G,11}$, $c_{G,22}$, 
$c_{G,12}$, $c_{G,21}$, $c_{G,13}$, and $c_{G,23}$ all vanish (exponentially 
rapidly), so that 
\beq
P(G_{pt,m_1,m_2},q) \sim c_{G,33}(-1)^{m_1+m_2} + c_{G,31}(-1)^{m_1}(q-2)^{m_2}
+ c_{G,32}(-1)^{m_1}(q-3)^{m_2} \quad {\rm for} \ \ m_1 \to \infty
\label{pgm1m2_for_m1inf}
\eeq
But we have shown above in (\ref{ci3orj3factor}) that $c_{G,i_1 i_2}$
contains the factor $q^2-3q+1$ if $i_1=3$ or $i_2=3$. Since this factor
vanishes at $q=\tau+1$ in this interval $[q_w,3)$, it follows that for
sufficiently large $m_1$, $P(G_{pt,m_1,m_2},q)$ has a real zero that approaches
$\tau+1$.  With obvious changes, a corresponding argument shows that for
sufficiently large $m_2$ and fixed $m_1$, $P(G_{pt,m_1,m_2},q)$ has a real 
zero that approaches $\tau+1$.  Clearly, the result also holds if both 
$m_1$ and $m_2$ get large.

\section{General Form of $P(G_{pt,\vec m},q)$}

The generalization of our structural results for two-parameter families of 
planar triangulations, $G_{pt,m_1,m_2}$ to $p$-parameter families is as
follows. Let $G_{pt,\vec m}$ be a family of planar triangulation graphs 
involving the three $\lambda_j$'s in (\ref{lamform}) and depending on the $p$
parameters ${\vec m}=(m_1,...,m_p)$ taking on integer values in the ranges
$m_i \ge (m_i)_{min}$, $i=1,...,p$. Then 
\beq
P(G_{pt,m_1,...m_p},q) =
\sum_{i_1=1}^3 \cdots \sum_{i_p=1}^3 \, c_{G_{pt},i_1...i_p} \, \bigg [
  \prod_{\ell=1}^p \, \lambda_{i_\ell}^{m_\ell} \bigg  ] \ .
\label{pgmvector}
\eeq
In general, there are $3^p$ terms involving products of the $\lambda$'s
(multipled by respective coefficients) in this sum.  

This general form is of considerable interest.  It shows that one can carry out
a $p$-fold sequence of edge proliferations, each of which involves arbitrarily
many additional edges, as indexed by the parameters $m_1,...,m_p$, with the
chromatic polynomial $P(G_{pt,\vec m},q)$ still retaining the rather simple
form (\ref{pgmvector}) with the same set of three $\lambda_i$'s given in
(\ref{lamform}).  This is a much simpler situation than that in previous
calculations of chromatic polynomials for multiparameter families of
graphs. For example, in \cite{nec}, Tsai and one of the present authors
calculated the chromatic polynomial $P(G_{e_1,e_2,e_g,m},q)$ for a certain
four-parameter family of cyclic chain graphs in which each subgraph on the
chain has $e_1$ edges above, and $e_2$ edges below, the main line, with $e_g$
edges between the subgraphs, and $m$ subgraphs in all.  Although the number
$N_\lambda$ of $\lambda_{G,j}$'s for this family has the fixed value of 2, the
$\lambda_{G,j}$'s have functional forms that depend on the parameters $e_1$,
$e_2$, and $e_g$ (as is the case for the full Potts model partition function
\cite{neca}). This property was also found to be true for (i) the chromatic
polynomials $P((Ch)_{k,m,cyc.},q)$ and $P((Ch)_{k,m,Mb.},q)$ of cyclic ($cyc.$)
and M\"obius ($Mb.$) strips depending on a homeomorphic expansion parameter $k$
and the strip length, $m$, where $N_\lambda=4$ and three of the
$\lambda_{Ch,j}$'s depended on $k$ \cite{pg}; and (ii) $P(H_{k,r},q)$ for a
family of ``hammock'' graphs $H_{k,r}$ with $r$ ``ropes'' (linear sets of
edges) joining two end vertices, with each rope having $k$ ``knots''
(vertices), where again the $N_\lambda=2$ terms $\lambda_{H,j}$ depended on $k$
and $r$ \cite{wa2,wa3}.  The remarkable simplicity of the form
(\ref{pgmvector}) is a result of the restrictive property that $G_{pt,\vec m}$
is a planar triangulation.  We know that this simple behavior does not obtain
even for the lowest case of one-parameter families for planar
near-triangulations, from the explicit calculation the chromatic polynomials
for free strips of the triangular lattice of length $m$ and width $L_y=2, \ 3$
\cite{strip} (which are near-triangulations), where it was found that the
$\lambda$'s changed with increasing width \cite{strip}. (Here a
near-triangulation is defined as a graph such that all faces except one are
triangles.)  We also know that it does not hold for nonplanar triangulations,
from explicit calculations of chromatic polynomials for the $L_y=2$ \cite{k},
$L_y=3$ \cite{t}, and $L_y=4$ \cite{tor} strips of the triangular lattice with
doubly periodic (toroidal) boundary conditions.  Thus, chromatic polynomials of
multiparameter families of planar triangulation are especially amenable to
exact analytic treatment.

The $P(G_{pt,m_1,...m_p},q)$ satisfy a $p$-dimensional recursion relation, for
$m_\ell \ge (m_\ell)_{min}+3$, $\ell=1,...,p$, namely
\beqs
& & P(G_{pt,m_1,...m_p},q) + \sum_{i_1=1}^3 \cdots \sum_{i_p=1}^3
b_{G_{pt},i_1...i_p} \, P(G_{pt,m_1-i_1,...,m_p-i_p},q) = 0 \cr\cr
& &
\label{pdimrecursionrel}
\eeqs
where the $b_{G_{pt},i_1...i_p}$ are given by
\beq
 1 + \sum_{i_1=1}^3 \cdots \sum_{i_p=1}^3 b_{G_{pt},i_1...i_p} \,
(\prod_{s=1}^p \, x_s^{i_s}) =
\prod_{\ell=1}^p \Bigg [ \prod_{i=1}^3 (1-\lambda_i \, x_\ell) \Bigg ] \ .
\label{blampdim}
\eeq

Using the same methods as for $p=2$, it is straightforward to generalize our
results to this case, including (i) the conditions on the coefficients
$c_{G_{pt}, \vec i}$ (where $\vec i \equiv (i_1...i_p)$) derived from the
evaluations $P(G_{pt,\vec m},q) = 0$ for $q=0, \ 1, \ 2$ and the Tutte upper
bound at $q=\tau+1$, and (ii) the results for $r(G_{pt,\vec m})$ and its limits
as one or more of the $m_i \to \infty$.  Clearly, our result on a real zero in
the interval $[q_w,3)$ that approaches $\tau+1$ also generalizes to this case
of families $G_{pt,\vec m}$ with $p \ge 3$.

As in the $p=2$ case, if one holds all but one of the $m_1,...,m_p$ fixed and
allows one to vary, then the general form (\ref{pgmvector}) reduces to
(\ref{pgfacform}) with $m$ being equal to the variable parameter, up to an
appropriate integer shift.

\section{The Two-Parameter Family $D_{m_1,m_2}$} 

\begin{figure}
  \begin{center}
    \includegraphics[height=6cm]{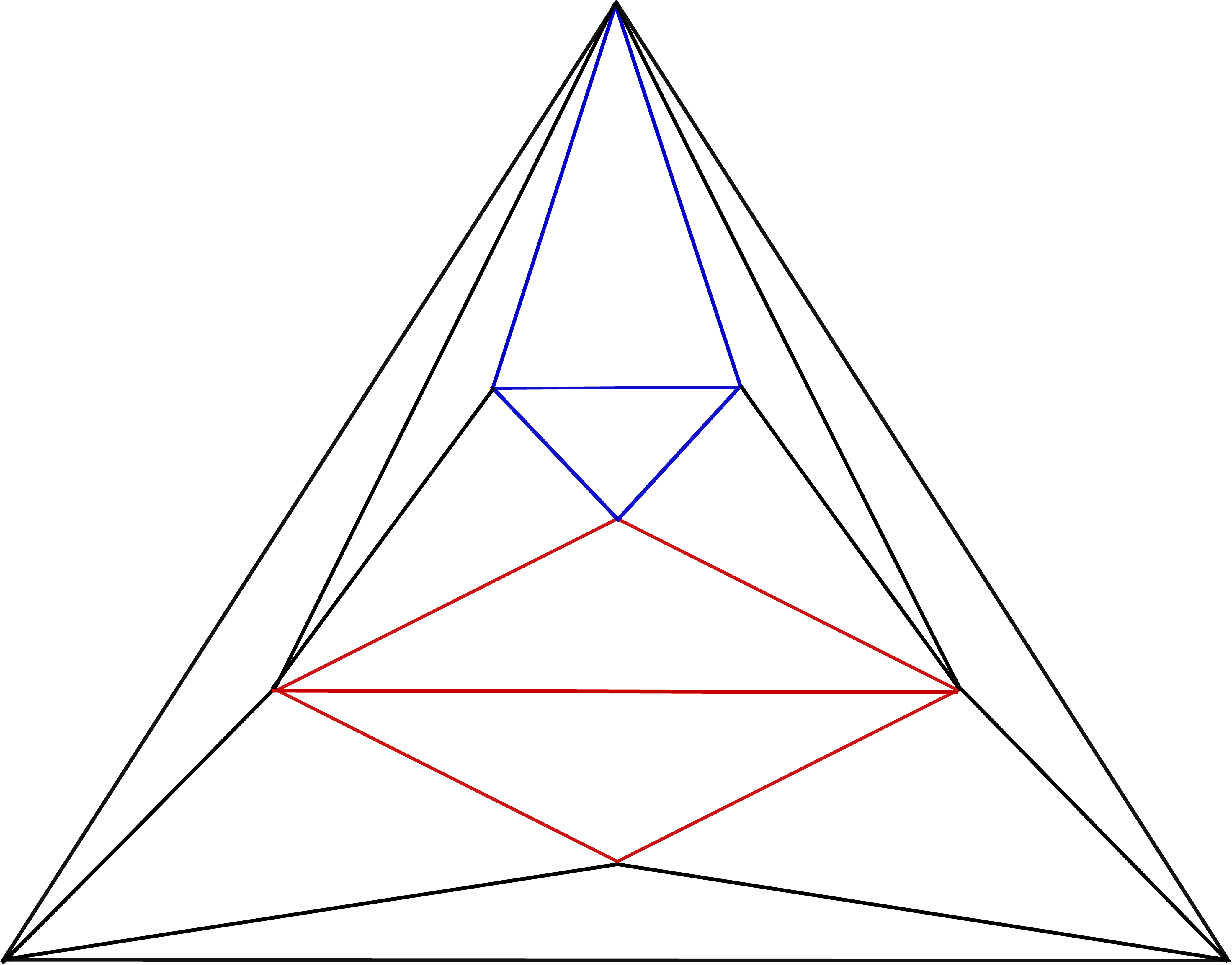}
  \end{center}
\caption{Graph $D_{0,0}$.}
\label{d00n9}
\end{figure}

\begin{figure}
  \begin{center}
    \includegraphics[height=6cm]{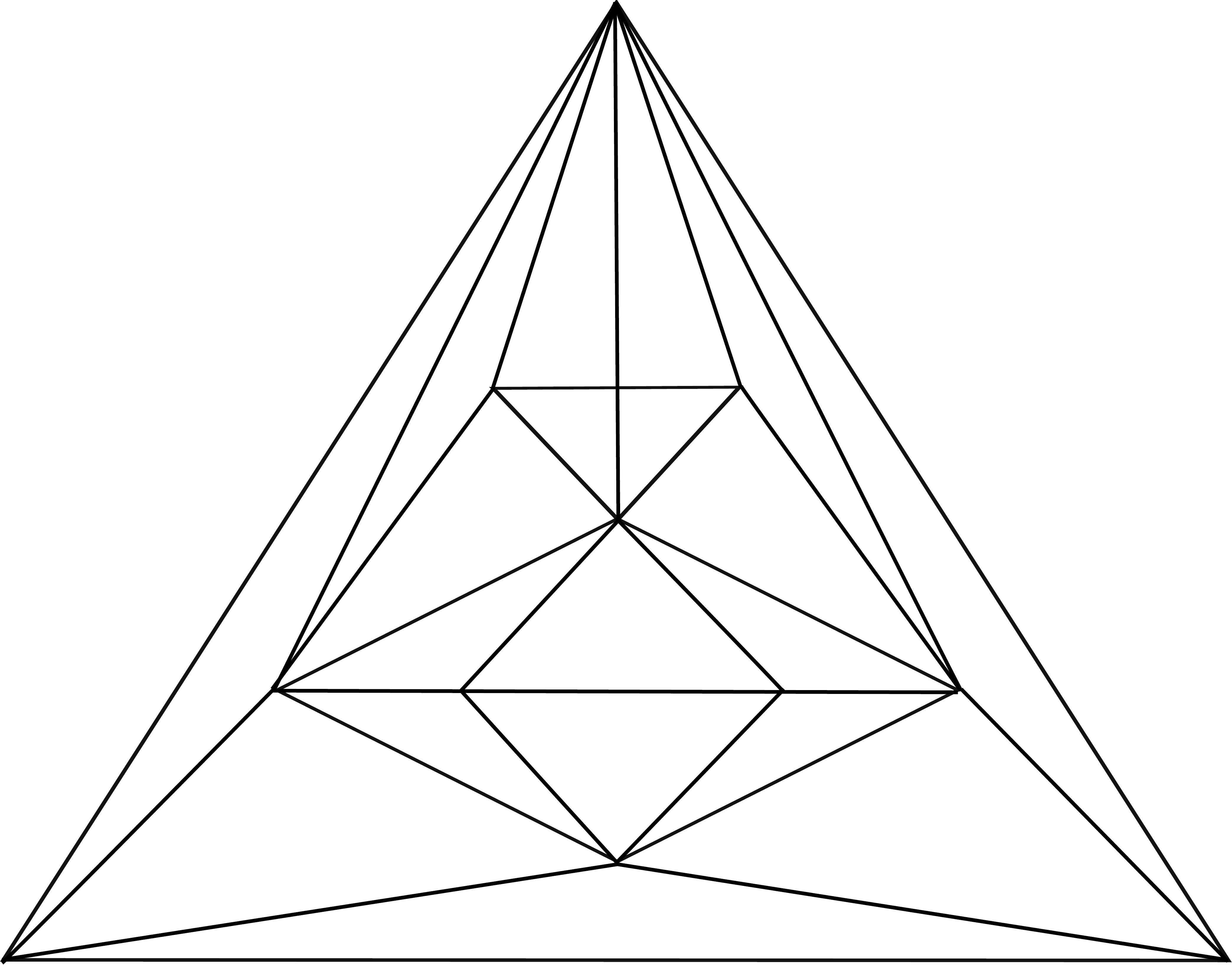}
  \end{center}
\caption{Graph $D_{1,2}$.}
\label{d12n12}
\end{figure}

\begin{figure}
  \begin{center}
    \includegraphics[height=6cm]{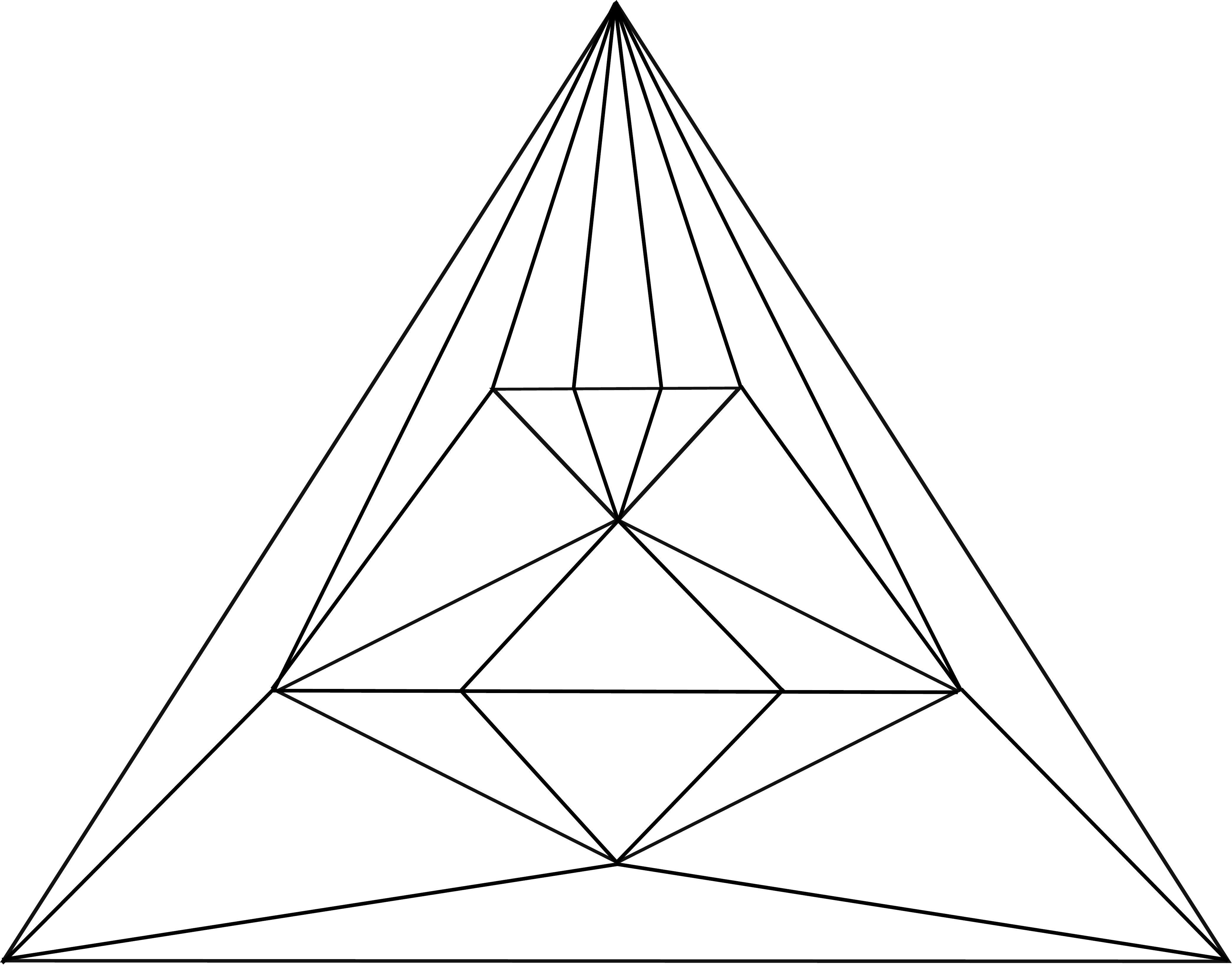}
  \end{center}
\caption{Graph $D_{2,2}$.}
\label{d22n13}
\end{figure}

We proceed to analyze our first explicit two-parameter family of planar
triangulations, denoted $D_{m_1,m_2}$ (where $D$ stands for the proliferation
of a double set of edges).  To explain the general method of construction of
this family, we show in Fig. \ref{d00n9} the lowest member of the series,
namely the graph $D_{0,0}$.  We now add $m_1$ inner edges joining the uppermost
vertex to the upper horizontal edge (thereby producing several such upper
horizontal edges) and, separately, add $m_2$ inner edges joining the central
vertex to the lower horizontal edge (thereby producing several such lower
horizontal edges), with corresponding edges connecting to the lower central
vertex.  Thus, in the $D_{m_1,m_2}$ graph, the uppermost vertex has degree
$6+m_1$, the central vertex has degree $m_1+m_2+4$, and the lower central
vertex has degree $4+m_2$. To illustrate this, we show the graphs $D_{1,2}$,
$D_{2,2}$ in Figs. \ref{d12n12} and \ref{d22n13}.


From inspection of these graphs, it is evident how to construct $D_{m_1,m_2}$
graphs with higher values of $m_1$ and $m_2$.  The number of vertices
in the graph $D_{m_1,m_2}$ is
\beq
n(D_{m_1,m_2}) = m_1 + m_2 + 9 \ .
\label{ndm1m2}
\eeq
For the chromatic number, we find that (i) if $m_2$ is odd, then
$\chi(D_{m_1,m_2})=4$, and (ii) if $m_2$ is even, then $\chi(D_{m_1,m_2})=3$ if
$m_1$ is even and $\chi(D_{m_1,m_2})=4$ if $m_1$ is odd.  That is, denoting
even as $e$ and odd as $o$, $\chi(D_{m_1,m_2})=3$ for $(m_1,m_2)=(e,e)$ and
$\chi(D_{m_1,m_2})=4$ for $(m_1,m_2)=(e,o)$, $(o,o)$, and $(o,e)$.  The
chromatic polynomials for the $D_{m_1,m_2}$ with $\chi=4$ contain a factor
$P(K_4,q)=q(q-1)(q-2)(q-3)$ (and some contain an additional factor of $(q-3)$).
Of course, all chromatic polynomials of triangulations have the factor
$P(K_3,q)=q(q-1)(q-2)$.

By means of an iterative use of the deletion-contraction relation, we
have calculated the chromatic polynomial $P(D_{m_1,m_2},q)$ for an arbitrary
graph in this general two-parameter $D_{m_1,m_2}$ family.
We find that $P(D_{m_1,m_2},q)$ has the form (\ref{pgm1m2})
with coefficients $c_{D,ij}$ that are rational functions of $q$. With the
definition (\ref{cijbar}), we calculate
\beq
\bar c_{D,11}=\frac{(q-2)^7}{q-1} \ , 
\label{c11bar}
\eeq
\beq
\bar c_{D,22}=\frac{(q-1)(q-3)^5(q^3-9q^2+30q-35)}{q-2} \ , 
\label{c22bar}
\eeq
\beq
\bar c_{D,33}=\frac{(q^5-11q^4+46q^3-88q^2+74q-23)(q^2-3q+1)}{(q-1)(q-2)} \ , 
\label{c33bar}
\eeq
\beq
\bar c_{D,12} = (q-2)^3(q-3)^4 \ , 
\label{c12bar}
\eeq
\beq
\bar c_{D,21} = (q-2)(q-3)^6 \ , 
\label{c21bar}
\eeq
\beq
\bar c_{D,13} = \frac{(q-2)^3(q^2-3q+1)}{q-1} \ , 
\label{c13bar}
\eeq
\beq
\bar c_{D,31} = \frac{(q-2)(q^2-3q+1)}{q-1} \ , 
\label{c31bar}
\eeq
\beq
\bar c_{D,23} = -\frac{(q-3)^4(q-5)(q^2-3q+1)}{q-2} \ , 
\label{c23bar}
\eeq
and
\beq
\bar c_{D,32} = -\frac{(q-3)^2(q-5)(q^2-3q+1)}{q-2} \ . 
\label{c32bar}
\eeq
It should be noted that the poles at $q=1$ and $q=2$ in certain of these
$c_{D,ij}$ coefficients are cancelled in the actual evaluation of
$P(D_{m_1,m_2},q)$, which is, as it must be, a polynomial in $q$.  Furthermore,
not only are these poles cancelled, but also the resultant $P(D_{m_1,m_2},q)$
vanishes at $q=0$, $q=1$, and $q=2$.  It is easily verified that the
coefficients $c_{D,ij}$ satisfy the requisite conditions
(\ref{pgm1m2q1conditions}) and (\ref{pgm1m2q2conditions}) for these zeros to
occur. 

Furthermore, since $c_{D,11}=c_{D,33}=c_{D,13}=c_{D,31}=3/2$ with the other
$c_{D,ij}=0$ at $q=3$, it follows that 
\beq
P(D_{m_1,m_2},3) = \frac{3}{2} \bigg [
1 + (-1)^{m_1+m_2} + (-1)^{m_2} +(-1)^{m_1} \bigg ] \ . 
\label{pdm1m2q3}
\eeq
This vanishes for $(m_1,m_2)=(e,o)$, $(o,o)$, and $(o,e)$, and is nonvanishing
for $(m_1,m_2)=(e,e)$ in our notation above, in agreement with our result on
the chromatic number $\chi(D_{m_1,m_2})$. 

We proceed to discuss the evaluation of the chromatic polynomial
$P(D_{m_1,m_2},q)$ at $q=\tau+1$ and the comparison with the Tutte upper
bound.  We have
\beqs
P(D_{m_1,m_2},\tau+1) & = & (9-4\sqrt{5} \ ) (\tau-1)^{m_1+m_2} + \bigg (
\frac{445-199\sqrt{5}}{2} \ \bigg ) (\tau-2)^{m_1+m_2} \cr\cr
& + & \bigg ( \frac{-38+17\sqrt{5}}{2} \bigg ) (\tau-1)^{m_1}(\tau-2)^{m_2}
\cr\cr
& + &  \bigg ( \frac{-199+89\sqrt{5}}{2} \bigg ) (\tau-2)^{m_1}(\tau-1)^{m_2}
\label{pdm1m2tp1}
\eeqs
Comparing this with the Tutte upper bound $(\tau-1)^{m_1+m_2+4}$, we have
\beqs
r(D_{m_1,m_2}) & = & \frac{3-\sqrt{5}}{2} + 
\bigg ( \frac{65-29\sqrt{5}}{2} \ \bigg ) 
\bigg ( \frac{1-\sqrt{5}}{2} \ \bigg )^{m_1+m_2} \crcr
& + & \bigg ( \frac{-11+5\sqrt{5}}{2} \ \bigg )  
      \bigg ( \frac{1-\sqrt{5}}{2} \ \bigg )^{m_2} \cr\cr
& + & \bigg ( \frac{-29+13\sqrt{5}}{2} \ \bigg ) 
      \bigg ( \frac{1-\sqrt{5}}{2} \ \bigg )^{m_1} \ . 
\label{rdm1m2}
\eeqs
We list a number of values of $r(D_{m_1,m_2})$ in Table \ref{dm1m2ratios}.  One
may investigate the behavior of $r(D_{m_1,m_2})$ as $m_1 \to \infty$ for fixed
$m_2$ and as $m_2 \to \infty$ for fixed $m_1$.  Because the quantity
$(1-\sqrt{5} \ )/2$ that is raised to the powers indicated in (\ref{rdm1m2}) is
negative, it follows that, if one keeps $m_2$ fixed and increases $m_1$, then
$r(D_{m_1,m_2})$ does not approach $r(D_{\infty,m_2})$ monotonically, although
the members of the subsequences $r(D_{m_1,m_2})$ with even (odd) $m_1$ approach
$r(D_{\infty,m_2})$ monotonically from above (below), respectively.  Similarly,
if one keeps $m_1$ fixed and increases $m_2$, then $r(D_{m_1,m_2})$ does not
approach $r(D_{m_1,\infty})$ monotonically, although the members of the
subsequences $r(D_{m_1,m_2})$ with even (odd) $m_2$ approach
$r(D_{m_1,\infty})$ monotonically from above (below), respectively.  The
results are
\beq
r(D_{\infty,m_2}) \equiv \lim_{m_1 \to \infty} r(D_{m_1,m_2}) = 
\frac{3-\sqrt{5}}{2} 
+ \bigg ( \frac{-11+5\sqrt{5}}{2} \ \bigg ) 
\bigg ( \frac{1-\sqrt{5}}{2} \ \bigg )^{m_2} 
\label{rm1infm2fixed}
\eeq
\beq
r(D_{m_1,\infty}) \equiv \lim_{m_2 \to \infty} r(D_{m_1,m_2}) = 
\frac{3-\sqrt{5}}{2} 
+ \bigg ( \frac{-29+13\sqrt{5}}{2} \ \bigg ) 
\bigg ( \frac{1-\sqrt{5}}{2} \ \bigg )^{m_1} 
\label{rm2infm1fixed}
\eeq
In the limit where one takes both $m_1$ and $m_2$ to $\infty$, one has
\beqs
r(D_{\infty,\infty}) & = &  
\lim_{m_1 \to \infty} \lim_{m_2 \to \infty} r(D_{m_1,m_2}) = 
\lim_{m_2 \to \infty} \lim_{m_1 \to \infty} r(D_{m_1,m_2}) \cr\cr
& = & \frac{3-\sqrt{5}}{2} = 2-\tau = 0.381966...
\label{rdinfinf}
\eeqs
As $m_2$ increases from 0 to $\infty$, $r(D_{\infty,m_2})$ decreases
(non-monotonically) from the value 
\beq
r(D_{\infty,0}) = -4+2\sqrt{5} = 0.4721359...
\label{rdinfzero}
\eeq
to the value in (\ref{rdinfinf}), and as $m_1$ increases from
0 to $\infty$, $r(D_{m_1,\infty})$ decreases (non-monotonically) from the value
\beq
r(D_{0,\infty}) = -13+6\sqrt{5} = 0.4164078...
\label{rdzeroinf}
\eeq
to the value in (\ref{rdinfinf}).  As a consequence of the relation 
(\ref{cdrel}) (see below), it follows that
\beq
r(D_{\infty,k+2}) = r(D_{k,\infty})
\label{rdrel}
\eeq
In general, the maximal value of $r(D_{m_1,m_2})$ occurs for the member of the
$D_{m_1,m_2}$ family with the minimal values of $m_1$ and $m_2$, namely for
$D_{0,0}$. This property is similar to the property that the maximum value of
$r(G_{pt})$ for all planar triangulations $G_{pt}$ occurs for the $G_{pt}$ with
the minimum number of vertices, namely the single triangle, $K_3$.

\begin{table}
\caption{\footnotesize{Values of the ratio $r(D_{m1,m2})$. The rows and columns
list $m_1$ and $m_2$, respectively, so that, for example, $r(D_{1,2})$ is the
entry 0.3769.}}
\begin{center}
\begin{tabular}{|c|c|c|c|c|c|c|c|c|}
\hline\hline
$m_1,m_2$ & 0   &   1    &   2    &   3    &   4   &    5    &    6   
& $\infty$ \\
\hline\hline
0     & 0.5836 & 0.3131 & 0.4803 & 0.3769 & 0.4408 & 0.4013 & 0.4257 & 0.4164 
\\
1     & 0.4033 & 0.3344 & 0.3769 & 0.3506 & 0.3669 & 0.3568 & 0.3631 & 0.3607 
\\
2     & 0.5147 & 0.3212 & 0.4408 & 0.3669 & 0.4126 & 0.3843 & 0.4018 & 0.3951
\\
3     & 0.4458 & 0.3293 & 0.4013 & 0.3568 & 0.3843 & 0.3673 & 0.3778 & 0.3738
\\
4     & 0.4884 & 0.3243 & 0.4257 & 0.36305& 0.4018 & 0.3778 & 0.3926 & 0.3870
\\
5     & 0.4621 & 0.3274 & 0.4106 & 0.3592 & 0.3910 & 0.3714 & 0.3835 & 0.3789
\\
6     & 0.4783 & 0.3255 & 0.4200 & 0.3616 & 0.3977 & 0.3754 & 0.38915& 0.3839
\\
$\infty$& 0.4721 & 0.3262 & 0.4164 & 0.3607 & 0.3951 & 0.3738 & 0.3870 & 0.3820 \\
\hline\hline
\end{tabular}
\end{center}
\label{dm1m2ratios}
\end{table}

We next show that our general form for $P(D_{m_1,m_2},q)$ reduces to
(\ref{pgfacform}) when either $m_2$ is held fixed and $m_1$ varies, or vice
versa.  If we keep $m_2$ fixed and vary $m_1$, then we can write 
$P(D_{m_1,m_2},q)$ as 
\beq
P(D_{m_1,m_2},q) = 
\sum_{i=1}^3 \, \bigg [ \sum_{j=1}^3 \, c_{D,ij} \,  \lambda_j^{m_2}\bigg ] \, 
\lambda_i^{m_1} \ . 
\label{pdm1withm2fixed}
\eeq
The sum $\sum_{j=1}^3 c_{D,ij} \, \lambda_j^{m_2}$ contains a factor 
$\lambda_i^4$, which we combine with the $\lambda_i^{m_1}$, to make
$\lambda_i^m$, where
\beq
m = m_1+4 \ . 
\label{mreldm1}
\eeq
This shows that $P(D_{m_1,m_2},q)$ has the form (\ref{pgfacform}) with $m$
given by (\ref{mreldm1}); explicitly, 
\beq
P(D_{m-4,m_2},q) = \sum_{i=1}^3 c_{D_{m_2(\ell)},i} \, \lambda_i^m \ , 
\label{pdmwithm2fixed}
\eeq
where
\beq c_{D_{m_2 (\ell)},i} = \lambda_i^{-4} \, \sum_{j=1}^3 c_{D,ij} \,
\lambda_j^{m_2} \ . 
\label{cdiwithm2fixed}
\eeq
Here, since these coefficients depend only on $m_2$, and not on $m_1$, we have
introduced the notation $D_{m_2 (\ell)}$ to refer to the entire family
$D_{m_1,m_2}$ with fixed $m_2$ and variable $m_1$, where $\ell$ indicates
that $m_2$ describes the edge proliferation in the \underline{l}ower
part of the graph.  Expressing (\ref{kappa1})-(\ref{kappa3}) in our 
notation (and suppressing the $q$ arguments), we have 
\beq
c_{D_{m_2(\ell)},1} = q \, \kappa_{D_{m_2(\ell)},1} \ , 
\label{cdkappa1withm2fixed}
\eeq
\beq
c_{D_{m_2(\ell)},2} = q(q-1) \, \kappa_{D_{m_2(\ell)},2} \ , 
\label{cdkappa2withm2fixed}
\eeq
and
\beq
c_{D_{m_2(\ell)},3} = q(q^2-3q+1)\, \kappa_{D_{m_2(\ell)},3} \ . 
\label{cdkappa3withm2fixed}
\eeq

Similarly, if we keep $m_1$ fixed and vary $m_2$, then we can write 
$P(D_{m_1,m_2},q)$ as 
\beq
P(D_{m_1,m_2},q) = 
\sum_{j=1}^3 \, \bigg [ \sum_{i=1}^3 \, c_{D,ij} \, \lambda_i^{m_1}\bigg ] \, 
\lambda_j^{m_2} \ . 
\label{pdm2withm1fixed}
\eeq
The sum $\sum_{i=1}^3 c_{D,ij} \, \lambda_i^{m_1}$ contains a factor 
$\lambda_j^2$, which we combine with the $\lambda_j^{m_2}$, to make
$\lambda_j^m$, where, for this one-parameter reduction, 
\beq
m = m_2+2 \ . 
\label{mreldm2}
\eeq
This shows that $P(D_{m_1,m_2},q)$ has the form (\ref{pgfacform}) with $m$
given by (\ref{mreldm2}); explicitly, 
\beq
P(D_{m_1,m-2},q) = \sum_{j=1}^3 c_{D_{m_1(u)},j} \, \lambda_j^m \ , 
\label{pdmwithm1fixed}
\eeq
where 
\beq
c_{D_{m_1 (u)},j} = \lambda_j^{-2} \, \sum_{i=1}^3 c_{D,ij} \, \lambda_i^{m_1} 
\ . 
\label{cdiwithm1fixed}
\eeq
Here again, since these coefficients depend only on $m_1$, and not on $m_2$, we
have introduced the notation $D_{m_1 (u)}$ to refer to the entire family
$D_{m_1,m_2}$ with fixed $m_1$ and variable $m_2$, where $u$ indicates
that $m_1$ describes the edge proliferation in the \underline{u}pper
part of the graph.  As before, we write
\beq
c_{D_{m_1(u)},1} = q \, \kappa_{D_{m_1(u)},1} \ , 
\label{cdkappa1withm1fixed}
\eeq
\beq
c_{D_{m_1(u)},2} = q(q-1) \, \kappa_{D_{m_1(u)},2} \ , 
\label{cdkappa2withm1fixed}
\eeq
and
\beq
c_{D_{m_1(u)},3} = q(q^2-3q+1)\, \kappa_{D_{m_1(u)},3} \ . 
\label{cdkappa3withm1fixed}
\eeq

We find that
\beq
c_{D_{k+2 \, (\ell)},i} = c_{D_{k \, (u)},i} \quad {\rm for} \ 
i = 1, \ 2, \ 3 
\label{cdrel}
\eeq
and thus
\beq
\kappa_{D_{k+2 \, (\ell)},i} = \kappa_{D_{k \, (u)},i} \quad {\rm for} \ 
i = 1, \ 2, \ 3  \ . 
\label{kapparel}
\eeq
However, we note that for arbitrary $q$, 
\beq
P(D_{m_1,m_2},q) \ne P(D_{m_2,m_1},q) \quad {\rm unless} \ \ m_1=m_2 \ . 
\label{pdneq}
\eeq

\begin{table}
\caption{\footnotesize{Values of the ratios $r(D_{m_1,\infty})$ and 
$r(D_{\infty,m_2})$. Note that $r(D_{k,\infty})=r(D_{\infty,k-2})$
for $k \ge 2$.}}
\begin{center}
\begin{tabular}{|c|c|c|}
\hline\hline
$r(D_{\infty,m_2})$, $r(D_{m_1,\infty})$ &  analytic             & numerical \\
\hline\hline
$r(D_{\infty,0})$                  & $-4+2\sqrt{5}$            & 0.472136 \\
$r(D_{\infty,1})$                  & $(-15+7\sqrt{5} \ )/2$    & 0.326238 \\
$r(D_{\infty,2})=r(D_{0,\infty})$  & $-13+6\sqrt{5}$           & 0.416408 \\
$r(D_{\infty,3})=r(D_{1,\infty})$  & $-22+10\sqrt{5}$          & 0.360680 \\
$r(D_{\infty,4})=r(D_{2,\infty})$  & $(-73+33\sqrt{5} \ )/2$   & 0.395122 \\
$r(D_{\infty,5})=r(D_{3,\infty})$  & $-60+27\sqrt{5}$          & 0.373835 \\
$r(D_{\infty,6})=r(D_{4,\infty})$  & $-98+44\sqrt{5}$          & 0.386991 \\
$r(D_{\infty,7})=r(D_{5,\infty})$  & $(-319+143\sqrt{5} \ )/2$ & 0.378860 \\
$r(D_{\infty,8})=r(D_{6,\infty})$  & $-259+116\sqrt{5}$        & 0.383885 \\
$r(D_{\infty,9})=r(D_{7,\infty})$  & $-420+188\sqrt{5}$        & 0.380780 \\
$r(D_{\infty,10})=r(D_{8,\infty})$ & $(-1361+609\sqrt{5}\ )/2$ & 0.382700 \\
$r(D_{\infty,\infty})$             & $(3-\sqrt{5} \ )/2$       & 0.381966 \\
\hline\hline
\end{tabular}
\end{center}
\label{rinfm2andrm1inf}
\end{table}

\section{The Family $D_{m-4,0}$}

We proceed to examine a number of different $D_{m_1,m_2}$ families of planar
triangulations, with $m_2$ held fixed.  Then, we will analyze
analogous families with $m_1$ held fixed, and finally, we will investigate
families in which both $m_1$ and $m_2$ vary together, and are related in a
linear manner.  For a given graph $D_{m_1,m_2}$, one can use either our general
result for $P(D_{m_1,m_2},q)$ above or either of the one-parameter reductions, 
(\ref{pdmwithm2fixed}) or (\ref{pdmwithm1fixed}). However, we shall be
interested in the limits $m_1 \to \infty$ with  $m_2$ fixed, and $m_2 \to
\infty$ with $m_1$ fixed, and, to study these, it is convenient to use the
one-parameter reductions of our general formula. 

We begin with a study of the family $D_{m_1,0} \equiv D_{m-4,0}$ with $m_1 \ge
0$, i.e., $m \ge 4$. From (\ref{ndm1m2}), we have $n(D_{m-4,0})=m+5$.  For the
coefficients that enter into the equation (\ref{pgfacform}), our general
formulas (\ref{cdiwithm2fixed})-(\ref{cdkappa3withm2fixed}) yield 
\beq
\kappa_{D_{0  (\ell)},1} = \kappa_{D_{0(\ell)},2} = \kappa_{D_{0(\ell)},3} = 
               q^3-9q^2+29q-32
\label{kd0}
\eeq
(equal to $\lambda_{TC}$).  Because these coefficients $\kappa_{D_{0(\ell)},j}$
are all the same, $\lambda_{TC}$ is a common factor, so for all $m$, the three
zeros of $\lambda_{TC}$ are zeros of $P(D_{m-4,0},q)$.  Of these, one is real,
namely $q_w$, given in (\ref{qw}).  In accordance with our general analysis
above, $P(D_{m-4,0},q)$ also has a zero, denoted $q_z$, that approaches
$\tau+1$ as $m$ increases. We list this zero for $m=1$ to $m=16$ in Table
\ref{pd0zerotable}.  As is evident from this table, for odd (even) $m$, this
zero is slightly above (below) $\tau+1$.

\begin{table}
\caption{\footnotesize{Location of zero $q_z$ of $P(D_{m-4,0},q)$ 
closest to $\tau+1$, as a function of the number of vertices, $n=m+5$.}}
\begin{center}
\begin{tabular}{|c|c|c|}
\hline\hline
 $n$ &  $q_z$     &  $q_z-(\tau+1)$    \\
\hline\hline
 10  &  2.677815  &    0.05978     \\
 11  &  2.594829  &  $-0.02321$    \\
 12  &  2.636118  &    0.01808     \\
 13  &  2.609130  &  $-0.8904$e-2  \\
 14  &  2.624356  &    0.6322e-2   \\
 15  &  2.614541  &  $-0.3493$e-2  \\
 16  &  2.620356  &    2.322e-3    \\
 17  &  2.616673  &  $-1.361$e-3   \\
 18  &  2.618905  &    0.8713e-3   \\
 19  &  2.617509  &  $-0.5254$e-3  \\
 20  &  2.618364  &    0.3301e-3   \\
 21  &  2.617832  &  $-2.017$e-4   \\
 22  &  2.618160  &    1.2560e-4   \\
 23  &  2.617957  &  $-0.7725$e-4  \\
 24  &  2.618082  &    0.4790e-4   \\
 25  &  2.618004  &  $-2.954$e-5   \\
\hline\hline
\end{tabular}
\end{center}
\label{pd0zerotable}
\end{table}

For the evaluation at $q=\tau+1$, we have 
\beq
P(D_{m-4,0},\tau+1) = (-4+2\sqrt{5} \ )(\tau-1)^m + (3-\sqrt{5} \ )(\tau-2)^m 
\ , 
\label{pd0tp1}
\eeq
so that 
\beq
r(D_{m-4,0}) = -4+2\sqrt{5} + (3-\sqrt{5} \ ) 
\bigg ( \frac{1-\sqrt{5}}{2} \ \bigg )^m \ . 
\label{rdm0}
\eeq
Hence, 
\beq
r(D_{\infty,0}) = -4+2\sqrt{5} = 0.4721359...
\label{rd0infty}
\eeq
with $a_{D_0(\ell)}=1$.

\section{The Family $D_{m-4,1}$}

We continue with a study of the family $D_{m_1,1} \equiv D_{m-4,1}$. 
From (\ref{ndm1m2}), we have $n(D_{m-4,1})=m+6$.  Our general formulas
(\ref{cdiwithm2fixed})-(\ref{cdkappa3withm2fixed}) give the coefficients 
$\kappa_{D_{1(\ell)},j}$ as 
\beq
\kappa_{D_{1(\ell)},1} = (q-3)(q^3-9q^2+30q-35) \ , 
\label{kd1j1}
\eeq
\beq
\kappa_{D_{1(\ell)},2} = q^4-12q^3+58q^2-133q+119 \ , 
\label{kd1j2}
\eeq
and
\beq
\kappa_{D_{1(\ell)},3} = -(q-3)(2q^2-14q+25) \ . 
\label{kd1j3}
\eeq
If $m$ is even, then $P(D_{m-4,1},q)$ has not only the factor
$q(q-1)(q-2)(q-3)$, but also an additional factor of $(q-3)$.  According to our
general analysis above, $P(D_{m-4,1},q)$ has a real zero that approaches
$\tau+1$ as $m \to \infty$. We also derived the result that for sufficiently
large $m$, a chromatic polynomial of the form (\ref{pgfacform}) has another
real zero in the interval $[q_w,3)$ if and only if $\kappa_{G_{pt},3}$ has a
zero in this interval. For the present family, $\kappa_{D_{1(\ell)},3}$ has
zeros at $q=3$ and the complex-conjugate pair $q=(7 \pm i)/2$, but does not
have a zero in the interval $[q_w,3)$, in accordance with the fact that
$P(D_{m-4,1},q)$ also does not have a zero in this interval.

For the evaluation at $\tau+1$, we compute 
\beq
P(D_{m-4,1},\tau+1) = \bigg ( \frac{25+11\sqrt{5}}{2} \ \bigg ) (\tau-1)^m
+ (-9+4\sqrt{5} \ )(\tau-2)^m   \ , 
\label{pdm1tp1}
\eeq
so that 
\beq
r(D_{m-4,1}) = \frac{-15+7\sqrt{5}}{2} + \bigg ( \frac{11-5\sqrt{5}}{2}\
\bigg ) \bigg ( \frac{1-\sqrt{5}}{2} \ \bigg )^m \ . 
\label{rdm1}
\eeq
and 
\beq
r(D_{\infty,1}) = \frac{-15+7\sqrt{5}}{2} = 0.3226238
\label{rd3infty}
\eeq
with $a_{D_{1(\ell)}}=1$. 

\section{The Family $D_{m-4,2}$}

We next study the family $D_{m_1,2} \equiv D_{m-4,2}$ with $m_1 \ge 0$, i.e.,
$m \ge 4$. Note that $D_{1,2}$ is the same as the graph denoted $G_{ce12}$ in
Fig. 8 of \cite{tub}. From (\ref{ndm1m2}), we have $n(D_{m-4,2})=m+7$.  
For this family our general results give 
\beq
\kappa_{D_{2(\ell)},1} = q^5-15q^4+94q^3-303q^2+498q-332 \ , 
\label{kd2j1}
\eeq
\beq
\kappa_{D_{2(\ell)},2} = q^5-15q^4+95q^3-317q^2+553q-398 \ , 
\label{kd2j2}
\eeq
and
\beq
\kappa_{D_{2(\ell)},3} = -(q^4-16q^3+91q^2-225q+206) \ . 
\label{kd2j3}
\eeq
In Table \ref{pd2zerotable} we list (real) zeros of $P(D_{m-4,2},q)$ in the
interval $q \in [q_w,3)$ as a function of $n$.  As proved above, one zero
approaches $\tau+1$ as $m \to \infty$.  In the same limit, our general analysis
above shows that $P(D_{m-4,2},q)$ has real zero(s) in the interval $[q_w,3)$
correponding to the zeros of $\kappa_{D_{2(\ell)},3}$ in this interval.  This
quartic polynomial has a zero at
\beq
q = 2.7227000945...
\label{q2p7227}
\eeq
together with one more real zero at $q = 6.955106..$, outside the interval
$[q_w,3)$, and a complex-conjugate pair.  Hence, $P(D_{m-4,2},q)$ has another
zero in the interval $[q_w,3)$, which is present for $m \ge 5$, and this
approaches the zero of $\kappa_{D_{2(\ell)},3}$ given in (\ref{q2p7227}) as $m
\to \infty$.  For even $m \ge 6$, i.e., odd $n \ge 13$, the real zero near to
this asymptotic value (\ref{q2p7227}) increases toward it, while for odd $m \ge
5$, i.e., even $n \ge 12$, the nearby real decreases toward the asymptotic
value. As noted above, the graph $D_{0,2}$ coincides with the graph $G_{CM,1}$
of \cite{tub}, for which there is no real zero close to $\tau+1$; instead, the
zeros closest to $\tau+1$ comprise a complex-conjugate pair at 
$q=2.641998 \pm 0.014795i$.  Correspondingly, for
$D_{0,2}$ there is no zero $q_z'$ in the interval $[q_w,3)$.  For all of the
$D_{m-4,2}$ with $m \ge 5$ in Table \ref{pd2zerotable}, the first real zero
$q_z$ in the interval $[q_w,3)$ is, in fact, the closest to $\tau+1$.

The evaluation at $q=\tau+1$ yields 
\beq
P(D_{m-4,2},\tau+1) = \bigg ( \frac{-69+31\sqrt{5}}{2} \ \bigg ) (\tau-1)^m
+ (27-12\sqrt{5} \ )(\tau-2)^m   \ , 
\label{pd2tp1}
\eeq
so that 
\beq
r(D_{m-4,2}) = -13+6\sqrt{5} + \bigg ( \frac{21-9\sqrt{5}}{2} \ \bigg ) 
\bigg ( \frac{1-\sqrt{5}}{2} \ \bigg )^m \ . 
\label{rd2m}
\eeq
Hence,
\beq
r(D_{\infty,2}) = -13+6\sqrt{5} = 0.41640786...
\label{rd2infty}
\eeq
with $a_{D_{2(\ell)}}=1$.

\begin{table}[htbp]
\caption{\footnotesize{Location of real zeros of $P(D_{m-4,2},q)$
in the interval $q \in [q_w,3)$, as a function of the number of
vertices, $n=m+7$. Here the notation nz means that there
is no second real zero in the interval $[q_w,3)$.}}
\begin{center}
\begin{tabular}{|c|c|c|}
\hline\hline
 $n$ &  $q_z$     &  $q_z'$    \\
\hline\hline
 11  &  c.c. pair &    nz      \\
 12  &  2.614614  &  2.818897  \\
 13  &  2.621801  &  2.689610  \\
 14  &  2.616506  &  2.762806  \\
 15  &  2.619226  &  2.705035  \\
 16  &  2.6174035 &  2.741044  \\
 17  &  2.618462  &  2.713055  \\
 18  &  2.617785  &  2.731543  \\
 19  &  2.618194  &  2.717464  \\
 20  &  2.617938  &  2.727100  \\
 21  &  2.618094  &  2.719886  \\
 22  &  2.617997  &  2.724931  \\
 23  &  2.618057  &  2.721202  \\
 24  &  2.618020  &  2.723845  \\
\hline\hline
\end{tabular}
\end{center}
\label{pd2zerotable}
\end{table}

\section{The Family $D_{m-4,3}$} 

\begin{figure}
  \begin{center}
    \includegraphics[height=6cm]{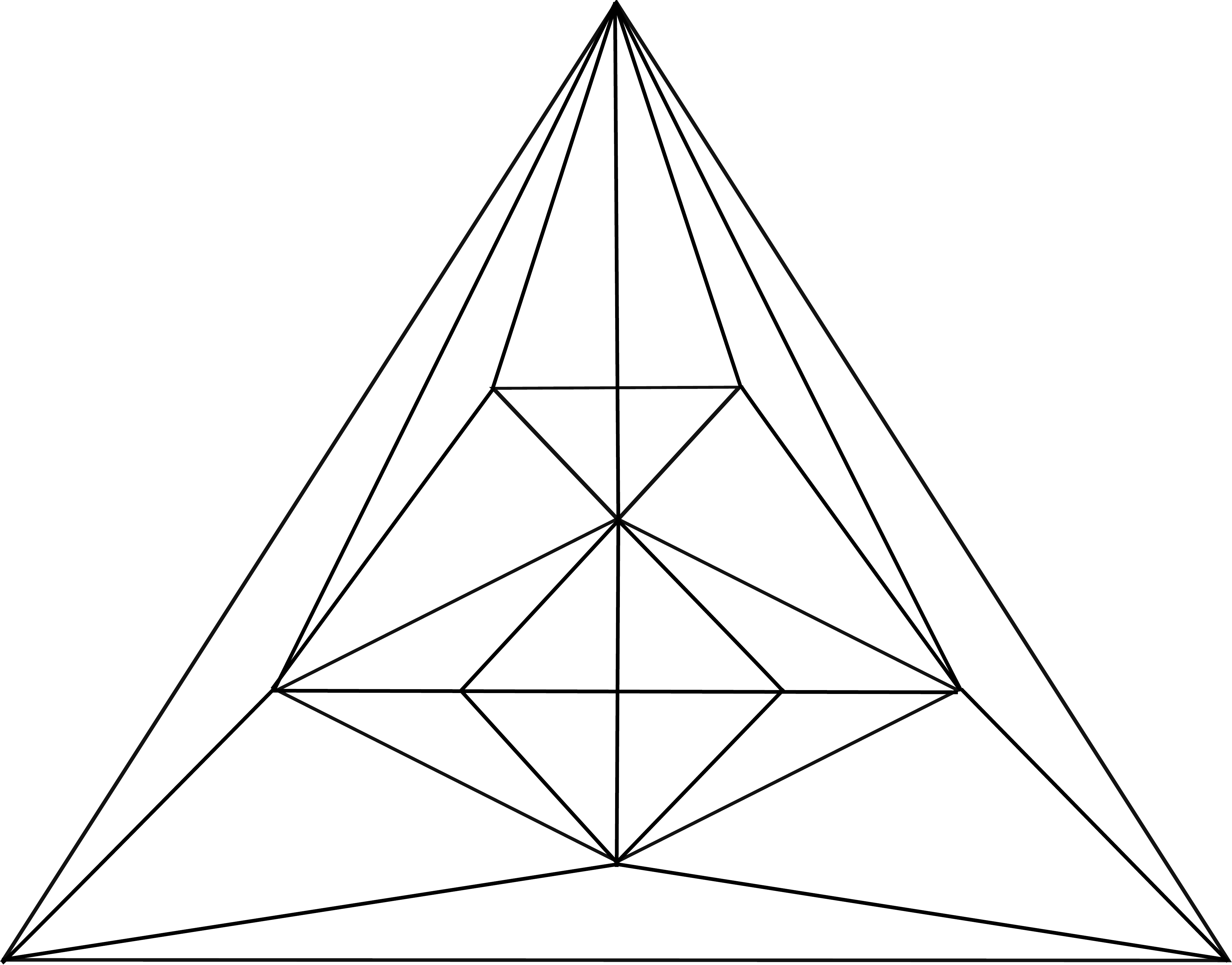}
  \end{center}
\caption{Graph $D_{1,3}$.}
\label{d13n13}
\end{figure}

\begin{figure}
  \begin{center}
    \includegraphics[height=6cm]{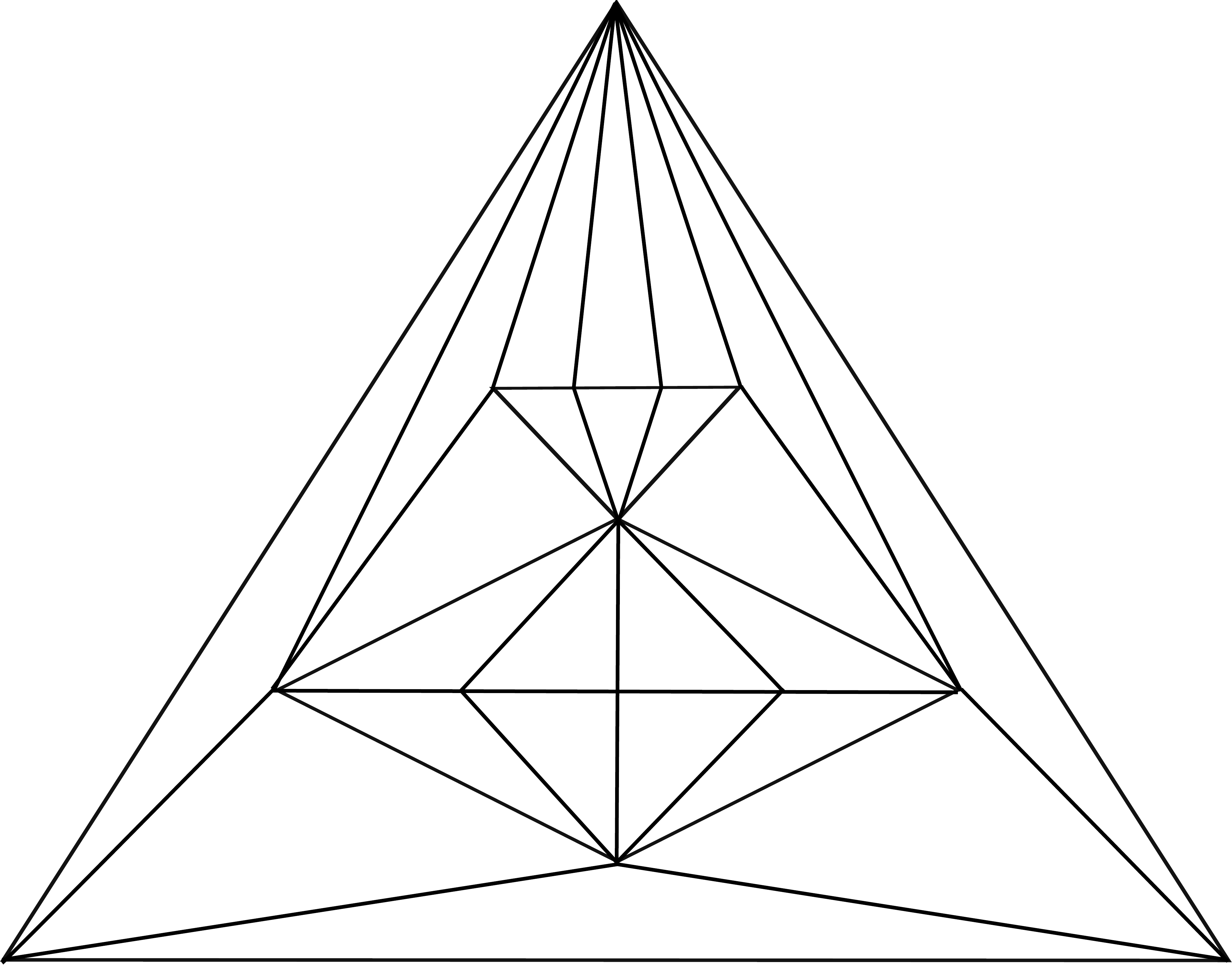}
  \end{center}
\caption{Graph $D_{2,3}$.}
\label{d23n14}
\end{figure}

The final family that we study in this series is $D_{m_1,3} \equiv D_{m-4,3}$,
with $m_1 \ge 0$, i.e., $m \ge 4$.  From (\ref{ndm1m2}), it follows that the
graph $D_{m-4,3}$ has $n(D_{m-4,3})=m+8$.  The graphs $D_{1,3}$ and $D_{2,3}$
are shown in Figs. \ref{d13n13} and \ref{d23n14}.

For the coefficients $\kappa_{D_{3(\ell)},j}$ we have 
\beq
\kappa_{D_{3(\ell)},1} = (q-3)(q^2-5q+7)(q^3-10q^2+38q-49)  \ , 
\label{kd3j1}
\eeq
\beq
\kappa_{D_{3(\ell)},2} = q^6-18q^5+141q^4-613q^3+1551q^2-2152q+1271 \ , 
\label{kd32}
\eeq
and
\beq
\kappa_{D_{3(\ell)},3} = -(q-3)^2(q^3-12q^2+48q-67) \ . 
\label{kd3j3}
\eeq

In Table \ref{pd3zerotable} we list real zeros of $P(D_{m-4,3},q)$ in the
interval $q \in [q_w,3)$.  As is evident in this table, in addition to the zero
that approaches $\tau+1$, there is a second real zero in this interval if and
only if $m$ (and hence $n$) is even.  We can prove that, for the subset of
$D_{m-4,3}$ with even $m$ where this second zero is present, it approaches
$q=3$ from below as $m \to \infty$.  The proof is as follows.  According to our
little theorem above on real chromatic zeros besides the one (or complex pair)
that approach $\tau+1$ as $m \to \infty$, there is a second zero in the
interval $q \in [q_w,3)$ if and only if $\kappa_{_{G_{pt}},3}$ has a real zero
in this interval.  Now for the $D_{m-4,3}$ family, $\kappa_{D_{3(\ell)},3}$ has
no real zero in the interval $[q_w,3)$.  (Its zeros are at $q=3$, with
multiplicity 2, at $q = 5.44224957...$ and at $q \simeq 3.278875 \pm
1.249025$.)  Hence, according to our little theorem, as $m \to \infty$ on even
integers, in addition to the real zero that is near to $\tau+1$, the other zero
must approach 3, so that in this limit, there is no other zero in the interval
$[q_w,3)$.  This family may thus be added to the two known families given (in
his Theorem 4) by Woodall in Ref. \cite{woodall97} as examples of one-parameter
families of graphs, each of which has a chromatic zero that approaches 3 from
below as the parameter ($m$ here) goes to infinity.

\begin{table}[htbp]
\caption{\footnotesize{Location of real zeros of $P(D_{m-4,3},q)$ 
in the interval $q \in [q_w,3)$, as a function of the number of
vertices, $n=m+8$. Notation nz means that there is no
second zero in this interval.}}
\begin{center}
\begin{tabular}{|c|c|c|}
\hline\hline
 $n$ &  $q_z$     &  $q_z'$    \\
\hline\hline
 12  &  2.614614  &  2.818897  \\
 13  &  2.619530  &    nz      \\
 14  &  2.616973  &  2.847527  \\
 15  &  2.618625  &    nz      \\
 16  &  2.617649  &  2.866268  \\
 17  &  2.618264  &    nz      \\
 18  &  2.617889  &  2.880165  \\
 19  &  2.618122  &    nz      \\
 20  &  2.617979  &  2.890985  \\
 21  &  2.618068  &    nz      \\
 22  &  2.618013  &  2.899700  \\
 23  &  2.618057  &    nz      \\
 24  &  2.618020  &  2.906905  \\
 25  &  2.618039  &    nz      \\
 26  &  2.618031  &  2.912980  \\
\hline\hline
\end{tabular}
\end{center}
\label{pd3zerotable}
\end{table}

The evaluation at $q=\tau+1$ yields 
\beq
P(D_{m-4,3},\tau+1) = (94-42\sqrt{5} \ ) (\tau-1)^m 
+ (-76+34\sqrt{5} \ )(\tau-2)^m \ , 
\label{pd3tp1}
\eeq
so that
\beq
r(D_{m-4,3}) = -22+10\sqrt{5} + (18-8\sqrt{5} \ )
\bigg (\frac{1-\sqrt{5}}{2} \ \bigg )^m  \ . 
\label{rd3m}
\eeq
Hence, 
\beq
r(D_{\infty,3}) =  -22+10\sqrt{5} = 0.36067977...
\label{rxminfty}
\eeq
with $a_{D_{3(\ell)}}=1$.

\section{The Family $D_{0,m-2}$}

As an example of families with $m_1$ fixed and variable $m_2$ we discuss the
family $D_{0,m_2} \equiv D_{0,m-2}$. A graph in this family has
$n(D_{0,m-2})=m+7$.  In accord with our result (\ref{kapparel}), 
\beq
\kappa_{D_{0(u)},i} = \kappa_{D_{2(u)},i} \quad {\rm for} \ i=1, \ 2, \ 3 
\ . 
\label{k1dim}
\eeq
$P(D_{0,m-2},q)$ has a real zero near to $\tau+1$, which approaches this point
as $m \to \infty$. Furthermore, since the coefficient $\kappa_{D_{0(u)},3}$ has
a real zero in the interval $[q_w,3)$, at the value in (\ref{q2p7227}), it
follows from our general analysis above that for sufficiently large $m$,
$P(D_{0,m-2},q)$ has a real zero that approaches this value.  These zeros
approach their respective values in a manner similar to that discussed for the
family $D_{m-4,2}$.

\section{A Symmetric Two-Parameter Family $S_{m_1,m_2}$} 

\begin{figure}
  \begin{center}
    \includegraphics[height=6cm]{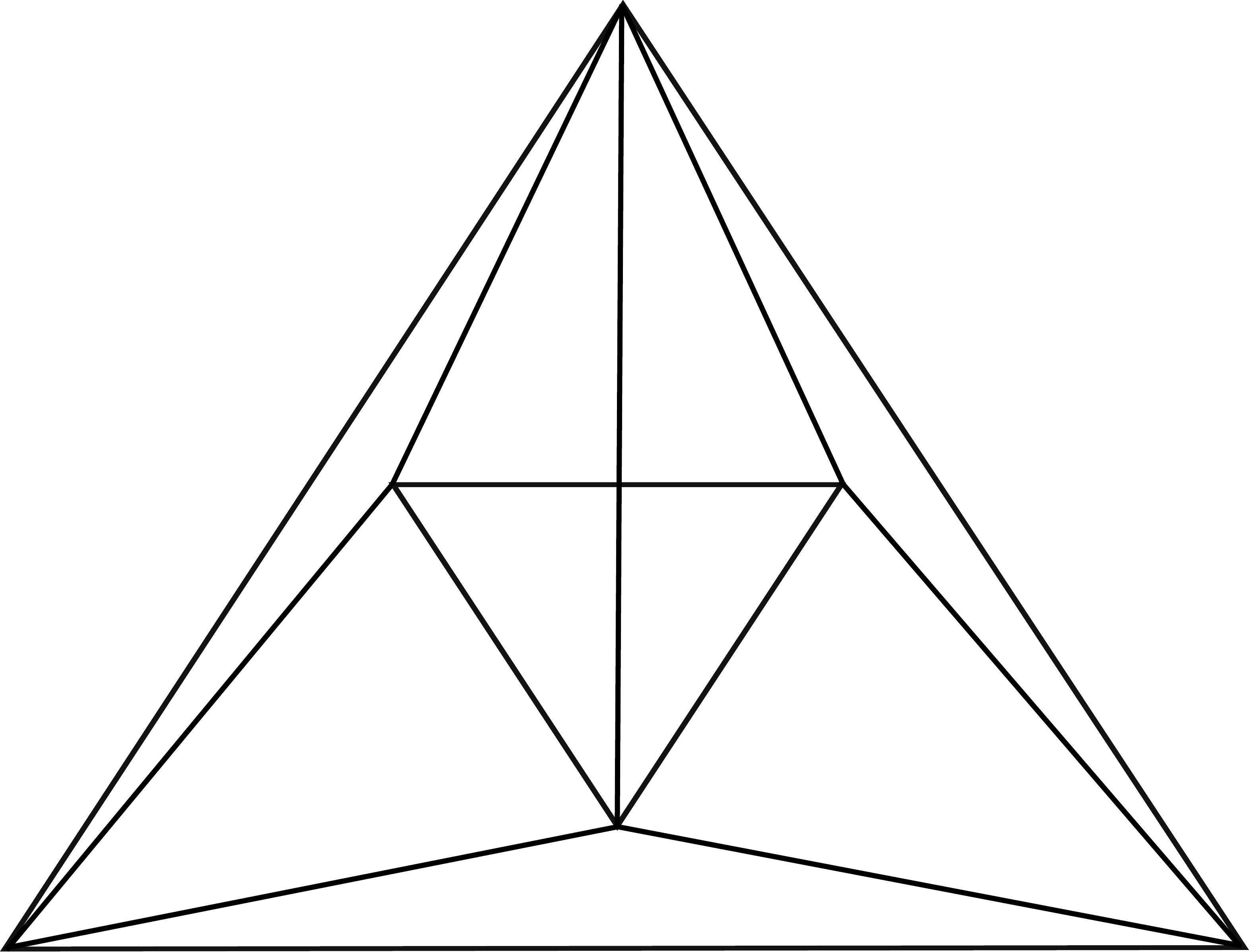}
  \end{center}
\caption{Graph $S_{0,0}$.}
\label{s00n7}
\end{figure}

\begin{figure}
  \begin{center}
    \includegraphics[height=6cm]{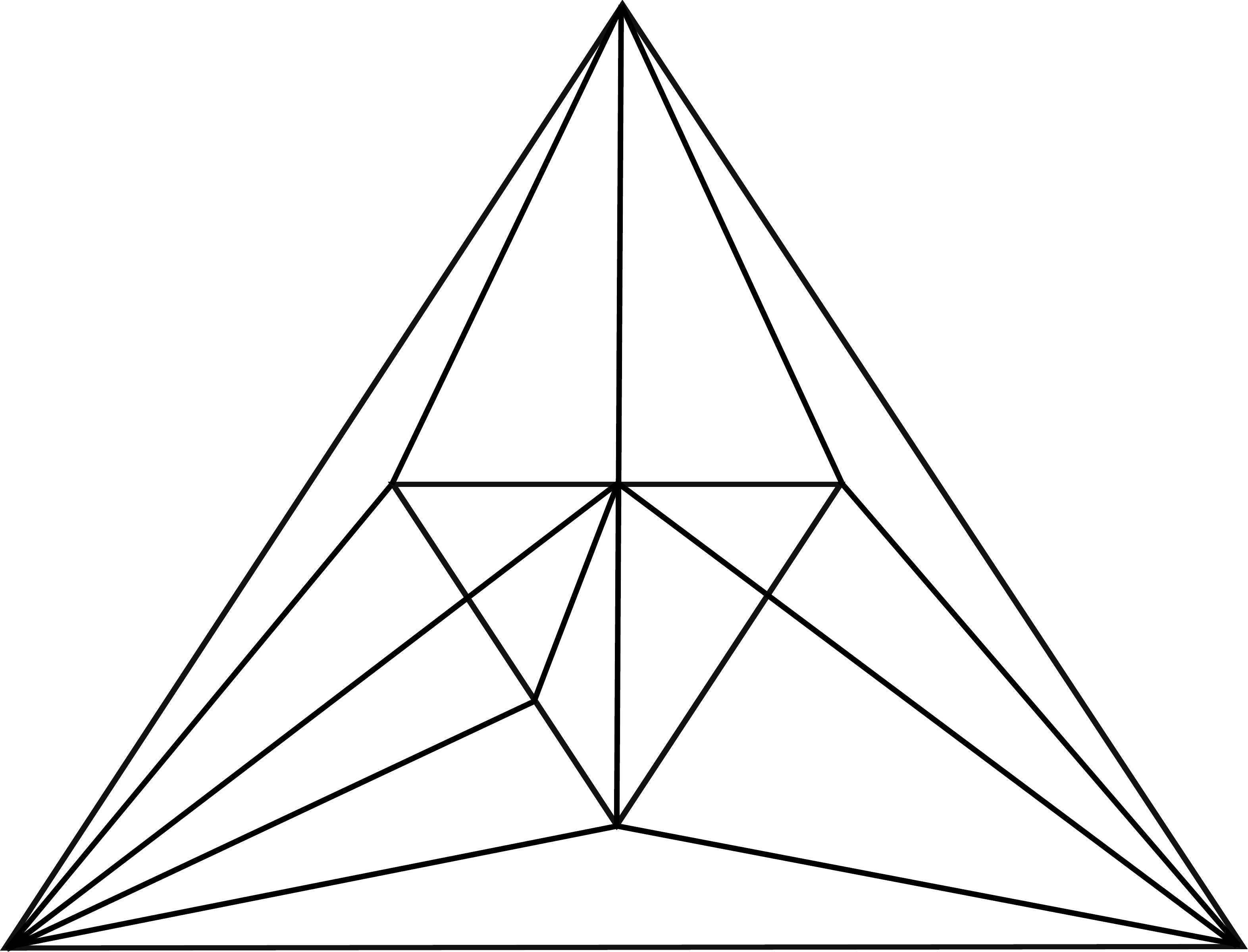}
  \end{center}
\caption{Graph $S_{2,1}$.}
\label{s21n10}
\end{figure}

In this section we study a two-parameter family of planar triangulations
$S_{m_1,m_2}$ which are symmetric under interchange of the parameters: 
\beq
S_{m_1,m_2} = S_{m_2,m_1} \ . 
\label{ssym}
\eeq
We show the lowest member of this family, $S_{0,0}$ in Fig. \ref{s00n7} and
another member, $S_{1,2}=S_{2,1}$ in Fig. \ref{s21n10}. From these it is clear
how to construct the general graph $S_{m_1,m_2}$ in this family.  We have
$n(S_{m_1,m_2}) = m_1+m_2+7$.  The chromatic number is $\chi(S_{m_1,m_2})=4$,
and $P(S_{m_1,m_2},q)$ contains the factor $P(K_4,q)=q(q-1)(q-2)(q-3)$.


We have calculated $P(S_{m_1,m_2},q)$ and find that it involves the same three
$\lambda$'s as in (\ref{lamform}), but with an interestingly different form
than $P(D_{m_1,m_2},q)$. Given the symmetry (\ref{ssym}), it follows that the
coefficients in $P(S_{m_1,m_2},q)$ satisfy 
\beq
c_{S,ij} = c_{S,ji} \ . 
\label{csijsym}
\eeq
Consequently, although there are nine terms of the form 
$\lambda_i^{m_1}\lambda_j^{m_2}$ in $P(S_{m_1,m_2},q)$, there are only six 
independent coefficients $c_{S,ij}$ to begin with, and we find that two of
these vanish, so that there are only four independent, nonvanishing
coefficients $c_{S,ij}$.  Explicitly, we calculate 
\beq
c_{S,ij} = q \bar c_{S,ij} \ , 
\label{csbar}
\eeq
with
\beq
c_{S,11} = c_{S,13} = c_{S,31} = 0 \ , 
\label{cszero}
\eeq
\beq
\bar c_{S,22} = \frac{(q-1)(q-3)^6}{q-2} \ , 
\label{cs22} 
\eeq
\beq
\bar c_{S,12} = \bar c_{S,21} = (q-2)^3(q-3)^2 \ , 
\label{cs12} 
\eeq
\beq
\bar c_{S,23} = \bar c_{S,32} = \frac{(q-3)^2(q^2-3q+1)}{q-2} \ , 
\label{cs23} 
\eeq
and
\beq
\bar c_{S,33} = \frac{(q-1)(q-3)(q^2-3q+1)}{q-2} \ . 
\label{cs33}
\eeq
so that
\beqs
& & P(S_{m_1,m_2},q)=c_{S,22}\lambda_2^{m_1+m_2}+c_{S,33}\lambda_3^{m_1+m_2} 
\cr\cr 
&+& c_{S,12}(\lambda_1^{m_1}\lambda_2^{m_2}+\lambda_2^{m_1} \lambda_1^{m_2}) 
+ c_{S,23}(\lambda_2^{m_1}\lambda_3^{m_2}+\lambda_3^{m_1} \lambda_2^{m_2}) 
\label{ps}
\eeqs
As before, the poles cancel in the calculation of $P(S_{m_1,m_2},q)$ and,
furthermore, these coefficients satisfy the requisite identities so that
$P(S_{m_1,m_2},q)=0$ for $q=1, \ 2, \ 3$.  These are special cases of 
(\ref{pgm1m2q1conditions}) and (\ref{pgm1m2q2conditions}) that incorporate 
the properties that $c_{S,ij}=c_{S,ji}$ and $c_{S,11}=c_{S,13}=c_{S,31}=0$, 
namely, 
\beq
c_{S,22} = 0, \quad c_{S,33} = 0, \quad c_{S,12} + c_{S,23} = 0 \quad {\rm at}
\ \ q=1 
\label{psq1condition}
\eeq
and
\beq
c_{S,22}+2c_{S,12}+c_{S,33}=0, \quad c_{S,12}=0  \quad {\rm at} \ \ q=2 \ .
\label{psq2condition}
\eeq
As a special case of (\ref{cijqtp1}) we also have 
\beq
c_{S,23} = c_{S,32} = c_{S,33} = 0 \quad {\rm at} \ \ q=\tau+1 \ . 
\label{psqtp1condition}
\eeq
Finally, the condition $P(S_{m_1,m_2},3)=0$ is equivalent to 
\beq
c_{S,33} = 0 \quad {\rm at} \ \ q=3
\label{psq3condition}
\eeq

For the comparison of $P(S_{m_1,m_2},q)$ at $q=\tau+1$ with the Tutte upper
bound $(\tau-1)^{m_1+m_2+2}$, we have
\beqs
& & r(S_{m_1,m_2}) = \bigg | (9-4\sqrt{5} \ ) 
\bigg ( \frac{1-\sqrt{5}}{2} \ \bigg )^{m_1+m_2} \cr\cr
& + & (-2+\sqrt{5} \ ) \bigg [ 
  \bigg ( \frac{1-\sqrt{5}}{2} \ \bigg )^{m_1} 
+ \bigg ( \frac{1-\sqrt{5}}{2} \ \bigg )^{m_2} \bigg ] \bigg | 
\label{rs}
\eeqs
This decreases (non-monotically) in magnitude as $m_1$ increases for fixed
$m_2$ and as $m_2$ increases for fixed $m_1$, approaching zero exponentially
rapidly as either of these parameters goes to infinity.  Thus,
\beq
\lim_{m_1 \to \infty} r(S_{m_1,m_2}) = \lim_{m_2 \to \infty} r(S_{m_1,m_2}) = 0
\ . 
\label{rsinf}
\eeq

It is also of interest to analyze the one-parameter reductions of
$P(S_{m_1,m_2},q)$ for variable $m_1$ and fixed $m_2$ and vice versa. These 
yield identical results, because of the symmetry (\ref{ssym}). Hence, 
without loss of generality we consider variable $m_1$ and fixed $m_2$ and find
that in this case $P(S_{m_1,m_2},q)$ reduces to (\ref{pgfacform}) with 
$m=m_1+2$, which we write as $P(S_{m-2,m_2},q)$. Since the coefficients 
only depend on $m_2$ and not $m$, we denote them by $c_{S_{m_2},i}$. They are 
given by 
\beq
c_{S_{m_2},i} = \lambda_i^{-2} \, \sum_{j=1}^3 c_{S,ij} \, \lambda_j^{m_2} 
\label{csiwithm1fixed}
\eeq
Thus, in terms of the corresponding $\kappa_{S_{m_2},i}$, 
\beq
\kappa_{S_0,1} = (q-2)(q-3)^2 \ , 
\label{cs1_m2eq0}
\eeq
\beq
\kappa_{S_0,2} = \lambda_{TC}=q^3-9q^2+29q-32 \ , 
\label{cs2_m2eq0}
\eeq
\beq
\kappa_{S_0,3} = 2(q-3) \ . 
\label{cs3_m2eq0}
\eeq
and so forth for higher values of $m_2$. 

This family exhibits a number of interesting properties.  Among these is the
fact that out of the possible $3^2$ terms in (\ref{pgmvector}) for $p=2$, some
may be absent because of vanishing coefficients $c_{G,ij}$.  In particular, the
term $c_{G,11}\lambda_1^{m_1+m_2}$ that would be dominant in the limit where 
the parameters $m_1 \to \infty$ and $m_2 \to \infty$, may be absent, so that
in this limit, $r(G_{\infty,\infty})$ may be zero.

\section{Families of the form $G_{pt,m_1,m_2}$ with $m_1=m_2$}

In previous sections we have analyzed the chromatic polynomials of special
cases of two-parameter families of planar triangulations $G_{pt,m_1,m_2}$ as a
function of $m_1$ with $m_2$ held fixed, and vice versa and shown how they
reduce to (\ref{pgform}) with $j_{max}=3$. A different type of special case in
which $G_{pt,m_1,m_2}$ reduces to a one-parameter family is obtained by
requiring that $m_1$ and $m_2$ be linearly related to each other. The simplest
such example of this type of reduction is the diagonal case obtained by
requiring that $m_1=m_2$. For general families $G_{pt,m_1,m_2}$ that satisfy
(\ref{nmmrel}) and for which $P(G_{pt,m_1,m_2},q)$ is of the form
(\ref{pgm1m2}), it follows that $n(G_{pt,k,k}) = 2k+\beta$ and that
$P(G_{pt,m_1,m_1},q)$ reduces, to the form (\ref{pgform}) with $j_{max}=6$. 
We use the shorthand $G_d$ to denote a generic $G_{pt,m_1,m_1}$.  We have 
\beq
P(G_{d,m_1,m_1},q) = \sum_{j=1}^6 c_{G_d,j} (\lambda_{G_d,j})^m
\label{pgdform}
\eeq
where $m = m_1 + \delta m$, with $\delta m$ depending on the family, and 
\beq
\lambda_{G_d,1}=\lambda_1^2 = (q-2)^2 \ , 
\label{lamgd1}
\eeq
\beq
\lambda_{G_d,2}=\lambda_2^2 = (q-3)^2 \ , 
\label{lamgd2}
\eeq
\beq
\lambda_{G_d,3}= \lambda_3^2 = 1 \ , 
\label{lamgd3}
\eeq
\beq
\lambda_{G_d,4}=\lambda_1 \lambda_2 = (q-2)(q-3) \ , 
\label{lamgd4}
\eeq
\beq
\lambda_{G_d,5}=\lambda_1\lambda_3 = -(q-2) \ , 
\label{lamgd5}
\eeq
and
\beq
\lambda_{G_d,6}=\lambda_2\lambda_3 = -(q-3) \ . 
\label{lamgd6}
\eeq
The corresponding coefficients are 
\beq
c_{G_d,j} = q \bar c_{G_d,j} \quad {\rm for} \ j=1,..,6, 
\label{cdjbar}
\eeq
with
\beq
c_{G_d,1}= c_{G,11} \ , 
\label{cgd1}
\eeq
\beq
c_{G_d,2}= c_{G,22} \ ,
\label{cgd2}
\eeq
\beq
c_{G_d,3}= c_{G,33} \ , 
\label{cgd3}
\eeq
\beq
c_{G_d,4} = c_{G,12}+c_{G,21} \ , 
\label{cgd4}
\eeq
\beq
c_{G_d,5} = c_{G,13}+c_{G,31} \ , 
\label{cgd5}
\eeq
and
\beq
c_{G_d,6} = c_{G,23}+c_{G,32} \ . 
\label{cgd6}
\eeq
The coefficients $c_{G_d,i}$, $i=1,...,6$ satisfy various conditions that 
follow from those that we have derived for the coefficients 
$c_{G,ij}$ in (\ref{cgijqfactor}), (\ref{pgm1m2q1conditions}), 
(\ref{pgm1m2q2conditions}), and (\ref{cijqtp1}).  These are
\beqs
& & c_{G_d,1}+c_{G_d,3}+c_{G_d,5}=0, \quad c_{G_d,2}=0, \cr\cr
& & c_{G_d,4}+c_{G_d,6}=0 \quad {\rm at} \ \ q=1 \ , 
\label{cgdq1conditions}
\eeqs
\beqs
& & c_{G_d,1}=0, \quad c_{G_d,2}+c_{G_d,3}+c_{G_d,6}=0, \cr\cr
& & c_{G_d,4}+c_{G_d,6}=0 \quad {\rm at} \ \ q=2 \ ,
\label{cgdq2conditions}
\eeqs
and
\beq
c_{G_d,3} = c_{G_d,5} = c_{G_d,6} = 0 \quad {\rm at} \ \ q=\tau+1 \ . 
\label{cdgtp1}
\eeq
Hence, 
\beq
c_{G_d,2} \quad {\rm contains \ the \ factor} \ \ q-1 \ , 
\label{cgd2qm1factor}
\eeq
\beq
c_{G_d,1} \quad {\rm contains \ the \ factor} \ \ q-2 \ , 
\label{cgd1qm2factor}
\eeq
and
\beq
c_{G_d,i} \quad {\rm contains \ the \ factor} \ \ q^2-3q+1 \ \ {\rm if} \ \ 
i=3, \ 5, \ 6 \ . 
\label{cgdtp1factor}
\eeq

\section{The Families $D_{m_1,m_2}$ and $S_{m_1,m_2}$ with $m_1=m_2$}

We now discuss two explicit examples of the diagonal special case of a
two-parameter planar triangulation family, namely $D_{m_1,m_2}$ and
$S_{m_1,m_2}$ with $m_1=m_2$. We shall use the shorthand notation $D_d$ and
$S_d$ to refer to these entire respective families.  From (\ref{ndm1m2}), we
have $n(D_{m_1,m_1})=2m_1+9$ and $n(S_{m_1,m_1})=2m_1+7$.

The chromatic polynomial $P(D_{m_1,m_1},q)$ has the form (\ref{pgdform}) with
$j_{max}=6$ and $\delta m =0$, i.e., $m=m_1$. The lowest member of this family,
the graph $D_{0,0}$, was shown in Fig. \ref{d00n9} and $D_{2,2}$ was shown in
Fig. \ref{d22n13}.  The chromatic number $\chi(D_{m,m})$ is 3 if $m$ is even
and 4 if $m$ is odd.  The coefficients are
\beq \bar c_{D_d,i}= \bar c_{D,ii}
\quad {\rm for} \ \ i=1, \ 2, \ 3,
\label{cd123}
\eeq
\beq
\bar c_{D_d,4} = \bar c_{12} + \bar c_{21} = (q-2)(q-3)^4(2q^2-10q+13) \ , 
\label{cd4}
\eeq
\beq
\bar c_{D_d,5} = \bar c_{13} + \bar c_{31} = 
\frac{(q-2)(q^2-4q+5)(q^2-3q+1)}{q-1} \ , 
\label{cd5}
\eeq
and
\beq
\bar c_{D_d,6} = \bar c_{23} + \bar c_{32} = 
-\frac{(q-3)^2(q-5)(q^2-6q+10)(q^2-3q+1)}{q-2} \ , 
\label{cd6}
\eeq
Since $\chi(D_{m,m})=4$ for odd $m$, it follows that 
\beq
c_{D_d,1}-c_{D_d,3}+c_{D_d,5}=0 \quad {\rm for} \ q=3 \ . 
\label{cdconditionoddm}
\eeq

We calculate 
\beqs 
& & P(D_{m,m},\tau+1) = (\tau+1)\Bigg [ \bigg (
\frac{47-21\sqrt{5}}{2} \ \bigg ) (\tau-1)^{2m} \cr\cr & + & \bigg (
\frac{1165-521\sqrt{5}}{2} \ \bigg )(\tau-2)^{2m} + (-360+161\sqrt{5} \
)[(\tau-1)(\tau-2)]^m \Bigg ] \cr\cr & &
\label{pdmtp1}
\eeqs
The ratio $r(D_{m,m})$ is 
\beqs
r(D_{m,m}) & = & \frac{3-\sqrt{5}}{2} + \bigg ( \frac{65-29\sqrt{5}}{2} \ 
\bigg ) \bigg ( \frac{1-\sqrt{5}}{2} \ \bigg )^{2m} 
 + (-20+9\sqrt{5} \ )\bigg ( \frac{1-\sqrt{5}}{2} \ \bigg )^m \ . \cr\cr
& & 
\label{rdmm}
\eeqs
Hence, defining $r(D_d,\infty) = \lim_{m \to \infty} r(D_{m,m})$, we have 
\beq
r(D_d,\infty) = \frac{3-\sqrt{5}}{2} = 0.381966...
\label{rdinf}
\eeq
and $a_{D_d}=1$.

In contrast, the chromatic polynomial $P(S_{m_1,m_1},q)$ has the form 
(\ref{pgform}) with $j_{max}=4$.  The
lowest member of this family, the graph $S_{0,0}$, was shown in
Fig. \ref{s00n7}. The chromatic number $\chi(S_{m,m})=4$.  The expression 
for $P(S_{m,m},q)$ follows immediately from (\ref{ps}) and has $j_{max}=4$, 
\beq
P(S_{m,m},q) = c_{S,22}(\lambda_2)^{2m} + c_{S,33} 
+ 2c_{S,12}(\lambda_1\lambda_2)^m + 2c_{S,23}(\lambda_2\lambda_3)^m \ , 
\label{psmm}
\eeq
where we used the fact that $(\lambda_3)^{2m}=1$.  The ratio $r(S_{m,m})$
follows from (\ref{rs}), with the result that 
$r(S_{\infty,\infty}) = 0$.

\section{The Family $F_m$} 

\begin{figure}
  \begin{center}
    \includegraphics[height=6cm]{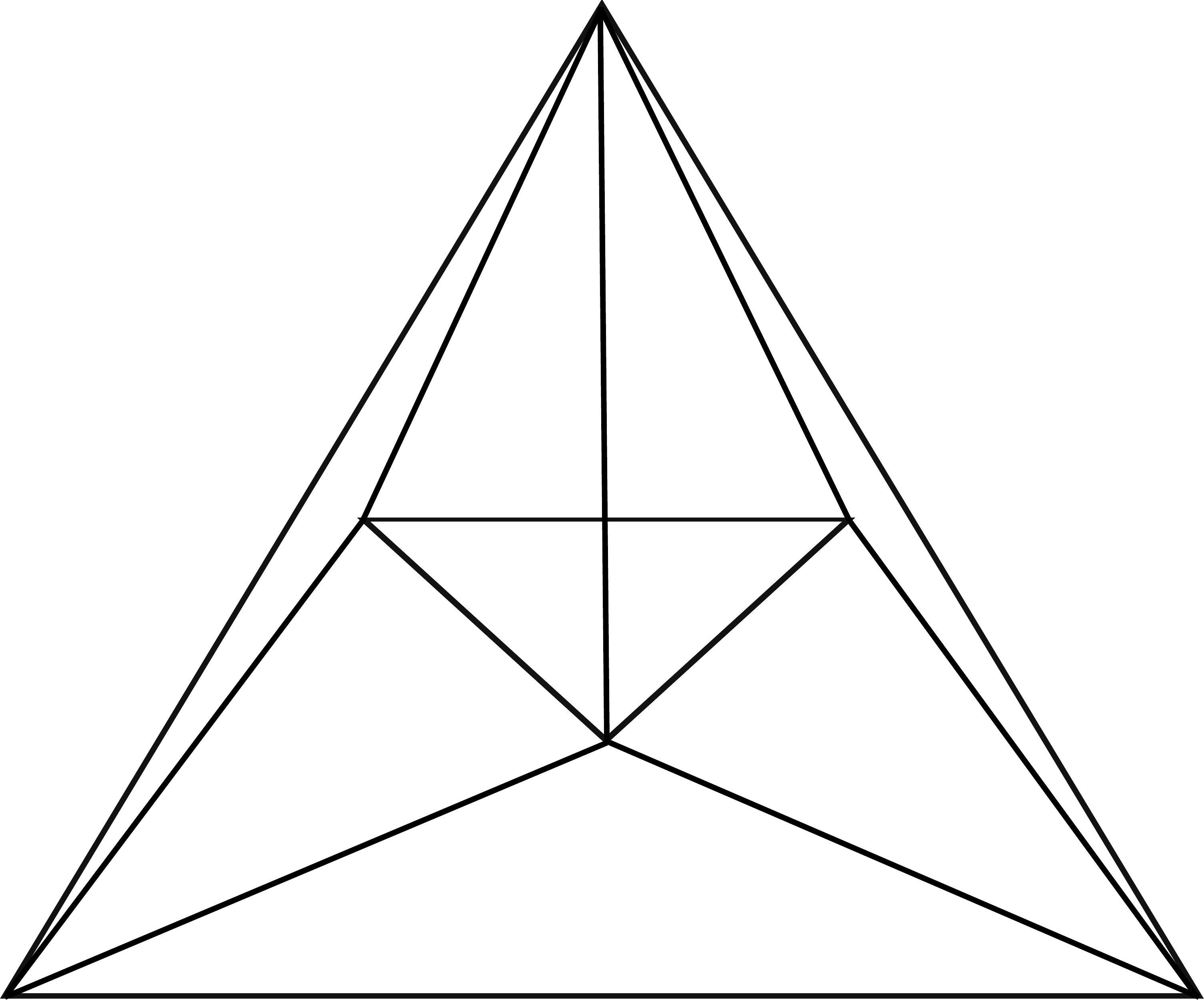}
  \end{center}
\caption{Graph $F_3$.}
\label{fm3n7}
\end{figure}

\begin{figure}
  \begin{center}
    \includegraphics[height=6cm]{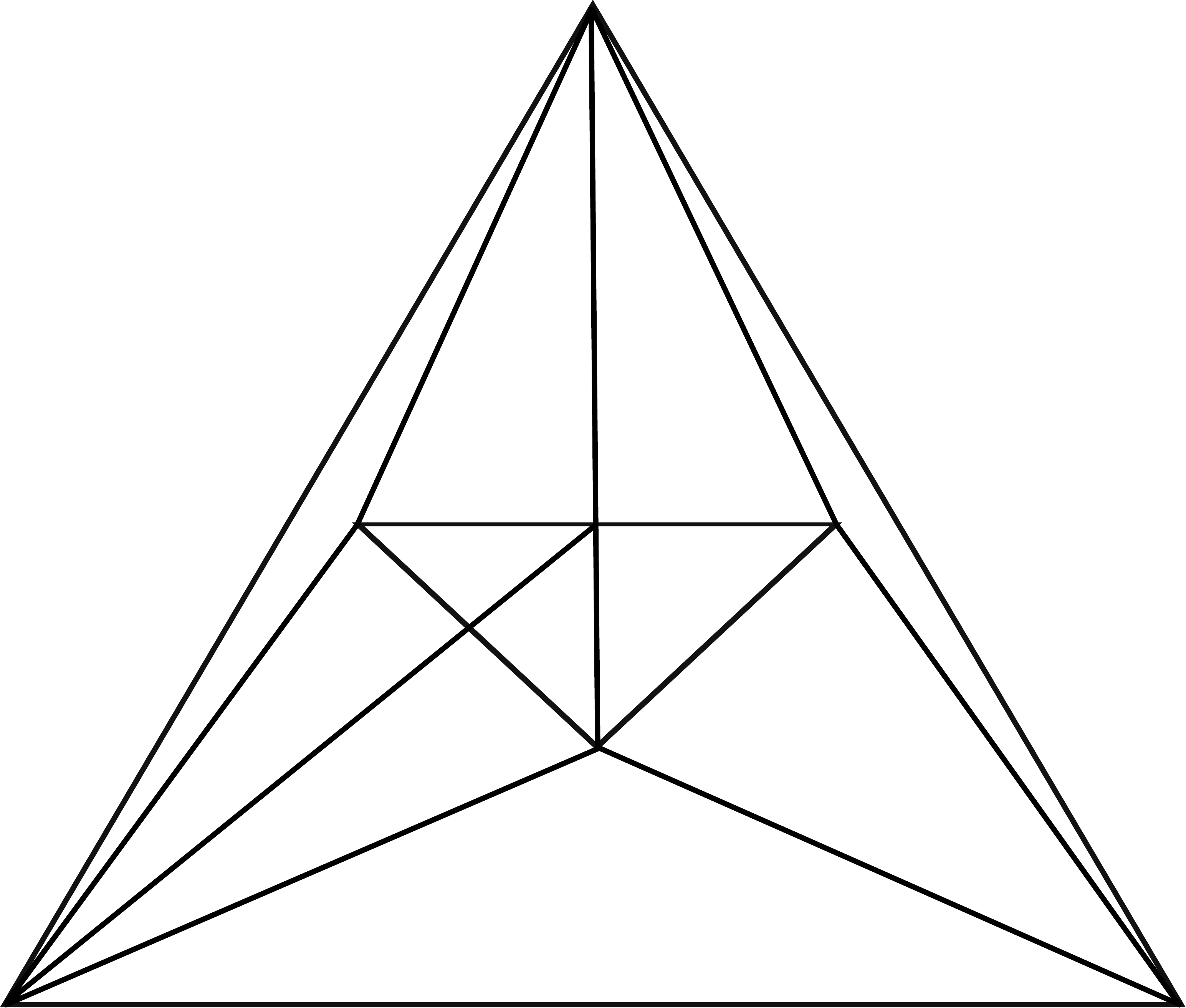}
  \end{center}
\caption{Graph $F_4$.}
\label{fm4n8}
\end{figure}

\begin{figure}
  \begin{center}
  \includegraphics[height=6cm]{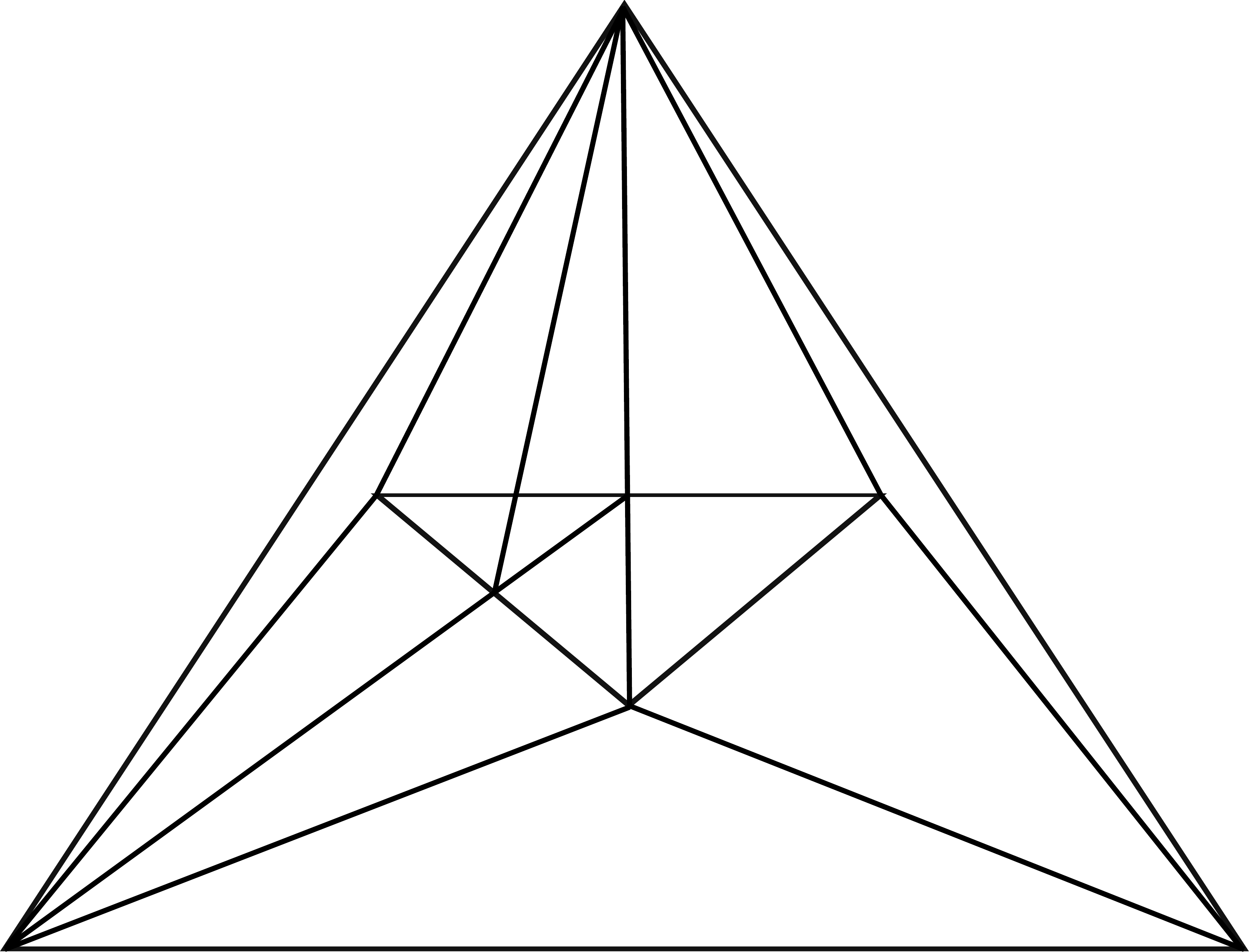}
  \end{center}
\caption{Graph $F_5$.}
\label{fm5n9}
\end{figure}

In this section we construct and study a family of planar triangulations,
denoted $F_m$, with the property that $P(F_m,q)$ has the form (\ref{pgform})
with $j_{max}=3$, but the $\lambda_{F,j}$ are not given by (\ref{lamform}), but
instead are roots of a certain cubic equation.  The number of vertices is
$n(F_m)=m+4$. This family is useful as a contrast to the other one-parameter
families of planar triangulations with chromatic polynomials of the form
(\ref{pgform}) that we have constructed.  The construction of members of this
family is somewhat more complicated than that of the other families analyzed in
this paper, and accordingly, for illustration we include several graphs, namely
$F_m$ with $m=3, \ 4, \ 5$, are shown in Figs.  \ref{fm3n7}-\ref{fm5n9}.  As
these show, starting from a given member $F_m$, one constructs the next higher
member $F_{m+1}$ in an interleaved manner, first adding a new set of edges one
of which emanates from the lower left-hand vertex of the graph, and then a new
set of edges one of which emanates from the uppermost vertex, and so forth.  We
note that in contrast to the previous planar triangulations with chromatic
polynomials of the form (\ref{pgfacform}), the degrees of the vertices remain
bounded as $m \to \infty$ for this family.
\begin{figure}
  \begin{center}
    \includegraphics[height=6cm]{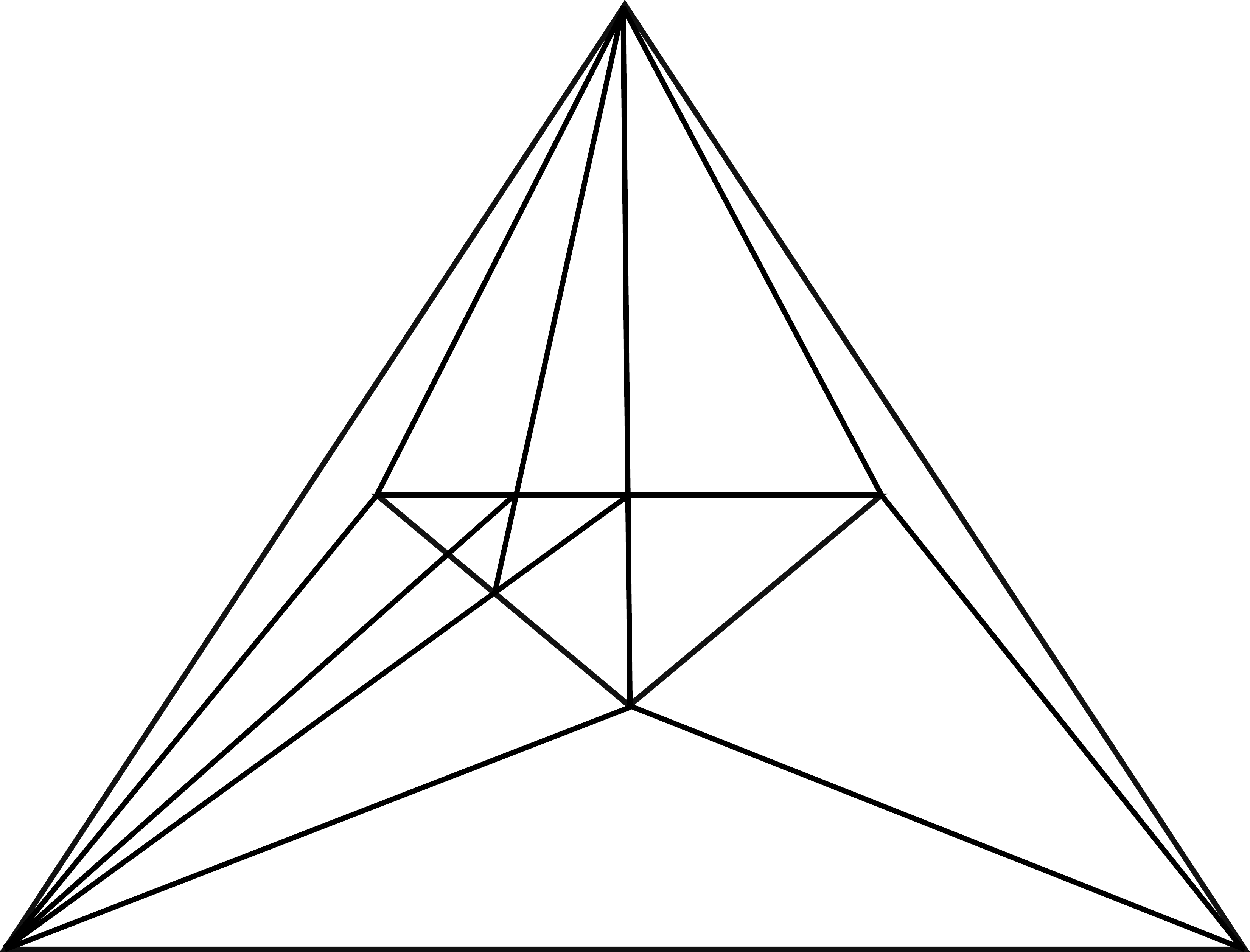}
  \end{center}
\caption{Graph $F_6$.}
\label{fm6n10}
\end{figure}
\begin{figure}
  \begin{center}
    \includegraphics[height=6cm]{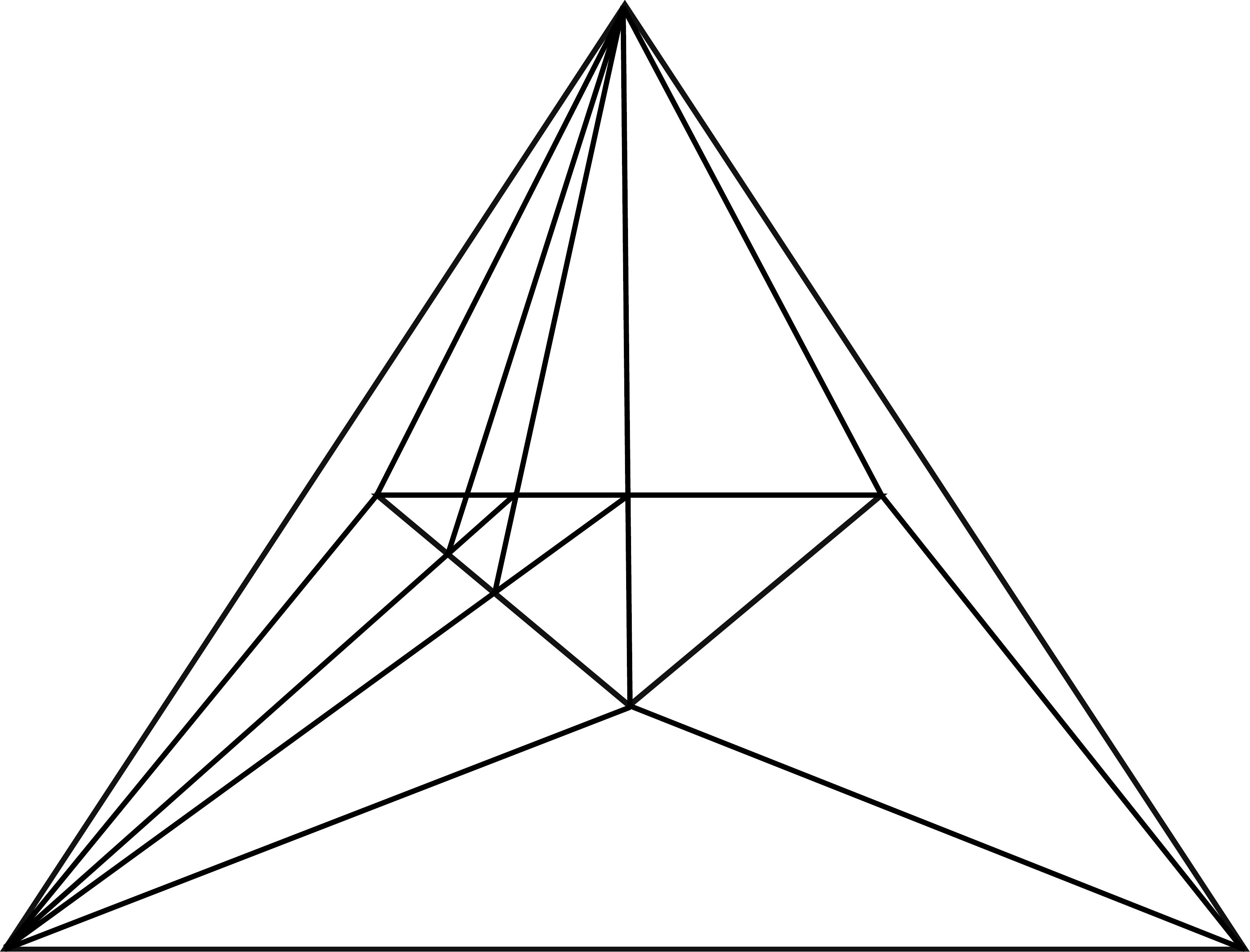}
  \end{center}
\caption{Graph $F_7$.}
\label{fm7n11}
\end{figure}

As in earlier works \cite{w,hs,a}, it is most convenient to express the 
$P(F_m,q)$ via a generating function, $\Gamma(F,q,x)$, which is a rational
function in $q$ and an auxiliary expansion variable $x$, of the form 
\beq
\Gamma(F,q,x) = \frac{{\cal N}(F,q,x)}{{\cal D}(F,q,x)} \ , 
\label{gammagen}
\eeq
where the numerator and denominator are
\beq
{\cal N}(F,q,x) = a_{F,0}+a_{F,1} \, x+a_{F,2} \, x^2
\label{gennum}
\eeq
and
\beq
{\cal D}(F,q,x) = 1 + b_{F,1} \, x+b_{F,2} \, x^2+b_{F,3} \, x^3 \ . 
\label{genden}
\eeq
with $a_{F,j}$ and $b_{F,j}$ being polynomials in $q$. The chromatic
polynomial $P(F_m,q)$ is then given as the coefficient in the Taylor series
expansion of this generating function:
\beq
\Gamma(F,q,x) = \sum_{m=0}^{\infty} \, P(F_{m+1},q) \, x^m
\label{gamtaylor}
\eeq

Using an iterative deletion-contraction method, we have determined this
generating function.  We find 
\beq
a_{F,0} = q(q-1)(q-2)(q-3)^2
\label{af0}
\eeq
\beq
a_{F,1} = q(q-1)(q-2)(2q-5)
\label{af1}
\eeq
\beq
a_{F,2} = q(q-1)(q-2)^2(q-3)^2
\label{af2}
\eeq
\beq
b_{F,1} = -(q-3) 
\label{bf1}
\eeq
\beq
b_{F,2} = q-3
\label{bf2}
\eeq
and
\beq
b_{F,3} = -(q-2)(q-3) \ . 
\label{bf3}
\eeq
The chromatic polynomial may also be expressed in the form of
Eq. (\ref{pgform}), with $j_{max}=3$, namely
\beq
P(F_m,q) = \sum_{j=1}^3 c_{F,j} \, (\lambda_{F,j})^m \ . 
\label{pfform}
\eeq
Using Eq (2.13) (or (2.15)) of Ref. \cite{hs}, one can calculate
the coefficients $c_{F,j}$ for $j=1,2,3$ from the generating function. 
Specifically, we have 
\beq
c_{F,1} = \frac{(a_{F,0}\lambda_{F,1}^2+a_{F,1}\lambda_{F,1}+a_{F,2})}
{(\lambda_{F,1}-\lambda_{F,2})(\lambda_{F,1}-\lambda_{F,3})}
\label{cf1}
\eeq
\beq
c_{F,2} = \frac{(a_{F,0}\lambda_{F,2}^2+a_{F,1}\lambda_{F,2}+a_{F,2})}
{(\lambda_{F,2}-\lambda_{F,1})(\lambda_{F,2}-\lambda_{F,3})}
\label{cf2}
\eeq
and
\beq
c_{F,3} = \frac{(a_{F,0}\lambda_{F,3}^2+a_{F,1}\lambda_{F,3}+a_{F,2})}
{(\lambda_{F,3}-\lambda_{F,1})(\lambda_{F,3}-\lambda_{F,2})} \ . 
\label{cf3} 
\eeq
As discussed before \cite{w,a,hs}, the $\lambda_{F,j}$s appear via the
factorized form of the denominator of the generating function,
\beq
{\cal D}(F,q,x) = \prod_{j=1}^3(1-\lambda_{F,j}x) \ . 
\label{facden}
\eeq
Equivalently, the $\lambda_{F,j}$s are determined from the equation
$\xi^3 + b_{F,1}\xi^2 + b_{F,2}\xi + b_{F,3}=0$, i.e., 
\beq
\xi^3 + (3-q)\xi^2 + (q-3)\xi -(q-2)(q-3) = 0 \ . 
\label{eqf}
\eeq
Let us define 
\beq
R_F = 3(4q^3-24q^2+76q-93) 
\label{rf}
\eeq
and
\beq
S_F=\bigg [4(q-3)\bigg ( 2q^2+6q-9 + 3\sqrt{R_F} \ \bigg ) \bigg ] ^{1/3} \ . 
\label{sf}
\eeq
With appropriate choices of branch cuts for the various fractional powers in
(\ref{sf}), we have 
\beq
\lambda_{F,1} = \frac{S_F}{6} + \frac{2(q-3)(q-6)}{3S_F} + \frac{q-3}{3} \ . 
\label{lamf1}
\eeq
The other $\lambda_{F,j}$, $j=2,3$ can be written explicitly in a similar
manner.  Thus, this family is valuable as an illustration of a family of planar
triangulation graphs with a chromatic polynomial of the form (\ref{pgform}) and
with $\lambda$ terms that are different from those in (\ref{lamform}) and,
indeed, are nonpolynomial, in contrast to the families with chromatic
polynomials of the form (\ref{pgform1}) or (\ref{pgfacform}).

With regard to the evaluation of $P(F_m,q)$ at $q=\tau+1$ (with 
an appropriate choice of branch cuts for the square and cube roots),
$\lambda_{F,1}$ and one of the other two roots of (\ref{eqf}) comprise the 
complex-conjugate pair
\beq
\frac{1}{4}\bigg [ -1+\sqrt{5} \ \pm (-38+18\sqrt{5} \, )^{1/2} \, i \bigg ]
\label{complexroots}
\eeq
with magnitude 0.485867.., while the third root of (\ref{eqf}) is equal to
$-1$.  Since 0.485867.. is less than $\tau-1=0.6180..$, the corresponding two
coefficients do not have to, and do not, vanish at $q=\tau+1$. Since the third
root has magnitude greater than $\tau-1$, its coefficient must vanish at
$q=\tau+1$ in order for $|P(F_m,\tau+1)|$ to obey the Tutte upper bound
(\ref{tub}).  With these values of the $\lambda_{F,j}$'s at $q=\tau+1$, the
ratio $r(F_m)$ vanishes (exponentially rapidly) as $m \to \infty$ and
$r(F_\infty)=0$.  This illustrates the general property that if $G_{pt,m}$ is a
family of planar triangulations with $P(G_{pt,m},q)$ of the form (\ref{pgform})
and $\alpha=1$ in (\ref{nmrel}), and if none of the $\lambda_{G_{pt},j}$ has
magnitude equal to $\tau-1$ when evaluated at $q=\tau+1$, then, since (i) the
$\lambda_{G_{pt},j}$ with $|\lambda_{G_{pt},j}| > \tau-1$ have coefficients
that must vanish, and (ii) the $\lambda_{G_{pt},j}$ with $|\lambda_{G_{pt},j}|
< \tau-1$ give zero contribution in the limit $m \to \infty$, it follows that
$r(G_{pt})=0$.  We calculate
\beq
a_F = 0.786151..
\label{af}
\eeq

The term $\lambda_{F,1}$ is real and positive and is dominant for $q > \tau+2 =
3.618...$.  In this interval, the other two roots, $\lambda_{F,j}$, $j=2,3$ are
complex, with smaller magnitudes.  At $q=\tau+2$, $\lambda_{F,1}=-1$ and
$|\lambda_{F,2}|=|\lambda_{F,3}|=1$, so all $\lambda_{F,j}$ are degenerate in
magnitude.  Hence, in the notation of \cite{w}, $q_c = \tau+2$ for this family.
At $q=3$, all $\lambda_{F,j}=0$, $j=1,2,3$, as is obvious from Eq.
(\ref{eqf}).

We exhibit the first few $P(F_m,q)$.  For $m=1$, $P(F_1,q)=a_{F,0}$, as given
above in (\ref{af0}). For $m=2$ to $m=6$,
\beq
P(F_2,q)=q(q-1)(q-2)(q^3-9q^2+29q-32)
\label{pf2n6}
\eeq
\beq
P(F_3,q)=q(q-1)(q-2)(q-3)(q^3-9q^2+30q-35)
\label{pf3n7}
\eeq
\beq
P(F_4,q)=q(q-1)(q-2)(q-3)(q^4-12q^3+58q^2-133q+119)
\label{pf4n8}
\eeq
\beqs
& & P(F_5,q)=q(q-1)(q-2)(q-3)(q^5-15q^4+95q^3-317q^2+553q-398) \cr\cr
& & 
\label{pf5n9}
\eeqs
\beqs
& & P(F_6,q)=q(q-1)(q-2)(q-3)^2(q^5-15q^4+96q^3-327q^2+591q-447) \cr\cr
& & 
\label{pf6n10}
\eeqs
As $m$ increases further, $P(F_m,q)$ has increasingly high powers of the factor
$(q-3)$.  

As with the other planar triangulation families, the $F_m$ family has chromatic
zeros near to $\tau+1$. We find that these approach $\tau+1$ as $m$ gets large.
Depending on the value of $m$, $P(F_m,q)$ also may have real zeros in the
interval $[q_w,3)$. The complex zeros of $P(F_m,q)$ form a complex-conjugate
arc, with arc endpoints at the complex zeros of $R_F$, namely $q, \ q^* \simeq
1.9111 \pm 2.6502i$.

\section{Some Implications for Statistical Physics}
\label{gsentropy}

One of the interesting aspects of the present work is its implications for
nonzero ground state entropy of the Potts antiferromagnet.  (For background on
the Potts model, see Refs. \cite{wurev,baxter87}, \cite{w}.) This stems from
the identity noted above, $P(G,q)= Z(G,q,T=0)_{PAF} = W_{tot}(G,q)$.  As above,
we denote the formal limit of a family of graphs $G$ as $n(G) \to \infty$ by
the symbol $\{ G \}$. We recall that the entropy per vertex is given by $S_0 =
k_B \ln W$, where $W$ is the degeneracy per vertex, related to the total
degeneracy of spin configurations of the zero-temperature Potts antiferromagnet
(or equivalently the number of proper $q$-colorings of the graph) by $W_{tot}$
by $W(\{ G \},q) = \lim_{n \to \infty} [W_{tot}(G,q)]^{1/n}$.  We refer the
reader to Ref. \cite{w} for a discussion of a subtlety in this definition
resulting from a certain noncommuativity that occurs for a special set of
values of $q$, denoted as $\{q_s \}$, namely
\beq
\lim_{q \to q_s} \lim_{n \to \infty} [P(G,q)]^{1/n} \ne
\lim_{n \to \infty} \lim_{q \to q_s} [P(G,q)]^{1/n} \ .
\label{wnoncom}
\eeq
For the one-parameter families of planar triangulations considered here, this
set of special values $\{q_s\}$ includes $q=0, \ 1, \ 2, \ \tau+1$ and, for
cases where the chromatic number is 4, also $q=3$.  Because of this
noncommutativity, it is necessary to specify the order of limits taken in
defining $W$. For a particular value $q=q_s$, we thus define
\beq
W_{qn}(\{ G \},q_s) = \lim_{q \to q_s} \lim_{n \to \infty} [P(G,q)]^{1/n}
\label{wqn}
\eeq
and
\beq
W_{nq}(\{ G \},q_s) = \lim_{n \to \infty} \lim_{q \to q_s} [P(G,q)]^{1/n} \ .
\label{wnq}
\eeq
For real $q \ge \chi(G_{pt,m})$, both of these definitions are equivalent, and
in this case we shall write $W_{qn}(\{ G \},q_s) = W_{nq}(\{ G \},q_s) \equiv
W( \{ G \},q_s)$.

We generalize our calculations in \cite{tub} as follows. First, for a family of
planar triangulations $G_{pt,m}$ with $P(G_{pt,m},q)$ having the form
({\ref{pgform}) with $j_{max}=1$, $W = [\lambda_{G_{pt}}]^{1/\alpha}$.  As an
example, consider the family
\beq
R_m = P_m + P_2 \ , 
\label{rm}
\eeq
with $n=m+2$ considered in \cite{tub}. Here $P_m$ is the path graph with $m$
vertices and the join $G+H$ of two graphs $G=(V_G,E_G)$ and $H=(V_H,E_H)$ is
defined as the graph with vertex set $V_{G+H}=V_G \cup V_H$ and edge set
$E_{G+H}$ comprised of the union of $E_G \cup E_H$ with the set of edges
obtained by connecting each vertex of $G$ with each vertex of $H$.  Thus,
$R_1=K_3$, $R_2=K_4$, etc.  The $W$ function for the $m \to \infty$ limit of
this family is $W(\{R \},q)=q-3$, and $S_0 > 0$ for $q > 4$.  For the family
$TC_m$
\beq
W(\{TC\},q)=(\lambda_{TC})^{1/3} \ . 
\label{wtc}
\eeq
where $\lambda_{TC}$ was given in (\ref{lamtc}). The function $\lambda_{TC}$ is
a monotonically increasing function of $q$, which passes through zero at
$q=2.54660...$ and increases through unity at $q=3$ so that (for the $m \to
\infty$ limit of this family) $S_0 > 0$ for $q > 3$.  For the family $I_m$ of
iterated icosahedra,
\beq
W(\{I\},q) = (\lambda_I)^{1/9}  \ , 
\label{wi}
\eeq
where
\beqs
\lambda_I & = & (q-3)(q^8-24q^7+260q^6-1670q^5+6999q^4-19698q^3 \cr\cr
          & + & 36408q^2 -40240q+20170) \ .
\label{lami}
\eeqs
The function $\lambda_I$ vanishes at three real values of $q$, namely
$q=2.618197..$ (i.e., slightly above $\tau+1$), at $q=3$, and at $q =
3.222458..$. This function is positive for $q > 3.222458..$ (as well as in an
interval $2.618.. < q < 3$) and increases through unity as $q$ increases
through the value 3.5133658.., so that in this latter interval, $S_0 > 0$.

A second general result is that for a family of planar triangulations
$G_{pt,m}$ with $P(G_{pt,m},q)$ having the form ({\ref{pgfacform}), it follows
that (i) $W_{qn}(\{ G_{pt} \},q) = q-2$ for $q > 3$; (ii) even in the presence
of noncommutativty, $W( \{ G_{pt} \},q) = q-2$ for $q \ge 4$, so that $S_0 > 0$
in this interval (and also in the interval $q > 3$ if one uses $W_{qn}(\{
G_{pt} \},q)$). This result applies, in particular, to the families 
$B_m$, $H_m$, $L_m$, $D_{m-4,2}$, and $D_{m-4,3}$. Although the family 
$P(D_{m,m},q)$ is of the form (\ref{pgform}) with 
$j_{max}=6$, the dominant term for
$q > \chi(D_{m,m})$ is again $q-2$, so that in this interval 
$W( \{ D_d \},q) = q-2$ for this family also.  In contrast, 
$P(S_{m_1,m_2},q)$ has $c_{S,11}=0$ and hence
lacks the term that would normally be dominant as $m_1$ or $m_2$ goes to
infinity. In this case, for $q \ge 4$ where there is no noncommutativity in
limits, we find $W( \{ S \},q) = \sqrt{(q-2)(q-3)}$ . 

For the family $F_m$, we find that $\lambda_1$ in Eq. (\ref{lamf1}) is dominant
for $q > q_c = \tau+2 = 3.618...$, so that in this region,
\beq
W(\{ F \}, q) = \lambda_{F,1}
\label{wf}
\eeq
Furthermore, since $\lambda_{F,1} > 1$ for real $q > \tau+2$, it follows that
$S_0 > 0$ for (the $m \to \infty$ limit of this family of graphs) for this
range $q > \tau+2$.

For a regular lattice graph $G$ it is of interest to investigate the dependence
of $W(\{ G \},q)$ on the vertex degree (coordination number) $d$.  This
study was carried out in \cite{w} \cite{w3,wn}, and it was shown that $W(\{ G
\},q)$ is a non-increasing function of $d$.  This is understood as being a
consequence of the fact that (except for tree graphs, which are not relevant
here), roughly speaking, increasing the vertex degree tends to increase the
constraints on a proper $q$-coloring of the vertices and therefore tends to
decrease $W(\{ G \},q)$. One is also motivated to investigate the same question
with the families of planar triangulations under study here.  However, since
$d_{eff}=6$ for a family of planar triangulation graphs one is limited to
a fixed $d_{eff}=6$ and hence cannot carry out the type of comparative
study involving a variation in $d_{eff}$ that was performed in \cite{w},
\cite{w3,wn}.  In \cite{tub} and the present work, we have found that families
of planar triangulations can have different $W(\{ G \},q)$ functions.  This is 
consistent with the results in \cite{w}-\cite{wn}.  Indeed, one has already
encountered examples of this.  For example, the square and kagom\'e $(3 \cdot 6
\cdot 3 \cdot 6$) lattices both have the same vertex degree, namely 4, but they
have different $W$ functions, and similarly, the honeycomb, $(3 \cdot 12^2)$,
and $(4 \cdot 8^2)$ lattices have the same vertex degree, namely 3, but again,
they have different $W$ functions \cite{w,w3,wn}.

\section{Comparative Discussion} 

In this section we give a comparative discussion of some limiting quantities
for the various families of planar triangulations that we have studied so
far. For one-parameter families of planar triangulations $G_{pt,m}$ for which
$P(G_{pt,m},q)$ is of the form (\ref{pgform1}) we have proved that
$r(G_{pt,\infty})=0$ and have investigated the various values of $a_{G_{pt}}$
defined in (\ref{a1}).  This constant is strictly less than unity, and it is of
interest to see which families yield larger and smaller values of $a_{G_{pt}}$.
We display the values that we have obtained in Table \ref{arproperties}. As is
evident, in the set of $j_{max}=1$ families of planar triangulations, the
family of cylindrical strips of the triangular lattice, $TC_m$ (equivalently,
iterated octahedra) yields the largest value of $a_{G_{pt}}$, which is within 9
\% of its upper bound of 1. In the $j_{max}=3$ families, the one that yields
the largest value of the limiting ratio $r(G_{pt,\infty})$ is the family,
$B_m$, with $r({B_\infty}) = 0.6180.$.  A second type of asymptotic limiting
function is $W(\{ G \},q)$.  We have given a comparative analysis of this in
the previous section.

We have also investigated the values of $P(G_{pt,m},q)$ at $q=\chi(G_{pt,m})$
for the families of planar triangulations that we have studied.  Recall the
definition that a graph $G$ is $k$-critical iff $\chi(G)=k$ and $P(G,k)=k!$.
We find a variety of behavior.  For example, (i) $\chi(R_m)=4$ and
$P(R_m,4)=4!$, so $R_m$ is 4-critical; (ii) $\chi(TC_m)=3$ and $P(TC_m,3)=3!$,
so $TC_m$ is 3-critical; but (iii) $\chi=4$ for $I_m$, $H_m$, $L_m$,
$D_{m-4,3}$, and $F_{m \ge 3}$ but none of these families is 4-critical.  For
other families $G_{pt,m}$, the chromatic number depends on whether $m$ is even
or odd.  For example, for even $m$, $\chi= 3$ for $B_m$ and $D_{m-4,2}$, and
these graphs are 3-critical, while for odd $m$, $\chi=4$ for $B_m$ and
$D_{m-4,2}$, but neither of these graphs is 4-critical.

\begin{table}[htbp]
\caption{\footnotesize{Some asymptotic limiting quantities for one-parameter
families of planar triangulations. The shorthand notation $3me,4mo$ means $\chi
= 3$ if $m$ is even and $\chi=4$ if $m$ is odd. Additional information for
$\chi$ values is $\chi(R_3)=3$, $\chi(F_1)=4$, and $\chi(F_2)=3$. Numerical values are quoted
to three significant figures.}}
\begin{center}
\begin{tabular}{|c|c|c|c|c|c|}
\hline\hline
$G_{pt,m}$ & $n(G_{pt,m})$  & $\chi(G_{pt,m})$ & $j_{max}$ 
& $r(G_{pt,\infty})$ & $a_{G_{pt}}$  \\
\hline\hline
$R_m$  & $m+2$ & 4 if $m\ge 2$ & 1 & 0 & $(-1+\sqrt{5})/2=0.618$  \\
$TC_m$ & $3m$  & 3             & 1 & 0 & $(3-\sqrt{5}\ )^{1/3}=0.914$ \\
$I_m$  & $9m+3$& 4             & 1 & 0 & $[(-315+141\sqrt{5} \ )/2]^{1/9}=
0.8055$ \\
$G_{CM,m}$ &$8m+3$ & 3         & 1 & 0 & $[(115-51\sqrt{5} \ )/2]^{1/8} = 
0.885$ \\
$F_m$ & $m+4$ &4 if $m\ge 3$   & 3 & 0                         & 0.786 \\
$B_m$  & $m+2$ & $3me,4mo$     & 3 & $(-1+\sqrt{5} \ )/2 = 0.618$ & 1 \\
$H_m$ & $m+5$ & 4        & 3 & $(7-3\sqrt{5} \ )/2 = 0.146$ & 1 \\
$L_m$ & $m+5$ & 4        & 3 & $-2+\sqrt{5}=0.236$          & 1 \\
$D_{m-4,0}$ & $m+5$ & $3me,4mo$     & 3 & $-4+2\sqrt{5}=0.472 $        & 1 \\ 
$D_{m-4,1}$ & $m+6$& 4              & 3 & $(-15+7\sqrt{5} \ )/2=0.326$ & 1 \\ 
$D_{0,m-2}$ & $m+7$ & $3me,4mo$     & 3 & $-13+6\sqrt{5}=0.416$        & 1 \\
$D_{1,m-2}$ & $m+8$ & 4             & 3 & $-22+10\sqrt{5}=0.361$       & 1 \\
$D_{m,m}$   & $2m+9$ &$3me,4mo$     & 6 & $(3-\sqrt{5} \ )/2 = 0.382$  & 1 \\
$S_{m,m}$   & $m+7$ & 4             & 4 & 0 & $(-1+\sqrt{5} \ )/2 = 0.618$ \\
\hline\hline
\end{tabular}
\end{center}
\label{arproperties}
\end{table}

There are a number of further directions in which the present research could be
extended.  One could, for example, study Tutte polynomials, or the equivalent,
Potts model partition functions $Z(G_{pt},q,v)$, for planar triangulation
graphs $G_{pt}$ (where $v$ is a temperature-like Boltzmann variable). However,
a number of the special properties that make the chromatic polynomials of these
graphs amenable to analysis do not generalize to the full Potts model partition
function.  These include, for example, the use of the complete-graph
intersection theorem and the property that the chromatiic polynomial has
$q(q-1)(q-2)$ as a factor.  For example, for the triangle graph $K_3$ itself,
the Potts model partition function $Z(K_3,q,v)=(q+v)^3+(q-1)v^3 =
q(q^2+3qv+3v^2+v^3)$ only has the factor $q$.  Another direction of
investigation would be to calculate weighted-set chromatic polynomials
\cite{ph,phs} and Potts model partition functions with an external magnetic
field that favors or disfavors a single value of $q$ or a set of $s$ such
values \cite{phs2} for planar triangulation graphs.  Work on this is underway.

\section{Conclusions}

In this paper we have presented an analysis of the structure and properties of
chromatic polynomials $P(G_{pt,\vec m},q)$ of families of planar triangulation
graphs $G_{pt,\vec m}$, where ${\vec m} = (m_1,...,m_p)$ is a vector of integer
parameters.  We have discussed a number of specific families with $p=1$ and
$p=2$. These planar triangulation graphs form a particularly attractive class
of graphs for the analysis of chromatic polynomials because of their special
properties.  One of these is the fact that when evaluated at $q=\tau+1$, the
chromatic polynomial of a planar triangulation graph satisfies the Tutte upper
bound (\ref{tub}).  We have studied the ratio of $|P(G_{pt,\vec m},\tau+1)|$ to
the Tutte upper bound $(\tau-1)^{n-5}$ and have calculated limiting values of
this ratio as $n \to \infty$ for various families of planar triangulations.  We
also have used our calculations to study zeros of these chromatic polynomials.
Among our results, we have shown that if $G_{pt,\vec m}$ is a planar
triangulation graph with a chromatic polynomial $P(G_{pt,\vec m},q)$ of the
form (\ref{pgmvector}), then (i) the coefficients $c_{_{G_{pt}},\vec i}$ must
satisfy a number of properties, which we have derived; and (ii) $P(G_{pt,\vec
m},q)$ has a real chromatic zero that approaches $(1/2)(3+\sqrt{5} \ )$ as one
or more $m_i \to\infty$.  We have constructed a $p=1$ family of planar
triangulations with real zeros that approach 3 from below as $m \to \infty$.  A
one-parameter family $F_m$ with $j_{max}=3$ and nonpolynomial $\lambda_{F,j}$
has been studied.  We have also presented results for a number of results for
chromatic polynomials of various two-parameter families of planar
triangulations. Implications for the ground-state entropy of the Potts
antiferromagnet are discussed.  Our results are of interest both from the point
of view of mathematical graph theory and statistical physics and further show
the fruitful connections between these fields.

\begin{acknowledgments}

This research was partially supported by the grant NSF-PHY-09-69739.  We thank
Prof. J. I. Brown for a discussion of \cite{jasonbrown98} and 
Prof. D. R. Woodall for the private communication \cite{woodallpriv}.

\end{acknowledgments}

\end{document}